\documentclass[11pt,a4paper]{article}
\pdfoutput=1

\usepackage{jheppub-nosort,amsthm}

\PassOptionsToPackage{svgnames}{xcolor}
\usepackage{xcolor}
\usepackage{amsmath,amssymb,amsthm,amscd}
\usepackage{physics}
\usepackage{tikz}
\usetikzlibrary{decorations.pathmorphing}
\tikzset{node distance=2cm, auto}
\tikzset{snake it/.style={decorate, decoration=snake}}
\usepackage{epsfig}
\usepackage{psfrag,comment}
\usepackage{graphicx}
\usepackage{caption}
\usepackage{subcaption}
\usepackage{mathrsfs}
\usepackage[normalem]{ulem}
\usepackage{bm}
\usepackage{lscape}
\usepackage{diagbox}
\usepackage{upgreek}

\usepackage{url}
\newcommand\doilink[1]{\href{http://dx.doi.org/#1}{#1}}
\newcommand\arxivlink[1]{\href{http://arxiv.org/abs/#1}{#1}}

\newcommand{\CA}{{\mathcal A}}

\newcommand{\CC}{{\mathcal C}}

\newcommand{\CF}{{\mathcal F}}

\newcommand{\CI}{{\mathcal I}}

\newcommand{\CN}{{\mathcal N}}
\newcommand{\CO}{{\mathcal O}}

\newcommand{\CT}{{\mathcal T}}

\newcommand{\CZ}{{\mathcal Z}}

\newcommand{\NCC}{{\mathscr C}}

\def\BZ{{\mathbb Z}}

\def\BC{{\mathbb C}}
\def\BP{{\mathbb P}}

\newcommand{\be}{\begin{equation}}
\newcommand{\ee}{\end{equation}}
\newcommand{\ba}{\begin{aligned}}
\newcommand{\ea}{\end{aligned}}
\newcommand{\bea}{\begin{eqnarray}}
\newcommand{\eea}{\end{eqnarray}}
\newcommand{\bean}{\begin{eqnarray*}}
\newcommand{\eean}{\end{eqnarray*}}

\def\r{\right\rangle}

\def\1{\mathbf{1}}
\def\0{|\1\r}

\newcommand{\rme}{{\mathrm{e}}}
\newcommand{\rmi}{{\mathrm{i}}}
\newcommand{\rmd}{{\mathrm{d}}}

\def\XXint#1#2#3{{\setbox0=\hbox{$#1{#2#3}{\int}$}
     \vcenter{\hbox{$#2#3$}}\kern-.5\wd0}}

\makeatletter
\newsavebox\myboxA
\newsavebox\myboxB
\newlength\mylenA

\newcommand*\widebar[2][0.75]{%
    \sbox{\myboxA}{$\m@th#2$}%
    \setbox\myboxB\null
    \ht\myboxB=\ht\myboxA%
    \dp\myboxB=\dp\myboxA%
    \wd\myboxB=#1\wd\myboxA
    \sbox\myboxB{$\m@th\overline{\copy\myboxB}$}
    \setlength\mylenA{\the\wd\myboxA}
    \addtolength\mylenA{-\the\wd\myboxB}%
    \ifdim\wd\myboxB<\wd\myboxA%
       \rlap{\hskip 0.8\mylenA\usebox\myboxB}{\usebox\myboxA}%
    \else
        \hskip -0.5\mylenA\rlap{\usebox\myboxA}{\hskip 0.5\mylenA\usebox\myboxB}%
    \fi}
\makeatother

\newdimen\tableauside\tableauside=1.0ex
\newdimen\tableaurule\tableaurule=0.4pt
\newdimen\tableaustep
\def\phantomhrule#1{\hbox{\vbox to0pt{\hrule height\tableaurule width#1\vss}}}
\def\phantomvrule#1{\vbox{\hbox to0pt{\vrule width\tableaurule height#1\hss}}}
\def\sqr{\vbox{%
  \phantomhrule\tableaustep
  \hbox{\phantomvrule\tableaustep\kern\tableaustep\phantomvrule\tableaustep}%
  \hbox{\vbox{\phantomhrule\tableauside}\kern-\tableaurule}}}
\def\squares#1{\hbox{\count0=#1\noindent\loop\sqr
  \advance\count0 by-1 \ifnum\count0>0\repeat}}
\def\tableau#1{\vcenter{\offinterlineskip
  \tableaustep=\tableauside\advance\tableaustep by-\tableaurule
  \kern\normallineskip\hbox
    {\kern\normallineskip\vbox
      {\gettableau#1 0 }%
     \kern\normallineskip\kern\tableaurule}%
  \kern\normallineskip\kern\tableaurule}}
\def\gettableau#1{\ifnum#1=0\let\next=\null\else
\squares{#1}\let\next=\gettableau\fi\next}
\definecolor{cottoncandy}{rgb}{1.0, 0.74, 0.85}
\definecolor{cornellred}{rgb}{0.7, 0.11, 0.11}
\definecolor{darktangerine}{rgb}{1.0, 0.66, 0.07}
\newcommand{\picscale}{0.8}

\newcommand{\ZPertpe}{\mathord{
\begin{tikzpicture}[baseline={([yshift=-0.6ex]current bounding box.center)}, line width=1, scale=\picscale]
\draw[->] (-1,0) -- (1,0);
\draw[->] (0,-1) -- (0,1);
\draw[gray] (-1,1) -- (1,1);
\draw[gray] (1,1) -- (1,-1);
\draw[gray] (1,-1) -- (-1,-1);
\draw[gray] (-1,-1) -- (-1,1);
\fill[red] (0,0) circle (1pt);
\fill[red] (1/2,0) circle (1pt);
\fill[red] (-1/2,0) circle (1pt);
\draw[blue] (-9/10, -9/10) to[bend right=25] (-1/2, -1/10);
\draw[blue] (-1/2, -1/10) -- (1/2, 1/10);
\draw[blue, ->] (1/2, 1/10) to[bend left=25] (9/10, 9/10);
\end{tikzpicture}}}

\newcommand{\ZPertme}{\mathord{
\begin{tikzpicture}[baseline={([yshift=-0.6ex]current bounding box.center)}, line width=1, scale=\picscale]
\draw[->] (-1,0) -- (1,0);
\draw[->] (0,-1) -- (0,1);
\draw[gray] (-1,1) -- (1,1);
\draw[gray] (1,1) -- (1,-1);
\draw[gray] (1,-1) -- (-1,-1);
\draw[gray] (-1,-1) -- (-1,1);
\fill[red] (0,0) circle (1pt);
\fill[red] (1/2,0) circle (1pt);
\fill[red] (-1/2,0) circle (1pt);
\draw[blue] (-1/2, 1/10) to[bend right=25] (-9/10, 9/10);
\draw[blue] (-1/2, 1/10) -- (1/2, -1/10);
\draw[blue, ->] (1/2, -1/10) to[bend right=25] (9/10, -9/10);
\end{tikzpicture}}}

\newcommand{\ZPertpipe}{\mathord{
\begin{tikzpicture}[baseline={([yshift=-0.6ex]current bounding box.center)}, line width=1, scale=\picscale]
\draw[->] (-1,0) -- (1,0);
\draw[->] (0,-1) -- (0,1);
\draw[gray] (-1,1) -- (1,1);
\draw[gray] (1,1) -- (1,-1);
\draw[gray] (1,-1) -- (-1,-1);
\draw[gray] (-1,-1) -- (-1,1);
\fill[red] (0,0) circle (1pt);
\fill[red] (1/2,0) circle (1pt);
\fill[red] (-1/2,0) circle (1pt);
\draw[blue, ->] (1/10, -9/10) -- (-1/10, 9/10);
\end{tikzpicture}}}

\newcommand{\ZPertpime}{\mathord{
\begin{tikzpicture}[baseline={([yshift=-0.6ex]current bounding box.center)}, line width=1, scale=\picscale]
\draw[->] (-1,0) -- (1,0);
\draw[->] (0,-1) -- (0,1);
\draw[gray] (-1,1) -- (1,1);
\draw[gray] (1,1) -- (1,-1);
\draw[gray] (1,-1) -- (-1,-1);
\draw[gray] (-1,-1) -- (-1,1);
\fill[red] (0,0) circle (1pt);
\fill[red] (1/2,0) circle (1pt);
\fill[red] (-1/2,0) circle (1pt);
\draw[blue, ->] (-1/10, -9/10) -- (1/10, 9/10);
\end{tikzpicture}}}

\newcommand{\ZPertthreepipe}{\mathord{
\begin{tikzpicture}[baseline={([yshift=-0.6ex]current bounding box.center)}, line width=1, scale=\picscale]
\draw[->] (-1,0) -- (1,0);
\draw[->] (0,-1) -- (0,1);
\draw[gray] (-1,1) -- (1,1);
\draw[gray] (1,1) -- (1,-1);
\draw[gray] (1,-1) -- (-1,-1);
\draw[gray] (-1,-1) -- (-1,1);
\fill[red] (0,0) circle (1pt);
\fill[red] (1/2,0) circle (1pt);
\fill[red] (-1/2,0) circle (1pt);
\draw[blue, <-] (1/10, -9/10) -- (-1/10, 9/10);
\end{tikzpicture}}}

\newcommand{\ZPertthreepime}{\mathord{
\begin{tikzpicture}[baseline={([yshift=-0.6ex]current bounding box.center)}, line width=1, scale=\picscale]
\draw[->] (-1,0) -- (1,0);
\draw[->] (0,-1) -- (0,1);
\draw[gray] (-1,1) -- (1,1);
\draw[gray] (1,1) -- (1,-1);
\draw[gray] (1,-1) -- (-1,-1);
\draw[gray] (-1,-1) -- (-1,1);
\fill[red] (0,0) circle (1pt);
\fill[red] (1/2,0) circle (1pt);
\fill[red] (-1/2,0) circle (1pt);
\draw[blue, <-] (-1/10, -9/10) -- (1/10, 9/10);
\end{tikzpicture}}}

\newcommand{\ZPerttwopipe}{\mathord{
\begin{tikzpicture}[baseline={([yshift=-0.6ex]current bounding box.center)}, line width=1, scale=\picscale]
\draw[->] (-1,0) -- (1,0);
\draw[->] (0,-1) -- (0,1);
\draw[gray] (-1,1) -- (1,1);
\draw[gray] (1,1) -- (1,-1);
\draw[gray] (1,-1) -- (-1,-1);
\draw[gray] (-1,-1) -- (-1,1);
\fill[red] (0,0) circle (1pt);
\fill[red] (1/2,0) circle (1pt);
\fill[red] (-1/2,0) circle (1pt);
\draw[blue, <-] (-9/10, -9/10) to[bend right=25] (-1/2, -1/10);
\draw[blue] (-1/2, -1/10) -- (1/2, 1/10);
\draw[blue] (1/2, 1/10) to[bend left=25] (9/10, 9/10);
\end{tikzpicture}}}

\newcommand{\ZPerttwopime}{\mathord{
\begin{tikzpicture}[baseline={([yshift=-0.6ex]current bounding box.center)}, line width=1, scale=\picscale]
\draw[->] (-1,0) -- (1,0);
\draw[->] (0,-1) -- (0,1);
\draw[gray] (-1,1) -- (1,1);
\draw[gray] (1,1) -- (1,-1);
\draw[gray] (1,-1) -- (-1,-1);
\draw[gray] (-1,-1) -- (-1,1);
\fill[red] (0,0) circle (1pt);
\fill[red] (1/2,0) circle (1pt);
\fill[red] (-1/2,0) circle (1pt);
\draw[blue, ->] (-1/2, 1/10) to[bend right=25] (-9/10, 9/10);
\draw[blue] (-1/2, 1/10) -- (1/2, -1/10);
\draw[blue] (1/2, -1/10) to[bend right=25] (9/10, -9/10);
\end{tikzpicture}}}

\newcommand{\ZSadpzero}{\mathord{
\begin{tikzpicture}[baseline={([yshift=-0.6ex]current bounding box.center)}, line width=1, scale=\picscale]
\draw[->] (-1,0) -- (1,0);
\draw[->] (0,-1) -- (0,1);
\draw[gray] (-1,1) -- (1,1);
\draw[gray] (1,1) -- (1,-1);
\draw[gray] (1,-1) -- (-1,-1);
\draw[gray] (-1,-1) -- (-1,1);
\fill[red] (0,0) circle (1pt);
\fill[red] (1/2,0) circle (1pt);
\fill[red] (-1/2,0) circle (1pt);
\draw[blue] (-9/10, -9/10) to[bend right=25] (-1/2, 0);
\draw[blue, ->] (-1/2, 0) to[bend right=25] (-9/10, 9/10);
\draw[blue, ->] (1/2, 0) to[bend left=25] (9/10, 9/10);
\draw[blue] (1/2, 0) to[bend right=25] (9/10, -9/10);
\end{tikzpicture}}}

\newcommand{\ZSuperSaddles}{\mathord{
\begin{tikzpicture}[baseline={([yshift=-0.6ex]current bounding box.center)}, line width=1, scale=\picscale]
\draw[->] (-1,0) -- (1,0);
\draw[->] (0,-1) -- (0,1);
\draw[gray] (-1,1) -- (1,1);
\draw[gray] (1,1) -- (1,-1);
\draw[gray] (1,-1) -- (-1,-1);
\draw[gray] (-1,-1) -- (-1,1);
\fill[red] (0,0) circle (1pt);
\fill[red] (1/2,0) circle (1pt);
\fill[red] (-1/2,0) circle (1pt);
\draw[blue] (-9/10, -9/10) to[bend right=25] (-1/2, -1/10);
\draw[blue] (-1/2, -1/10) -- (1/2, 1/10);
\draw[blue, ->] (1/2, 1/10) to[bend left=25] (9/10, 9/10);
\draw[blue, ->] (-1/2, 1/10) to[bend right=25] (-9/10, 9/10);
\draw[blue] (-1/2, 1/10) -- (1/2, -1/10);
\draw[blue] (1/2, -1/10) to[bend right=25] (9/10, -9/10);
\end{tikzpicture}}}

\newcommand{\ZPertBarme}{\mathord{
\begin{tikzpicture}[baseline={([yshift=-0.6ex]current bounding box.center)}, line width=1, scale=\picscale]
\draw[->] (-1,0) -- (1,0);
\draw[->] (0,-1) -- (0,1);
\draw[gray] (-1,1) -- (1,1);
\draw[gray] (1,1) -- (1,-1);
\draw[gray] (1,-1) -- (-1,-1);
\draw[gray] (-1,-1) -- (-1,1);
\draw[black] (-9/10, 11/10) -- (9/10, 11/10);
\fill[red] (0,0) circle (1pt);
\fill[red] (1/2,0) circle (1pt);
\fill[red] (-1/2,0) circle (1pt);
\draw[orange, ->] (-1/10, -9/10) -- (1/10, 9/10);
\end{tikzpicture}}}

\newcommand{\ZSadBarpe}{\mathord{
\begin{tikzpicture}[baseline={([yshift=-0.6ex]current bounding box.center)}, line width=1, scale=\picscale]
\draw[->] (-1,0) -- (1,0);
\draw[->] (0,-1) -- (0,1);
\draw[gray] (-1,1) -- (1,1);
\draw[gray] (1,1) -- (1,-1);
\draw[gray] (1,-1) -- (-1,-1);
\draw[gray] (-1,-1) -- (-1,1);
\draw[black] (-9/10, 11/10) -- (9/10, 11/10);
\fill[red] (0,0) circle (1pt);
\fill[red] (1/2,0) circle (1pt);
\fill[red] (-1/2,0) circle (1pt);
\draw[orange, <-] (-9/10, 1/20) to[bend right=5] (-1/4, 1/10);
\draw[orange] (-1/4, 1/10) to[bend right=10] (-1/8, 9/10);
\draw[orange] (1/4, -1/10) to[bend left=5] (9/10, -1/20);
\draw[orange, ->] (1/4, -1/10) to[bend right=10] (1/8, -9/10);
\end{tikzpicture}}}

\newcommand{\ZSadBarme}{\mathord{
\begin{tikzpicture}[baseline={([yshift=-0.6ex]current bounding box.center)}, line width=1, scale=\picscale]
\draw[->] (-1,0) -- (1,0);
\draw[->] (0,-1) -- (0,1);
\draw[gray] (-1,1) -- (1,1);
\draw[gray] (1,1) -- (1,-1);
\draw[gray] (1,-1) -- (-1,-1);
\draw[gray] (-1,-1) -- (-1,1);
\draw[black] (-9/10, 11/10) -- (9/10, 11/10);
\fill[red] (0,0) circle (1pt);
\fill[red] (1/2,0) circle (1pt);
\fill[red] (-1/2,0) circle (1pt);
\draw[orange, <-] (-9/10, -1/20) to[bend left=5] (-1/4, -1/10);
\draw[orange] (-1/4, -1/10) to[bend left=10] (-1/8, -9/10);
\draw[orange] (1/4, 1/10) to[bend right=5] (9/10, 1/20);
\draw[orange, ->] (1/4, 1/10) to[bend left=10] (1/8, 9/10);
\end{tikzpicture}}}

\newcommand{\ZSadBarpipe}{\mathord{
\begin{tikzpicture}[baseline={([yshift=-0.6ex]current bounding box.center)}, line width=1, scale=\picscale]
\draw[->] (-1,0) -- (1,0);
\draw[->] (0,-1) -- (0,1);
\draw[gray] (-1,1) -- (1,1);
\draw[gray] (1,1) -- (1,-1);
\draw[gray] (1,-1) -- (-1,-1);
\draw[gray] (-1,-1) -- (-1,1);
\draw[black] (-9/10, 11/10) -- (9/10, 11/10);
\fill[red] (0,0) circle (1pt);
\fill[red] (1/2,0) circle (1pt);
\fill[red] (-1/2,0) circle (1pt);
\draw[orange, <-] (-9/10, -9/10) to[bend right=25] (-1/2, 0);
\draw[orange] (-1/2, 0) to[bend right=25] (-9/10, 9/10);
\draw[orange] (1/2, 0) to[bend left=25] (9/10, 9/10);
\draw[orange, ->] (1/2, 0) to[bend right=25] (9/10, -9/10);
\end{tikzpicture}}}

\newcommand{\ZPertBarpipe}{\mathord{
\begin{tikzpicture}[baseline={([yshift=-0.6ex]current bounding box.center)}, line width=1, scale=\picscale]
\draw[->] (-1,0) -- (1,0);
\draw[->] (0,-1) -- (0,1);
\draw[gray] (-1,1) -- (1,1);
\draw[gray] (1,1) -- (1,-1);
\draw[gray] (1,-1) -- (-1,-1);
\draw[gray] (-1,-1) -- (-1,1);
\draw[black] (-9/10, 11/10) -- (9/10, 11/10);
\fill[red] (0,0) circle (1pt);
\fill[red] (1/2,0) circle (1pt);
\fill[red] (-1/2,0) circle (1pt);
\draw[orange, <-] (-9/10, -9/10) to[bend right=25] (-1/2, -1/10);
\draw[orange] (-1/2, -1/10) -- (1/2, 1/10);
\draw[orange] (1/2, 1/10) to[bend left=25] (9/10, 9/10);
\end{tikzpicture}}}

\newcommand{\ZPertBarpime}{\mathord{
\begin{tikzpicture}[baseline={([yshift=-0.6ex]current bounding box.center)}, line width=1, scale=\picscale]
\draw[->] (-1,0) -- (1,0);
\draw[->] (0,-1) -- (0,1);
\draw[gray] (-1,1) -- (1,1);
\draw[gray] (1,1) -- (1,-1);
\draw[gray] (1,-1) -- (-1,-1);
\draw[gray] (-1,-1) -- (-1,1);
\draw[black] (-9/10, 11/10) -- (9/10, 11/10);
\fill[red] (0,0) circle (1pt);
\fill[red] (1/2,0) circle (1pt);
\fill[red] (-1/2,0) circle (1pt);
\draw[orange, ->] (-1/2, 1/10) to[bend right=25] (-9/10, 9/10);
\draw[orange] (-1/2, 1/10) -- (1/2, -1/10);
\draw[orange] (1/2, -1/10) to[bend right=25] (9/10, -9/10);
\end{tikzpicture}}}

\title{New Instantons for Matrix Models}

\author[a]{Marcos~Mari\~no,}
\affiliation[a]{D\'epartement de Physique Th\'eorique et Section de Math\'ematiques,\\ Universit\'e de Gen\`eve, CH-1211 Gen\`eve, Switzerland\\}
\emailAdd{marcos.marino@unige.ch}

\author[b,c]{Ricardo~Schiappa,}
\affiliation[b]{Isaac Newton Institute for Mathematical Sciences,\\ University of Cambridge, Cambridge CB3 0EH, United Kingdom\\}
\affiliation[c]{CAMGSD, Departamento de Matem\'atica, Instituto Superior T\'ecnico,\\ Universidade de Lisboa, 1049-001 Lisboa, Portugal\\}
\emailAdd{ricardo.schiappa@}

\author[b,c]{Maximilian~Schwick\,}
\emailAdd{maximilian.schwick@tecnico.ulisboa.pt}

\abstract{The complete, nonperturbative content of random matrix models is described by resurgent-transseries---general solutions to their corresponding string-equations. These transseries include exponentially-suppressed multi-instanton amplitudes obtained by eigenvalue tunneling, but they also contain exponentially-enhanced and mixed instanton-like sectors with no known matrix model interpretation. This work shows how these sectors can be also described by eigenvalue tunneling in matrix models---but on the \textit{non-physical sheet} of the spectral curve describing their large-$N$ limit. This picture further explains the full resurgence of random matrices via analysis of all possible eigenvalue integration-contours. How to calculate these ``anti'' eigenvalue-tunneling amplitudes is explained in detail and in various examples, such as the cubic and quartic matrix models, and their double-scaling limit to Painlev\'e~I. This further provides direct matrix-model derivations of their resurgent Stokes data, which were recently obtained by different techniques.}


\arxivnumber{2210.13479}

\begin{document}

\maketitle

\vfill

\eject

\allowdisplaybreaks

\section{Introduction and Summary}\label{sec:intro}

In quantum theories, perturbation series in the coupling constant are usually insufficient to describe generic observables, and sometimes they even qualitatively fail to capture the relevant physics. The $1/N$ expansion (see, \textit{e.g.}, \cite{bw94}) is not any different in that respect, and it is in fact expected to have corrections which are exponentially small in $N$. In some cases, these corrections are of the instanton type, \textit{i.e.}, they are based on saddle-point approximations, and for this reason they are sometimes called large $N$ instantons (see, \textit{e.g.}, \cite{m15} for an introductory survey). 

The study of large $N$ instantons in general quantum theories is very difficult. The situation improves in zero dimensions, where the path integral reduces to either a vector or a matrix integral. The case of vector models is simple \cite{hb79} and will not be further addressed in this paper. In the case of matrix models, large $N$ instantons are much richer but still tractable. In many cases they may be obtained via a generalization of saddle-point analysis for one-dimensional integrals: once the integration over matrices has been reduced to an integration over eigenvalues, different instanton sectors are obtained by considering a large number of eigenvalues around a reference saddle, and a small number of eigenvalues around a different (nonperturbative) saddle. For this reason, these instantons are often said to arise via \textit{eigenvalue tunneling} \cite{d91, d92}. 

Across the years, the study of large $N$ instantons in matrix models has led to many interesting applications. These nonperturbative effects describe the large-order behavior of the $1/N$ expansion \cite{m12}. They can be also used to study large deviations in statistical ensembles of random matrices \cite{ms13, az13} and in the study of phase transitions in unitary matrix models \cite{n81, m08}.  

Large $N$ instantons also play an important role in holographic dualities between matrix models and string/gravity theories. The oldest dualities of this type involve double-scaled hermitian matrix-models, which describe non-critical strings \cite{gm90a, ds90, bk90, d90, gm90b}. In this case, large $N$ instantons obtained by eigenvalue tunneling correctly capture the large-order growth of the corresponding perturbative sectors \cite{gz90b, gz91, ez93}, and correspond to nonperturbative effects in non-critical string theory realized by D-branes. These models are in addition ``solvable'' in terms of differential or difference equations, sometimes called string equations, which yield their full perturbative data (see, \textit{e.g.}, \cite{g91, gm93, dgz93} for reviews). In some cases, the string equations reduce to a nonlinear ordinary differential equation (ODE) (\textit{e.g.}, in the case of pure gravity one finds the famous Painlev\'e~I equation). These ODE's have solutions of the transseries form, \textit{i.e.}, they include a perturbative, factorially divergent part, and an infinite series of exponentially small corrections that are expected to describe multi-instantons. However, some of the nonperturbative sectors appearing in these transseries are exponentially large and do not have a known matrix model realization. This is the problem that we want to address in this paper. Let us explain this in some detail. 

It was first pointed out in \cite{gikm10} that the transseries solution of the Painlev\'e~I equation is \textit{resonant}, \textit{i.e.}, the corresponding instanton actions always appear in \textit{symmetric} pairs. If we denote such action by $A$, one finds in fact a doubly-infinite set of multi-instantons, labelled by two non-negative integers $n$ and $m$, and the corresponding transseries has transmonomial components of the form 
\be
\sim \exp \left\{ -\left(n-m\right) \frac{A}{g_{\text{s}}} \right\}.
\ee
\noindent
Herein, $g_{\text{s}}$ is the expansion variable (it can be identified with the string coupling constant or with $1/N$). These multi-instantons also appear off-criticality, by using the string equation for the original hermitian matrix model \cite{asv11}. The multi-instanton sectors with $m=0$ correspond to the conventional large $N$ instantons obtained by eigenvalue tunneling, but there is no matrix model interpretation for the generic instanton sectors with $m\not=0$. The same phenomenon appears in many other examples \cite{sv13, as13, gs21} (in fact, resonance seems to be a generic feature of multicritical and minimal string models \cite{gs21}). The main question we want to address in this paper is: do the generic multi-instanton sectors discovered in \cite{gikm10} have a direct matrix model description, and if so, can we use this description to calculate them? Some generic ideas that they might correspond to ``different types'' of eigenvalues were put forward in \cite{kmr10}, and an attempt to compute them as ``anti-eigenvalues'' in a supermatrix model was done in \cite{mp09}---albeit neither of these calculations properly worked at the time, it turns out they were however morally correct as we shall now show and make completely precise in our present work.

Before we describe the contents of this paper, there is one more ingredient to mention. The multi-instanton sectors described above arise as formal solutions to an ODE or finite-difference equation, but the full resurgent structure of the theory involves an infinite number of Stokes coefficients \cite{gikm10, asv11, sv13, abs18}. In the aforementioned Painlev\'e~I example appearing in \cite{gikm10}, originally one Stokes coefficient was analytically known. This coefficient has a nice interpretation as the one-loop coefficient around the one-instanton sector associated to eigenvalue tunneling, and was simply computed via this interpretation in \cite{d92, msw07} (and for all multicritical and minimal string models in, \textit{e.g.}, \cite{gs21}). Some of the additional Stokes coefficients were first found numerically, and then many (empirical) relations in-between them were uncovered \cite{gikm10, asv11, sv13, as13}. Rather recently, this infinite set of (transcendental) numbers was finally found in analytical form via a detailed analysis of resonant resurgence \cite{bssv22}. It seems reasonable to expect that a semiclassical description of \textit{all} multi-instanton sectors in our transseries may also allow for a direct matrix-model calculation of resonant Stokes data---much in analogy with the calculation which computed ``standard'' Stokes data from eigenvalue tunneling. As we shall see, the matrix-model semiclassical description which we will find in this paper directly computes these transcendental numbers, and therefore decodes the full resonant structure of the resurgent transseries.

A complete understanding of what exactly is going-on with these resonant sectors clearly entails properly understanding the matrix model before anything else---and this is what we shall do in this work. Let us then describe the contents of our paper. Section~\ref{sec:ProblemAndProposal} overviews our above questions in a precise fashion, and qualitatively describes our proposed solution. Subsection~\ref{subsec:resurgenttranssriesandresonance} overviews the resonant properties of the resurgent transseries arising from matrix models, followed by a discussion of how a na\"\i ve matrix-model eigenvalue tunneling picture cannot account for all resonant sectors in subsection~\ref{subsec:matrixintegralsfail}. Our solution is presented in subsection~\ref{subsec:MatrixIntegralsWork}, where we describe how resonance emerges from tunneling of matrix-model eigenvalues into the involuted or non-physical sheets of the spectral curve (and how they may be regarded as ``holes'' or anti-eigenvalues on the physical sheet, which is somewhat reminiscent of the Dirac sea picture). Interestingly enough, this picture leads to integrals which also arise in supermatrix models (along the lines first conjectured in \cite{mp09}) and this is described in subsection~\ref{subsec:CommentOnSuperMatrixModel}. Having outlined our qualitative picture, we move to the quantitative calculations in section~\ref{sec:ResurgenceInMatrixModels}. Standard matrix model technology is briefly reviewed in subsection~\ref{subsec:matrixsetup} to set the stage and, armed with these tools, we discuss matrix integrals whose eigenvalues populate both sheets of our elliptic spectral curves in subsection~\ref{subsec:DeterminantCorrelators}. For these matrix integrals to become fully rigorous one still needs to discuss their integration contours, which is done in subsection~\ref{subsec:eigenvalueTunnelingContours}. As we shall see, the precise integration contours follow from our picture of generalized eigenvalue tunneling into all possible sheets, alongside resurgence considerations. Fully detailed calculations of diverse transseries sectors follow. First, in subsection~\ref{subsec:edgesofresurgentlattice}, we compute sectors on the ``edges'' of the resurgent-transseries lattice, and then, in subsection~\ref{subsec:bulkoneonesector}, we compute sectors in the ``bulk'' of this lattice. Particular detail is given to the $(1|1)$ configuration, with one tunneled eigenvalue and one tunneled anti-eigenvalue. One harder configuration, the $(2|1)$ transseries sector, with associated non-trivial Stokes data and transseries logarithm sectors, is addressed in subsection~\ref{subsec:resonanceandlogs}. The integrals at play are sometimes quite non-trivial and we have included toy examples of these types of integrals in appendix~\ref{app:DerivativeTrick}. A strategy for computing arbitrary configurations is finally outlined in subsection~\ref{subsec:partitionfunction}. Some subtleties concerning the construction of transseries for matrix-model free energy and partition function are addressed in appendix~\ref{app:resurgenceZvsF}, and further subtleties on Stokes data are discussed in appendix~\ref{app:HigherOrderStokes}. Up to this point our proposal is always presented depending on spectral-curve data alone, and we naturally move to examples in section~\ref{sec:ExamplesAndChecks}. These include the cubic matrix model, in subsection~\ref{subsec:CubicMatrixModelChecks}; the quartic matrix model, in subsection~\ref{subsec:quarticexample}; and further details on their respective double-scaling limits in subsection~\ref{subsec:DSLexample}. These examples further serve as non-trivial checks on our proposal, as they explicitly show how our matrix integral calculations precisely match against resurgent transseries results, and they further analytically compute non-trivial Stokes data (matching against \cite{bssv22}). Section~\ref{sec:ExamplesAndChecks} focuses on the general structure of these examples and on the results, with details on the data used therein to be found in appendix~\ref{app:TransseriesData}.

If it is true that the present work finally answers many questions which had been lingering in the resurgence framework for a long time, it also opens-up many other directions for future research. Because our formulae are model-independent---only depending on spectral curve data---many other examples could be addressed. For instance, in upcoming work we will report on the examples of minimal strings (somewhat following work in \cite{gs21}) and of the matrix model associated to Jackiw--Teitelboim gravity \cite{sss19}. These allow for a first-principles derivation of their transseries structure and resurgent Stokes data.

In recent years, it has been conjectured that topological string theories in non-compact target spaces are sometimes described by matrix models away from criticality \cite{dv02}. In particular, a proposal for matrix-model duals of topological strings on generic toric Calabi--Yau (CY) threefolds was made in \cite{ghm14, mz15, kmz15, z18}. Via localization and the AdS/CFT correspondence \cite{m97}, it was also found that some matrix models describe special observables for superstrings in full AdS backgrounds (see, \textit{e.g.}, \cite{m16}). Matrix models describing topological strings on toric CY's and superstrings in AdS are more complicated, and their large $N$ instantons are largely unexplored, but they should provide dual descriptions of nonperturbative effects in (topological) string theory and quantum gravity. As such, the work of this paper could likely be extended to topological strings, specially following a considerable amount of previous work focusing on constructing topological-string transseries in many examples; \textit{e.g.}, \cite{m06, msw07, m08, msw08, ps09, kmr10, dmp11, cesv13, gmz14, cesv14, ars14, c15, csv16, cms16, gm21, gm22}. We shall also return to these issues in the near future. Extensions to matrix models describing superstrings in AdS backgrounds via localization are also possible. Understanding the role of our calculations within the framework of the topological recursion \cite{eo07a} would also be of great interest. Finally, we would of course like to translate our present matrix model calculations into analogous statements in the D-brane language, with the hope of extending our results to generic string theories.

\section{One Open Problem and a Proposal for a Solution}\label{sec:ProblemAndProposal}

We begin by outlining our proposal to compute the full, resonant, nonperturbative contributions of hermitian matrix models. In particular, discussing resonance itself; a brief review of (standard) eigenvalue tunneling, how it seems to work and how it seems to fail; and  putting forward our proposal of populating all branched sheets of the spectral curve to obtain the complete answer. 

\subsection{Matrix Model Resurgent Transseries and Their Resonance}\label{subsec:resurgenttranssriesandresonance}

Let us briefly set-up our matrix model notation following \cite{m04} (to where we refer the reader for further details). With $M$ an hermitian $N \times N$ matrix and $V(x)$ the potential, the matrix-model partition-function\footnote{These matrix models are denoted either ``hermitian'' or ``normal'' in the literature. We mostly use ``hermitian''.} is ($g_{\text{s}}$ being the string coupling and $\text{vol} \left( \text{U}(N) \right)$ the volume of the gauge group) 
\bea
\label{eq:partitionfunctionhermitianmatrix}
\mathcal{Z}_{N} (g_{\text{s}}) &=& \frac{1}{\text{vol}\left(\text{U}(N)\right)} \int \text{d}M\, \text{e}^{-\frac{1}{g_{\text{s}}}\,\text{Tr}\, V(M)} = \\
\label{eq:partitionfunctioneigenvalues}
&=& \frac{1}{N!} \int \prod\limits_{i=1}^{N} \frac{\text{d}\lambda_i}{2\pi}\, \Delta_{N}(\lambda)^2\, \text{e}^{-\frac{1}{g_{\text{s}}} \sum\limits_{i=1}^{N} V(\lambda_i)},
\eea
\noindent
where we have fixed eigenvalue $\left\{ \lambda_i \right\}$ diagonal-gauge introducing the Vandermonde determinant $\Delta_{N}(\lambda) = \prod_{1 \leq i<j \leq N} \left(\lambda_i-\lambda_j\right)$. In this expression, the (so-far unspecified) eigenvalue integration contours may be classified as steepest-descent contours, to which we shall return later.

There are two standard ways to evaluate matrix-model partition functions in their large-$N$ 't~Hooft limit \cite{th74}. One is based upon exponentiation of the Vandermonde determinant in the above expression, giving rise to the holomorphic effective potential, $V_{\text{h;eff}}(x)$, essentially dictating the aforementioned paths of steepest descent for the matrix integral. Via $V_{\text{h;eff}}'(x) = y(x)$ one is then led to the introduction of the spectral curve $y(x)$ \cite{bipz78, ackm93}, from where eventually the full matrix-model free energy $\CF = \log \CZ$ follows as (herein $t = g_{\text{s}} N$ is the 't~Hooft coupling)
\be
\label{eq:matrix-integral-F}
\CF \simeq \sum_{g=0}^{\infty} \CF_g (t)\, g_{\text{s}}^{2g-2},
\ee
\noindent
via the topological-recursion construction \cite{eo07a}. The alternative to spectral curve and topological recursion methods is to use instead the so-called string equations \cite{biz80}. These come about by first introducing orthogonal polynomials $\left\{ p_n (x) \right\}$ (with respect to the natural, positive-definite measure inherited from the matrix integral) such that $p_n (x) = x^n + \cdots$. These polynomials satisfy recursion relations with recursion coefficients $\left\{ r_{n} \right\}$ dictated by the matrix-model potential \cite{biz80}. In the 't~Hooft limit, the ``continuous'' recursion coefficients then obey finite-difference string equations, for 
\be
\label{eq:rn-to-Rtgs}
r_n \mapsto R(t, g_{\text{s}}).
\ee
\noindent
Let us immediately alert the reader that throughout this paper it is important to keep in mind that we shall refer to (direct) matrix-integral quantities using curly notation, \textit{e.g.}, $\CZ$, whereas transseries quantities arising from string-equations shall be denoted regularly, \textit{e.g.}, $Z$. For example, the string-equation free energy\footnote{We will follow standard practice of normalization against the Gaussian weight \cite{m04}, $F-F_{\text{G}} = \log \left(\frac{Z}{Z_{\text{G}}}\right)$.} follows from, \textit{e.g.}, \cite{biz80, m04, m08, asv11},
\begin{equation}
\label{eq:freeenergyfromstringequation}
F \left(t+g_{\text{s}}, g_{\text{s}}\right) - 2 F \left(t, g_{\text{s}}\right) + F \left(t-g_{\text{s}}, g_{\text{s}}\right) = \log \frac{R(t, g_{\text{s}})}{t}.
\end{equation}

Starting with the work in \cite{m08, gikm10}, resurgent transseries solutions to matrix-model string-equations have been further addressed in, \textit{e.g.}, \cite{msw07, msw08, asv11, sv13, as13, ars14, csv15, gs21, bssv22}. In all these solutions, resonance \cite{gikm10} is generic \cite{asv11, sv13, as13, gs21, bssv22}: from matrix models to Painlev\'e equations, from multicritical to minimal string theories. The examples we shall focus upon later in this work, the cubic matrix model with large-$N$ string-equation \eqref{eq:cubicstringequationthooftlimit} in subsection~\ref{subsec:CubicMatrixModelChecks}, the quartic matrix model with large-$N$ string-equation \eqref{eq:quarticstringequationthooftlimit} in subsection~\ref{subsec:quarticexample}, and the Painlev\'e~I equation \eqref{eq:PainleveI} in subsection~\ref{subsec:DSLexample}, all abide to the same structure: a two-parameter, resonant, resurgent transseries (see, \textit{e.g.}, \cite{m12, s14, abs18} for reviews). What this means is that their solutions (and herein we are not distinguishing between the string-equation solution in \eqref{eq:rn-to-Rtgs}, or its corresponding free energy from \eqref{eq:freeenergyfromstringequation}, or even the partition function\footnote{Note that while the corresponding transseries for the matrix-model partition function still has the form \eqref{eq:transseriesexpansionGeneric}, it is however usually written with an additional global pre-factor in comparison to \eqref{eq:partitionfunctionhermitianmatrix}. Explicitly,
\begin{equation}
\mathcal{Z} = \rme^{-\frac{1}{g_{\text{s}}^2}\, F_{-2}^{(0,0)}}\, Z,
\end{equation}
\noindent
with $F_{-2}^{(0,0)}$ the lowest-genus perturbative free-energy coefficient, and $Z$ the partition-function transseries as \eqref{eq:transseriesexpansionGeneric}.}) are of the schematic form
\begin{align}
\label{eq:transseriesexpansionGeneric}
\Phi(t, g_{\text{s}}; \sigma_1,\sigma_2) &= \sum\limits_{n=0}^{\infty} \sum\limits_{m=0}^{\infty} \sigma_1^n \sigma_2^m\, \text{e}^{-\left(n-m\right)\frac{A(t)}{g_{\text{s}}}}\, \Phi^{(n,m)} (t, g_{\text{s}}), \\
\label{eq:transseriesexpansionGeneric-sectors}
\Phi^{(n,m)}(t, g_{\text{s}}) &= \sum\limits_{k=0}^{\min(n, m)} \Phi^{(n,m)[k]}(t, g_s)\, \log^k \frac{f(t)}{g_{\text{s}}^2},
\end{align}
\noindent
where $A(t)$ (the instanton action) and $f(t)$ are problem-specific functions of the 't~Hooft coupling $t$, and where each perturbative or nonperturbative sector is asymptotic,
\begin{align}
\label{eq:phinm-beta-starting-g}
\Phi^{(n,m)[k]}(t, g_{\text{s}}) \simeq \sum\limits_{g=\beta_{nm}^{[k]}}^{\infty} \Phi^{(n,m)[k]}_g (t)\, g_{\text{s}}^{g},
\end{align}
\noindent
with coefficients growing factorially fast $\sim g!$ and $\beta_{nm}^{[k]}$ the starting genus. It is immediately noticeable in \eqref{eq:transseriesexpansionGeneric} that the instanton action arises in a symmetric pair $(A,-A)$, which is one hallmark of resonance. Another, more problem-specific, pointer of resonance is the appearance of logarithms in the transseries expansions. These logarithms are however a somewhat mild feature as resonance further implies that one can relate different $[k]$-logarithm sectors to each other,
\begin{align}
\label{eq:Phinmk=alphaPhinm0}
\Phi^{(n,m)[k]}(t, g_\text{s}) = \frac{1}{k!}\, \Big( \alpha \left(n-m\right) \Big)^k\, \Phi^{(n-k,m-k)[0]}(t, g_{\text{s}}).
\end{align}
\noindent
Here $\alpha$ is a number depending on the explicit setting. 

\begin{figure}
\centering
	\begin{tikzpicture}[
	grayframe/.style={
		rectangle,
		draw=gray,
		text width=10em,
		align=center,
		rounded corners,
		minimum height=2em
	},
	longgrayframe/.style={
		rectangle,
		draw=gray,
		text width=12.5em,
		align=center,
		rounded corners,
		minimum height=2em
	}, scale =1, line width=2
	]
	\node[longgrayframe] (a) at (-5,3) {$\Phi^{(n, m)}(g_{\text{s}})$\\Divergent Series};
	\node (ab) at (-5,1.5) {\color{gray}\small Borel Transformation $\mathcal{B}$};
	\node[longgrayframe] (b) at (-5,0) {$\mathcal{B}\left[\Phi^{(n, m)}\right](s)$\\ (converges in disk $D \subset\mathbb{C}$)};
	\draw[blue, ->] (a) -- (b);
	\node (bc) at (0.5,0.5) {\color{gray} Analytic Continuation};
	\node[grayframe] (c) at (5,0) {Function in $\mathbb{C}$\\ (with singularities)};
	\draw[blue, ->] (b) -- (c);
	\node[grayframe] (d) at (5,3) {$\mathcal{S}_\theta\Phi^{(n,m)}(g_{\text{s}})$\\Borel Resummation};
	\draw[blue,->] (c) -- (d);
	\node (cd) at (5, 1.5) {\small \color{gray} Resummation $\mathcal{S}_\theta$};
	\end{tikzpicture}
\caption{Pictorial description of Borel resummation of a divergent series into a function. Start with the left-downward arrow performing a Borel transform, which yields a power-series convergent on a disk. Moving with the rightward arrow implements analytic continuation beyond the radius of convergence of the power-series. This results in a function on the complex plane, albeit with singularities. The final step with the right-upward arrow is resummation via Laplace transform (inverse Borel transform); albeit this is ill-defined when crossing any of the aforementioned singularities upon the complex Borel plane (sitting along rays known as Stokes lines).}
\label{fig:innerworkingsofresurgence}
\end{figure}
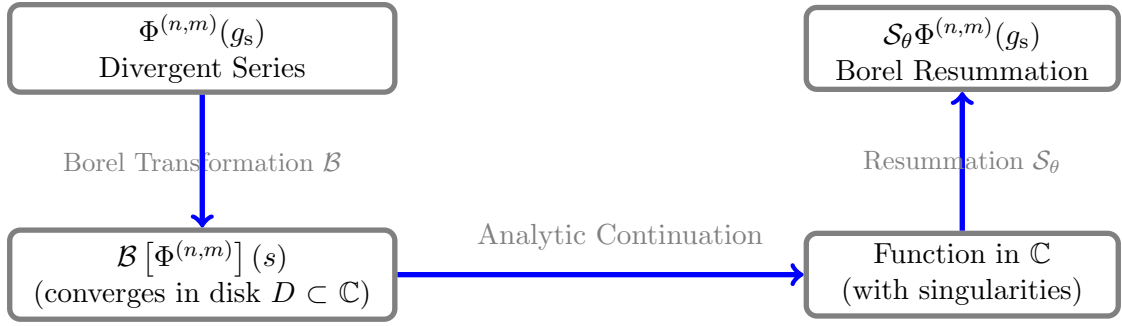

\begin{figure}
\centering
\begin{tikzpicture}[
	 scale =1, line width=2
	]
	\draw[->] (-5,0) -- (5, 0);
	\draw[->] (0,-2) -- (0, 2);
	\draw (-5, 1.6) -- (-4.6, 1.6);
	\draw (-4.6, 1.6) -- (-4.6, 2);
	\node at (-4.8, 1.8) {$g_{\text{s}}$};
	\draw[blue] (0,0) -- (4.5, -0.5);
	\draw[blue] (0,0) -- (4.5, 0.5);
	\draw[->, blue] (4, -0.4) to[bend right =30] (4, 0.4);
	\node[blue] at (5, 0.5) {$\underline{\mathfrak{S}}_0$};
	\draw[color=orange] (0,0) -- (-4.5, -0.5);
	\draw[color=orange] (0,0) -- (-4.5, 0.5);
	\draw[->, color=orange] (-4, 0.4) to[bend right =30] (-4, -0.4);
	\node[color=orange] at (-5, -0.5) {$\underline{\mathfrak{S}}_{\pi}$};
	\end{tikzpicture}
\caption{The two Stokes lines of the resonant transseries \eqref{eq:transseriesexpansionGeneric}. Depicted along $\theta=0$ is the ``forward'' Stokes automorphism ({\color{blue}blue}), arising from the instanton action $A$; and along $\theta=\pi$ the ``backward'' Stokes automorphism ({\color{orange}orange}), associated to the instanton action $-A$ (where we are assuming that the value for $A$ is real and positive).}
\label{fig:PictureTwoAutomorphisms}
\end{figure}
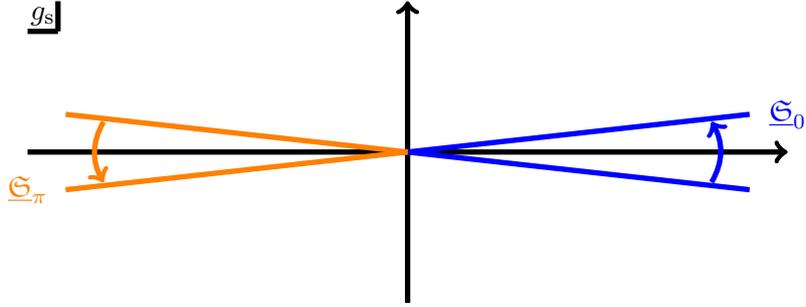

Resurgent analysis of the above transseries starts with Borel resummation, as illustrated in figure~\ref{fig:innerworkingsofresurgence} (again, see, \textit{e.g.}, \cite{m12, s14, abs18} for reviews). This entails the three steps illustrated in the figure: a Borel transform $\mathcal{B} \left[ g_{\text{s}}^{\alpha+1} \right] := \frac{s^{\alpha}}{\Gamma(\alpha+1)}$ from asymptotic series $\mathbb{C}[[g_{\text{s}}]]$ to convergent power-series $\mathbb{C}\lbrace s\rbrace$ in a disk on the complex Borel plane; an analytic continuation throughout $\BC$; and, upon picking a ray of integration $\theta$ on the Borel $s$-plane, the final inverse Borel transform (a Laplace transform) yielding the Borel $\theta$-resummation of the initial asymptotic power-series,
\begin{equation}
\label{eq:resummationandLaplacetransform}
\mathcal{S}_{\theta}\Phi(g_{\text{s}}) = \int_{0}^{\text{e}^{\rmi \theta} \infty}\text{d}s\, \mathcal{B} \left[\Phi\right] (s)\, \text{e}^{-\frac{s}{g_{\text{s}}}}.
\end{equation}
\noindent
One could now think that this might be the end of the story but quite the opposite; it is just the beginning. Whenever $\theta$ is such that the ray of integration hits a singularity of the integrand, the last step in the above procedure becomes ill-defined---and such rays are known as Stokes lines on the Borel plane. Hence, on top of resummation, resurgent analysis of the above transseries further requires a proper understanding of Stokes phenomena (\textit{e.g.}, \cite{m12, s14, abs18}). For our resonant case \eqref{eq:transseriesexpansionGeneric} there are instanton actions $\boldsymbol{A} = (A, -A)$, implying the existence of two Stokes lines along symmetric directions. They are both illustrated in figure~\ref{fig:PictureTwoAutomorphisms}. Avoiding the precise rays along these Stokes lines leads to the natural definition of lateral Borel resummations, $\mathcal{S}_{\theta^\pm}\Phi$ for $\theta=0,\pi$. They are however related by the action of the Stokes automorphism $\underline{\mathfrak{S}}_{\theta}$ at $\theta=0,\pi$, introduced via alien calculus in \cite{e81} as a tool to understand the singular structure of Borel transforms without actually having to revert to explicit analytic continuation. This relation is given by
\begin{equation}
\label{eq:definitionStokesAutomorphism}
\mathcal{S}_{\theta^{+}} = \mathcal{S}_{\theta^{-}} \circ \underline{\mathfrak{S}}_{\theta}.
\end{equation} 
\noindent
It may also be equivalently rewritten as a discontinuity relation
\begin{equation}
\label{eq:discontinutiystokesautomorphism}
\mathcal{S}_{\theta^{+}} - \mathcal{S}_{\theta^{-}} =: \mathcal{S}_{\theta^{-}} \circ \text{Disc}_{\theta}.
\end{equation}

\begin{figure}
\centering
	\begin{tikzpicture}[
	grayframe/.style={
		rectangle,
		draw=gray,
		text width=4em,
		align=center,
		rounded corners,
		minimum height=2em
	}, scale =1, line width=2
	]
	\foreach \n in {0,...,3}{
	  \foreach \m in {0,..., 2}{
	\node[grayframe] (\n_\m) at (4*\n,3*\m) {$\Phi^{(\n, \m)}(g_{\text{s}})$};
	}}
	\draw[color=ForestGreen, ->] (0_0) -- (1_0) node[midway, above]{$\mathsf{S}_{(0,0)\to(1,0)}$};
	\draw[blue, ->] (1_0) -- (2_0)node[midway, above]{$\mathsf{S}_{(1,0)\to(2,0)}$};
	\draw[color=DarkOrchid, ->] (2_0) -- (2_1)node[midway, right]{$\mathsf{S}_{(2,0)\to(2,1)}$};
		\draw[color=DarkOrchid, ->] (2_0) to [out=135, in=45, looseness=1.3] (1_0);
		\node[color=DarkOrchid] at (6, 1.7){$\mathsf{S}_{(2,0)\to(1,0)}$};
			\draw[color=Maroon, ->] (2_1) -- (2_2)node[midway, right]{$\mathsf{S}_{(2,1)\to(2,2)}$};
	\draw[color=Maroon, ->] (2_1) to [out=135, in=45, looseness=1.3] (1_1);
	\node[color=Maroon] at (6, 4.7){$\mathsf{S}_{(2,1)\to(1,1)}$};	
			\draw[color=Maroon, ->, dashed] (2_1) to [out=200, in=45, looseness=1.3] (0_0);
		\node[color=Maroon] at (1.1, 1.8){$\mathsf{S}_{(2,1)\to(0,0)}$};
		\draw[color=Maroon, ->] (1_0) -- (1_1)node[midway, below left]{$\mathsf{S}_{(1,0)\to(1,1)}$};
	\draw[color=Maroon, ->, dashed] (1_0) to [out=235, in=315, looseness=1.3] (0_0);
	\node[color=Maroon] at (2, -1.7){$\mathsf{S}_{(1,0)\to(0,0)}$};			
		\draw[dashed, brown] (0_0) -- (1_1);
		\draw[dashed, brown] (1_1) -- (2_2);
		\draw[dashed, brown] (2_2) -- (8.8, 6.6);
		\draw[dashed, brown] (1_0) -- (2_1);
		\draw[dashed, brown] (2_1) -- (3_2);
		\draw[dashed, brown] (3_2) -- (12.8, 6.6);	
		\draw[dashed, brown] (2_0) -- (3_1);
		\draw[dashed, brown] (3_1) -- (12.8, 3.6);
		\draw[dashed, brown] (3_0) -- (12.8, 0.6);
		\draw[dashed, brown] (0_1) -- (1_2);
		\draw[dashed, brown] (1_2) -- (4.8, 6.6);
		\draw[dashed, brown] (0_2) -- (0.8, 6.6);
	\end{tikzpicture}
\caption{The transseries sectors in \eqref{eq:transseriesexpansionGeneric} represented as a two-dimensional semi-positive lattice. Resurgence relates different sectors to each other, essentially via the Stokes automorphisms \eqref{eq:forwardStokesAutomorphism}-\eqref{eq:backwardStokesAutomorphism}. The arrows in the figure symbolize Borel residues $\mathsf{S}_{\boldsymbol{n}\to\boldsymbol{m}}$ and exemplify how to reach, for instance, the $(2,2)$-sector via single-step moves on this lattice. We have colored each step differently (as {\color{ForestGreen}first}, {\color{blue}second}, {\color{DarkOrchid}third}, and {\color{Maroon}fourth} steps). The dashed {\color{brown}brown} lines symbolize resonant directions, whereas the dashed arrows have associated vanishing Borel residues. Resurgence relations are fairly simple as long as we only move along the edges of the lattice; but as soon as one moves inwards, resonant contributions start appearing. For example, the {\color{DarkOrchid}third} step reaches two distinct endpoints, implying that on the {\color{Maroon}fourth} step there will be Borel residues starting at both these $(2,1)$ and $(1,0)$ sectors. Upon iterated use of Stokes transitions, one hence finds a multitude of different sectors that are spawned by resonance. Throughout this paper we shall refer to similar diagrams in order to better understand the resurgence relations at play.}
\label{fig:walkingonesteppathsintro}
\end{figure}
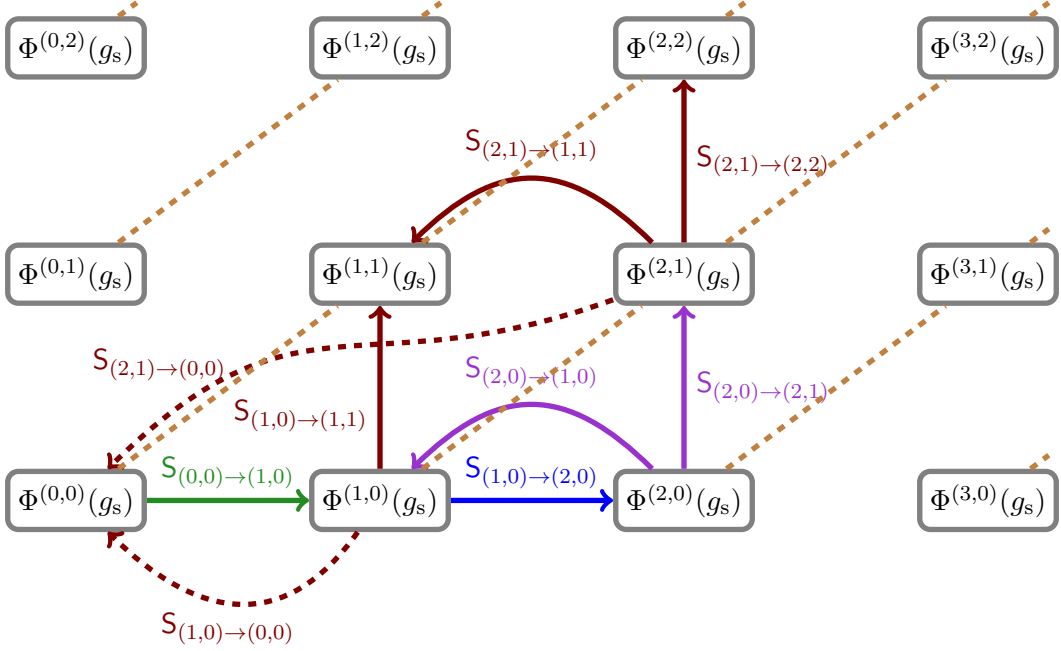

For our resonant transseries \eqref{eq:transseriesexpansionGeneric} the Stokes automorphisms read (see \cite{bssv22} for a discussion)
\begin{align}
\label{eq:forwardStokesAutomorphism}
\underline{\mathfrak{S}}_0 \Phi^{(n,m)} &= \Phi^{(n,m)} - \sum_{\ell=1}^{\infty} \rme^{-\ell \frac{A}{g_{\text{s}}}} \sum_{p=0}^{\min(n+\ell,m)} \mathsf{S}_{(n,m)\to(n+\ell-p,m-p)}\, \Phi^{(n+\ell-p,m-p)}, \\
\label{eq:backwardStokesAutomorphism}
\underline{\mathfrak{S}}_\pi \Phi^{(n,m)} &= \Phi^{(n,m)} - \sum_{\ell=1}^{\infty} \rme^{+\ell \frac{A}{g_{\text{s}}}} \sum_{p=0}^{\min(n,m+\ell)} \mathsf{S}_{(n,m)\to(n-p,m+\ell-p)}\, \Phi^{(n-p,m+\ell-p)}.
\end{align}
\noindent
Herein, the sets of numbers $\mathsf{S}_{(n,m)\to (n+\ell-p, m-p)}$ and $\mathsf{S}_{(n,m)\to (n-p, m+\ell-p)}$ are called forward and backward Borel residues, respectively (note how they amount to distinct sets of numbers). Very recently there has been progress in tackling these numbers and they have all been analytically computed for both the Painlev\'e~I and Painlev\'e~II problems (which equivalently amounts to analytically computing their full Stokes data) \cite{bssv22}. They are also directly related to the singularities on the Borel plane for simple resurgent functions \cite{abs18}. Our point here is that \eqref{eq:forwardStokesAutomorphism} and \eqref{eq:backwardStokesAutomorphism} fully encode the resurgence of our transseries \eqref{eq:transseriesexpansionGeneric}, \textit{i.e.}, encode how distinct transseries sectors relate to each other upon Stokes phenomenon. This essentially directly follows via all non-vanishing Borel residues in \eqref{eq:forwardStokesAutomorphism}-\eqref{eq:backwardStokesAutomorphism}, as illustrated in figure~\ref{fig:walkingonesteppathsintro}.

\begin{figure}
\centering
\begin{tikzpicture}[
	blueframe/.style={
		rectangle,
		draw=blue,
		fill=blue!20,
		text width=4em,
		align=center,
		rounded corners,
		minimum height=2em
	},
	redframe/.style={
		rectangle,
		draw=orange,
		fill=orange!20,
		text width=4em,
		align=center,
		rounded corners,
		minimum height=2em
	}, xscale =0.6,yscale =0.5, line width=2
	]
	\draw[->] (0, -10.5) -- (0, 4.5);
	\draw[->] (-14.5, 0) -- (6.5, 0);
\node[blueframe] (1_0) at (4,0) {$\boldsymbol{S}_{(1, 0)}$};
\node[blueframe] (1_0) at (4,-3) {$\boldsymbol{S}_{(1, -1)}$};
\node[blueframe] (1_0) at (4,-6) {$\boldsymbol{S}_{(1, -2)}$};
\node[blueframe] (1_0) at (4,-9) {$\boldsymbol{S}_{(1, -3)}$};
\node[blueframe] (1_0) at (0,-3) {$\boldsymbol{S}_{(0, -1)}$};
\node[blueframe] (1_0) at (0,-6) {$\boldsymbol{S}_{(0, -2)}$};
\node[blueframe] (1_0) at (0,-9) {$\boldsymbol{S}_{(0, -3)}$};
\node[blueframe] (1_0) at (-4,-6) {$\boldsymbol{S}_{(-1, -2)}$};
\node[blueframe] (1_0) at (-4,-9) {$\boldsymbol{S}_{(-1, -3)}$};
\node[blueframe] (1_0) at (-8,-9) {$\boldsymbol{S}_{(-2, -3)}$};
\node[redframe] (1_0) at (0,3) {$\boldsymbol{S}_{(0, 1)}$};
\node[redframe] (1_0) at (-4,3) {$\boldsymbol{S}_{(-1, 1)}$};
\node[redframe] (1_0) at (-8,3) {$\boldsymbol{S}_{(-2, 1)}$};
\node[redframe] (1_0) at (-12,3) {$\boldsymbol{S}_{(-3, 1)}$};
\node[redframe] (1_0) at (-4,0) {$\boldsymbol{S}_{(-1, 0)}$};
\node[redframe] (1_0) at (-8,0) {$\boldsymbol{S}_{(-2, 0)}$};
\node[redframe] (1_0) at (-12,0) {$\boldsymbol{S}_{(-3, 0)}$};
\node[redframe] (1_0) at (-8,-3) {$\boldsymbol{S}_{(-2, -1)}$};
\node[redframe] (1_0) at (-12,-3) {$\boldsymbol{S}_{(-3, -1)}$};
\node[redframe] (1_0) at (-12,-6) {$\boldsymbol{S}_{(-3, -2)}$};
\end{tikzpicture}
\caption{Stokes vectors for a two-parameter resonant transseries may be conveniently arranged on a two-dimensional lattice that extends infinitely along two negative directions but is bounded upon the positive ones \cite{bssv22}. These vectors naturally group themselves into two sets ({\color{orange}orange} and {\color{blue}blue}), separated by an empty diagonal, exactly related to the forward and backward Stokes automorphisms of figure~\ref{fig:PictureTwoAutomorphisms}. We shall later compute some of these entries from matrix integrals.}
\label{fig:stokeslattice}
\end{figure}
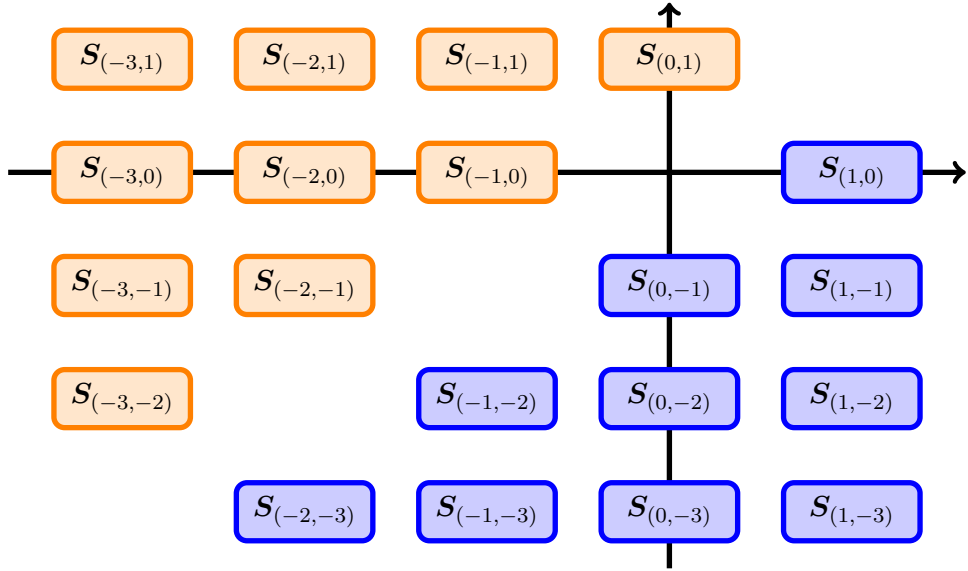

To close this swift review, let us list a brief set of relations between Borel residues and Stokes data, in particular a set which will be of relevance later on (see \cite{abs18, bssv22} for more details). Note that---albeit equivalent information---it is sometimes useful to work interchangeably with both, as Borel residues $\mathsf{S}_{\boldsymbol{n}\to\boldsymbol{n}+\boldsymbol{\ell}}$ appear directly in the Stokes automorphism while Stokes vectors $\boldsymbol{S}_{\boldsymbol{\ell}}$ relate to the underlying alien algebra via alien derivatives $\Delta_{\boldsymbol{\ell} \cdot \boldsymbol{A}} \Phi_{\boldsymbol{n}} = \boldsymbol{S}_{\boldsymbol{\ell}} \cdot \left( \boldsymbol{n}+\boldsymbol{\ell} \right) \Phi_{\boldsymbol{n}+\boldsymbol{\ell}}$ \cite{abs18, bssv22}. For our particular two-parameter resonant transseries \eqref{eq:transseriesexpansionGeneric} Stokes data may be conveniently organized on a two-dimensional lattice of vectors, as shown in figure~\ref{fig:stokeslattice}, and one finds, \textit{e.g.}, \cite{abs18, bssv22}
\begin{align}
\label{eq:forwardbackwardBorelResidues}
\mathsf{S}_{(0,0)\to(k,0)} &= -\frac{1}{k!}\, \prod\limits_{\ell=1}^{k} \left( \boldsymbol{S}_{(1,0)} \cdot
\left[\begin{array}{c}
k\\
0
\end{array}\right] \right), &
\mathsf{S}_{(0,0)\to(0,k)} &= -\frac{1}{k!}\, \prod\limits_{\ell=1}^{k} \left( \boldsymbol{S}_{(0,1)} \cdot
\left[\begin{array}{c}
0\\
k
\end{array}\right] \right), \\
\label{eq:BorelResidue01-11,11-01sector}
\mathsf{S}_{(0,1)\to(1,1)} &= - \boldsymbol{S}_{(1,0)} \cdot \left[\begin{array}{c}
1\\
1
\end{array}\right], &
\mathsf{S}_{(1,1)\to(1,0)} &= - \boldsymbol{S}_{(0,-1)} \cdot \left[\begin{array}{c}
1\\
0
\end{array}\right], \\
\label{eq:BorelResidue01-21,11-21sector}
\mathsf{S}_{(1,1)\to(2,1)} &= - \boldsymbol{S}_{(1,0)} \cdot \left[\begin{array}{c}
2\\
1
\end{array}\right],& \mathsf{S}_{(0,1)\to(2,1)} &= - \frac{1}{2} \left( \boldsymbol{S}_{(1,0)} \cdot \left[\begin{array}{c}
1\\
1
\end{array}\right] \right) \left( \boldsymbol{S}_{(1,0)} \cdot \left[\begin{array}{c}
2\\
1
\end{array}\right] \right), \\
\label{eq:BorelResidue01to10sector}
\mathsf{S}_{(0,1)\to(1,0)} &= - \boldsymbol{S}_{(1,-1)} \cdot \left[\begin{array}{c}
1\\
0
\end{array}\right]
- \frac{1}{2} \left( \boldsymbol{S}_{(0,-1)}\cdot
\left[\begin{array}{c}
1\\
0
\end{array}\right] \right) \left( \boldsymbol{S}_{(1,0)} \cdot \left[\begin{array}{c}
1\\
1
\end{array}\right] \right).\span\span
\end{align}
\noindent
We shall use these relations later in section~\ref{sec:ExamplesAndChecks}.

Tracing Stokes data back to the matrix model we started-off with, it was shown in, \textit{e.g.}, \cite{d91, msw07} how \textit{forward} resurgence relations are related to matrix-model \textit{eigenvalue tunneling}. This means one may walk \textit{forward} along the resurgence lattice by essentially using the eigenvalue integral \eqref{eq:partitionfunctioneigenvalues}. But walking \textit{backwards} was left unanswered in those references and this is what we must turn to next if to access the full lattice of figure~\ref{fig:walkingonesteppathsintro}.

\subsection{How Matrix Integrals Seem to Fail}
\label{subsec:matrixintegralsfail}

\begin{figure}
\centering
	\begin{tikzpicture}
	\draw[fill=LightBlue,fill opacity=0.2, line width=1pt] (0,0) to [out=90,in=95] (4,0)
	to [out=85,in=0] (1,2.1)
    to [out=180,in=90] (-2,0)
    to [out=270, in=180] (-1, -0.5)
    to [out=0, in=270] cycle;
    \draw[fill=darktangerine,fill opacity=0.2, line width=1pt] (-2,0)
    to [out=270,in=180] (1,-2.1)
    to [out=0,in=275] (4,0)
    to [out=265,in=270] (0,0)
    to [out=270, in=0] (-1, -0.5)
    to [out=180, in=270] cycle;
    \draw[color=ForestGreen, line width=2pt] (-2,0) to [out=270, in=180] (-1, -0.5)
    to [out=0, in=270] (0,0);
    \draw[dashed, color=ForestGreen, line width=2pt] (-2,0) to [out=90, in=180] (-1, 0.5)
    to [out=0, in=90] (0,0);
    \draw[blue, line width=1.5pt, ->] (0,0) to [out=90, in=180] (1.8, 1.6);
    \draw[blue, line width=1.5pt] (1.8, 1.6) to [out=0, in=90] (4, 0);
\draw[ForestGreen, fill=ForestGreen] (-2,0) circle (.7ex);
\draw[ForestGreen, fill=ForestGreen] (0,0) circle (.7ex);
\draw[cornellred, fill=cornellred] (4,0) circle (.7ex);
\node at (-2.3, 0) {$a$}; 
\node at (0.3, 0) {$b$};
\node at (4.4, 0) {$x^{\star}$};  
	\end{tikzpicture}
\caption{Illustration of the spectral curve for the cubic matrix model. There are two sheets separated by the cut ({\color{ForestGreen}green}) which further touch at the pinched cycle $x^{\star}$: the physical sheet ({\color{LightBlue}blue}) containing the integration contour for the action ({\color{blue}blue}), and the non-physical sheet ({\color{darktangerine}orange}).}
\label{fig:cubicspectralcurvejustphysicalaction}
\end{figure}
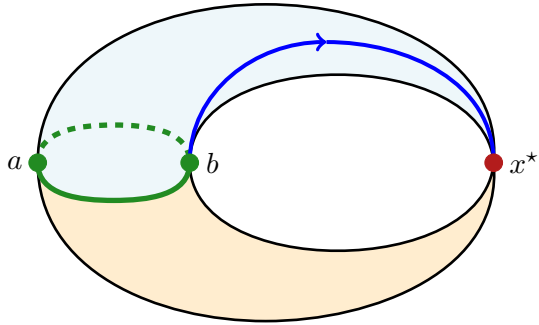

Let us reconsider our matrix integral \eqref{eq:partitionfunctionhermitianmatrix} with perturbative saddle corresponding to a one-cut spectral-curve configuration as in figure~\ref{fig:cubicspectralcurvejustphysicalaction}. Instanton corrections to this perturbative saddle are usually computed via eigenvalue tunneling out from the one-cut $\NCC = (a,b)$ into the nonperturbative saddle at $x^{\star}$ \cite{d91, gz91, d92, ez93, akk03, ss03, hhikkmt04, st04, iy05, iky05, msw07, msw08}. Let us briefly readdress the standard picture of eigenvalue tunneling in the following, to make contact with our previous Stokes automorphism discussion and, in particular, to make clear how only one of the two instanton actions in \eqref{eq:transseriesexpansionGeneric} appears and how this will hence only give rise to \textit{forward} resurgence relations.

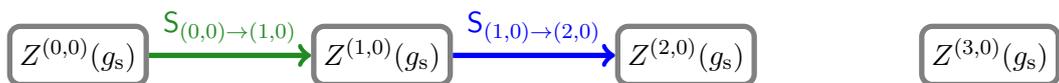
\begin{figure}
\centering
	\begin{tikzpicture}[
	grayframe/.style={
		rectangle,
		draw=gray,
		text width=4em,
		align=center,
		rounded corners,
		minimum height=2em
	}, scale =1, line width=2
	]
	\foreach \n in {0,...,3}{
	\node[grayframe] (\n_0) at (4*\n,3*0) {$Z^{(\n, 0)}(g_{\text{s}})$};
	}
	\draw[color=ForestGreen, ->] (0_0) -- (1_0) node[midway, above]{$\mathsf{S}_{(0,0)\to(1,0)}$};
	\draw[blue, ->] (1_0) -- (2_0)node[midway, above]{$\mathsf{S}_{(1,0)\to(2,0)}$};
	\end{tikzpicture}
\caption{Visualization of the resurgent lattice which may be reached via standard eigenvalue tunneling. Comparing to figure~\ref{fig:walkingonesteppathsintro} shows how we can only see one edge of the full resonant lattice.}
\label{fig:walkingonesteppathsintroLowestRow}
\end{figure}

Consider the partition function \eqref{eq:partitionfunctioneigenvalues} with integration contours for the eigenvalues $\left\{ \lambda_i \right\}$ given by perturbative steepest-descent contours\footnote{Explicitly, these will be defined in the upcoming formula \eqref{eq:steepestdescentconditionV}.} of the potential $V(x)$. What is the behavior of this perturbative partition-function as one rotates the phase of $g_{\text{s}}$? Let us illustrate what happens to the steepest-descent contours in the example of the quartic matrix model (with potential \eqref{eq:quarticPotential}, to be later discussed in subsection~\ref{subsec:quarticexample}), where they precisely change as
\begin{align}
\label{eq:rotationperturbativecontour}
\underset{\theta=0^{+}}{\ZPertpe}\,&\rightarrow\,\underset{\theta=\pi^{-}}{\ZPertpime}\,\rightarrow\,\underset{\theta=\pi^{+}}{\ZPertpipe}\,\rightarrow\,\underset{\theta=2\pi^{-}}{\ZPerttwopime}\,\overset{\footnotesize\text{Stokes}}{\longrightarrow}\,\underset{\theta=2\pi^{+}}{\ZPerttwopipe}\,\rightarrow\nonumber\\
&\rightarrow\,\underset{\theta=3\pi^{-}}{\ZPertthreepime}\,\rightarrow\,\underset{\theta=3\pi^{+}}{\ZPertthreepipe}\,\rightarrow\,\underset{\theta=4\pi^{-}}{\ZPertme}\,\overset{\footnotesize\text{Stokes}}{\longrightarrow}\,\underset{\theta=4\pi^{+}=0^{+}}{\ZPertpe}
\end{align}
\noindent
as we rotate\footnote{Notice how the contour exhibits a double-sheeted behavior in $g_{\text{s}}$, with return to the original configuration only after two full rotations. This is natural to expect as single eigenvalue-tunneling to the nonperturbative saddle comes with a factor of $\sqrt{g_{\text{s}}}$ in the one-instanton matrix integral \cite{msw07}.} $\theta = \arg  g_{\text{s}}$. When crossing the Stokes lines at $\theta=0, 2\pi$, the contours jump as
\begin{equation}
\label{eq:quarticexampleeigenvaluetunnelingscetch}
\ZPertme\,\rightarrow\,\ZPertpe\,-\,\ZSadpzero\,.
\end{equation}
\noindent
It is with this jump that we say ``an eigenvalue has tunneled to a nonperturbative saddle''. From the transseries \eqref{eq:transseriesexpansionGeneric} viewpoint this amounts to spelling out the forward Stokes transition \eqref{eq:forwardStokesAutomorphism} up to first instanton order:
\begin{equation}
\label{eq:Z00toZ10}
Z^{(0,0)}(g_{\text{s}}) - \rme^{- \frac{A}{g_{\text{s}}}}\, \mathsf{S}_{(0,0)\to(1,0)}\, Z^{(1,0)}(g_{\text{s}}) + \cdots.
\end{equation}
\noindent
Connecting eigenvalue and transseries pictures in this way (a relation we shall make fully precise later in subsection~\ref{subsec:eigenvalueTunnelingContours}), shows how standard eigenvalue tunneling is giving rise to part, \textit{but not all}, of the transseries: indeed, it only gives rise to what is illustrated in figure~\ref{fig:walkingonesteppathsintroLowestRow}, which is but a small fraction of the full resonant transseries depicted in figure~\ref{fig:walkingonesteppathsintro}.

Being slightly quantitative, consider the one-cut $\NCC = (a,b)$ solution with spectral curve
\begin{equation}
\label{eq:hermitianspectralcurve}
y(x) = M(x)\, \sqrt{\left(x-a\right) \left(x-b\right)},
\end{equation}
\noindent
where $M(x)$ is the moment function (see, \textit{e.g.}, \cite{m04}). The nonperturbative saddles correspond to zeroes of this moment function, $M(x^\star)=0$, and their corresponding one-instanton action appearing in \eqref{eq:Z00toZ10} follows via \cite{msw07} 
\begin{equation}
\label{eq:InstantonActionmsw07}
A (t) = V_{\text{h;eff}} (x^\star;t) - V_{\text{h;eff}} (b;t) = \int_{b}^{x^{\star}} \text{d}x\, y(x),
\end{equation}
\noindent
where we explicitly consider the integration path to be on the physical sheet of the spectral curve as illustrated in figure~\ref{fig:cubicspectralcurvejustphysicalaction}. In the example of the cubic matrix model (to be addressed in detail in subsection~\ref{subsec:CubicMatrixModelChecks}, but see \cite{msw08} as well) the moment function is
\begin{equation}
\label{eq:MCubicExample}
M_{\text{cubic}} (x) = \frac{1}{4} \left(4-\lambda\left(a+b+2x\right)\right),
\end{equation}
\noindent
with single nonperturbative saddle $x^{\star}=\frac{2}{\lambda}-\frac{1}{2}\left(a+b\right)$. The spectral curve hence seems not to see any additional saddle-point which would produce the resonant pair to the instanton action $A(t)$. Yet its need is clearly predicted from the resurgent analysis of the matrix-model string equation.

So where is the full resonant lattice in figure~\ref{fig:walkingonesteppathsintro} hiding? What saddle points or what eigenvalue tunneling may give rise to the $-A(t)$ pair, and hence the backward Stokes automorphism?

\subsection{How Matrix Integrals Actually Work}\label{subsec:MatrixIntegralsWork}

Our proposal is quite simple and straightforward (to be backed up by detailed calculations in section~\ref{sec:ResurgenceInMatrixModels}). It amounts to considering \textit{both} sheets of the spectral curve \eqref{eq:hermitianspectralcurve}, rather than just the physical sheet as in the previous subsection, and allowing eigenvalues to \textit{tunnel on both sheets}.

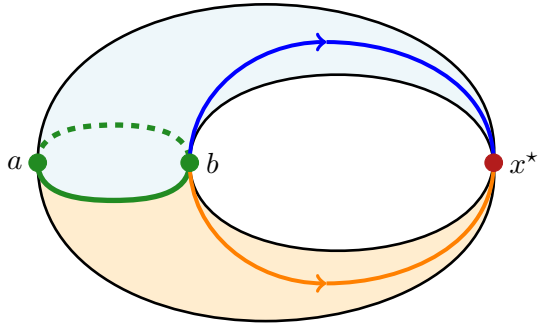
\begin{figure}
\centering
	\begin{tikzpicture}
	\draw[fill=LightBlue,fill opacity=0.2, line width=1pt] (0,0) to [out=90,in=95] (4,0)
	to [out=85,in=0] (1,2.1)
    to [out=180,in=90] (-2,0)
    to [out=270, in=180] (-1, -0.5)
    to [out=0, in=270] cycle;
    \draw[fill=darktangerine,fill opacity=0.2, line width=1pt] (-2,0)
    to [out=270,in=180] (1,-2.1)
    to [out=0,in=275] (4,0)
    to [out=265,in=270] (0,0)
    to [out=270, in=0] (-1, -0.5)
    to [out=180, in=270] cycle;
    \draw[color=ForestGreen, line width=2pt] (-2,0) to [out=270, in=180] (-1, -0.5)
    to [out=0, in=270] (0,0);
    \draw[dashed, color=ForestGreen, line width=2pt] (-2,0) to [out=90, in=180] (-1, 0.5)
    to [out=0, in=90] (0,0);
    \draw[blue, line width=1.5pt, ->] (0,0) to [out=90, in=180] (1.8, 1.6);
    \draw[blue, line width=1.5pt] (1.8, 1.6) to [out=0, in=90] (4, 0);
    \draw[orange, line width=1.5pt, ->] (0,0) to [out=270, in=180] (1.8, -1.6);
    \draw[orange, line width=1.5pt] (1.8, -1.6) to [out=0, in=270] (4, 0);
\draw[ForestGreen, fill=ForestGreen] (-2,0) circle (.7ex);
\draw[ForestGreen, fill=ForestGreen] (0,0) circle (.7ex);
\draw[cornellred, fill=cornellred] (4,0) circle (.7ex);
\node at (-2.3, 0) {$a$}; 
\node at (0.3, 0) {$b$};
\node at (4.4, 0) {$x^{\star}$};  
	\end{tikzpicture}
\caption{Illustration of the spectral curve of the cubic matrix model. There are two sheets separated by the cut ({\color{ForestGreen}green}): the physical sheet ({\color{LightBlue}blue}) containing the ``old'' integration contour for the action ({\color{blue}blue}), and the non-physical sheet ({\color{darktangerine}orange}) now containing the ``new'' integration contour giving rise to the (resonant) symmetric action ({\color{orange}orange}).}
\label{fig:cubicspectralcurvebothactions}
\end{figure}

To make it immediately precise, let us uniformize the spectral curve \eqref{eq:hermitianspectralcurve} with a parameter $\upzeta$ on the Riemann sphere, as (see, \textit{e.g.}, \cite{ekr15, bo18})
\begin{align}
\label{eq:uniformizationtransformation}
x(\upzeta) &= \frac{a+b}{2} + \frac{a-b}{4} \left(\upzeta+\frac{1}{\upzeta}\right),\\
y(\upzeta) &= M (x(\upzeta))\, \frac{a-b}{4} \left(\upzeta-\frac{1}{\upzeta}\right),
\end{align}
\noindent
describing a branched double-cover of the complex $x$-plane with two ramification points at $\upzeta = \pm 1$. Swapping sheets is implemented by the involution map $\sigma(\upzeta)$ as
\begin{equation}
\label{eq:SwitchingBetweenTwoSheets}
\upzeta \mapsto \sigma(\upzeta) = \frac{1}{\upzeta},
\end{equation}
\noindent
such that
\begin{equation}
\label{eq:flippingspectralcurve}
x(\sigma(\upzeta)) = x(\upzeta), \qquad y(\sigma(\upzeta)) = - y(\upzeta).
\end{equation}
\noindent
Let us now revisit the calculation of the action \eqref{eq:InstantonActionmsw07} via uniformization. If our nonperturbative saddle $x^\star$ is uniformized by $\upzeta^\star$ as $x^\star = x (\upzeta^\star)$, then there exists a \textit{second point}\footnote{Note that in spite of slightly conflicting notation, in this work the bar \textit{does not} denote complex conjugation.} on the Riemann sphere, $\bar{\upzeta}^\star = \sigma (\upzeta^\star)$, which is also mapped to $x^\star$ via $x (\bar{\upzeta}^\star)$. Performing the instanton-action integral \eqref{eq:InstantonActionmsw07} via uniformization then leads to these corresponding two possibilities:
\begin{align}
\label{eq:naiveminussignforaction}
A &= \int_{b}^{x^{\star}} \text{d}x\,y(x) = \int_{-1}^{\upzeta^{\star}} \text{d}\upzeta\, \frac{\partial x}{\partial \upzeta}\, y(\upzeta),\\
\bar{A} &= \int_{-1}^{\bar{\upzeta}^{\star}} \text{d}\upzeta\, \frac{\partial x}{\partial \upzeta}\, y(\upzeta) = \int_{-1}^{\upzeta^{\star}} \text{d}\upzeta\, \frac{\partial x}{\partial \upzeta}\, y(\sigma(\upzeta)) = - \int_{-1}^{\upzeta^{\star}} \text{d}\upzeta\, \frac{\partial x}{\partial \upzeta}\, y(\upzeta) \equiv -A
\end{align}
\noindent
(where in the second line we changed variables via \eqref{eq:SwitchingBetweenTwoSheets} in the second equality). This shows how a very simple argument gives rise to resonant instanton actions, exactly as predicted by resurgence. Of course non-trivial checks are still due. For the moment let us work under this assumption, and let us further refer to eigenvalues on the physical sheet as simply \textit{eigenvalues}, whereas eigenvalues on the involuted sheet will be denoted by \textit{anti-eigenvalues} (the reason for this will be made clear below). Let us also accordingly update figure~\ref{fig:cubicspectralcurvejustphysicalaction} into our new figure~\ref{fig:cubicspectralcurvebothactions}.

This na\"\i ve picture is consistent with our earlier discussion of the resonant Stokes automorphisms in figure~\ref{fig:PictureTwoAutomorphisms}. As explained therein, ``standard'' eigenvalue tunneling (on the physical sheet) leads to the forward Stokes automorphism $\underline{\mathfrak{S}}_0$, \textit{e.g.}, the \eqref{eq:quarticexampleeigenvaluetunnelingscetch} jump at $\theta=0$. But that same example saw no traces of any backward Stokes transition $\underline{\mathfrak{S}}_\pi$, \textit{e.g.}, there was no $\theta=\pi$ jump throughout \eqref{eq:rotationperturbativecontour}. Should we now add anti-eigenvalues to the discussion, things change. Indeed, via \eqref{eq:flippingspectralcurve}, the holomorphic effective potential will acquire a minus sign when switching sheets, hence anti-eigenvalue perturbative-contours would behave exactly like eigenvalue perturbative-contours \textit{except that rotated by $\pi$}---which is exactly what is needed to materialize $\underline{\mathfrak{S}}_\pi$. This is illustrated in figure~\ref{fig:eigenvaluesandholes}. Such backward Stokes transition \eqref{eq:backwardStokesAutomorphism} would result in (up to first instanton order; compare with \eqref{eq:Z00toZ10}):
\begin{equation}
Z^{(0,0)}(g_{\text{s}}) - \rme^{+ \frac{A}{g_{\text{s}}}}\, \mathsf{S}_{(0,0)\to(0,1)}\, Z^{(0,1)}(g_{\text{s}}) + \cdots.
\end{equation}

\definecolor{deepsaffron}{rgb}{1.0, 0.8, 0.6}
\begin{figure}
\centering
	\begin{tikzpicture}
	\draw[->, line width=2pt] (-5,0) -- (5, 0);
	\draw[->, line width=2pt] (0,-2) -- (0, 2);
	\draw[line width=2pt] (-5, 1.6) -- (-4.6, 1.6);
	\draw[line width=2pt] (-4.6, 1.6) -- (-4.6, 2);
	\node at (-4.8, 1.8) {$g_{\text{s}}$};
	\draw[blue, line width=2pt] (0,0) -- (4.5, -0.5);
	\draw[blue, line width=2pt] (0,0) -- (4.5, 0.5);
	\draw[->, blue, line width=2pt] (4, -0.4) to[bend right =30] (4, 0.4);
	\node[blue] at (5, 0.5) {$\underline{\mathfrak{S}}_0$};
	\draw[color=orange, line width=2pt] (0,0) -- (-4.5, -0.5);
	\draw[color=orange, line width=2pt] (0,0) -- (-4.5, 0.5);
	\draw[->, color=orange, line width=2pt] (-4, 0.4) to[bend right =30] (-4, -0.4);
	\node[color=orange] at (-5, -0.5) {$\underline{\mathfrak{S}}_{\pi}$};
	\begin{scope}[scale=0.7, shift={({-7},{-6})}]
	\draw[color=ForestGreen, line width=2pt] (-1.2,0.9) -- (1.55,0.9);
	\draw[color=ForestGreen, line width=1pt] (-1.03,0.6) -- (1.25,0.6);
	\draw[color=ForestGreen, line width=1pt] (-0.7,0.3) -- (0.85,0.3);
	 \draw[scale=2, domain=-1.3:3, smooth, variable=\x, LightBlue, line width=2pt] plot ({\x}, {\x*\x-2/6*\x*\x*\x});
	 \draw[scale=2, domain=1.55/2:3, smooth, variable=\x, deepsaffron, line width=2pt] plot ({\x}, {-\x*\x+2/6*\x*\x*\x+0.9});
	 \draw[scale=2, domain=-1.3:-0.6, smooth, variable=\x, deepsaffron, line width=2pt] plot ({\x}, {-\x*\x+2/6*\x*\x*\x+0.9});
	 \fill[orange, line width=2pt] (4,-0.89) circle (1.8ex);
	 \draw[orange, line width=2pt] (4,8/3) circle (1.8ex);
	 \draw[orange, line width=2pt, ->] (1.7, 0.3) to[out=310, in=190] (3.8, -1.3);
	\draw[ForestGreen, fill=ForestGreen] (-1.2,0.9) circle (1.1ex);
	 \draw[ForestGreen, fill=ForestGreen] (1.55,0.9) circle (1.1ex);
	 \node[ForestGreen] at (0,1.7) {$N+1$};
	\end{scope}
	\begin{scope}[scale=0.7, shift={({5},{-6})}]
	\draw[color=ForestGreen, line width=2pt] (-1.2,0.9) -- (1.55,0.9);
	\draw[color=ForestGreen, line width=1pt] (-1.03,0.6) -- (1.25,0.6);
	\draw[color=ForestGreen, line width=1pt] (-0.7,0.3) -- (0.85,0.3);
	 \draw[scale=2, domain=-1.3:3, smooth, variable=\x, LightBlue, line width=2pt] plot ({\x}, {\x*\x-2/6*\x*\x*\x});
	 \draw[scale=2, domain=1.55/2:3, smooth, variable=\x, deepsaffron, line width=2pt] plot ({\x}, {-\x*\x+2/6*\x*\x*\x+0.9});
	 \draw[scale=2, domain=-1.3:-0.6, smooth, variable=\x, deepsaffron, line width=2pt] plot ({\x}, {-\x*\x+2/6*\x*\x*\x+0.9});
	\draw[ForestGreen, fill=ForestGreen] (-1.2,0.9) circle (1.1ex);
	 \draw[ForestGreen, fill=ForestGreen] (1.55,0.9) circle (1.1ex);
	 \draw[blue, fill=blue] (4,8/3) circle (1.8ex);
	 \draw[blue, line width=2pt, ->] (1.7, 1.5) to[out=60, in=190] (3.8, 8/3+0.5);
	 \node[ForestGreen] at (0,1.7) {$N-1$};
	\end{scope}
	\end{tikzpicture}
\caption{The mechanics of (anti) eigenvalue tunneling, at leading instanton order and starting from the perturbative sector. {\color{blue}Eigenvalues} are tunneled via the {\color{blue}forward} Stokes automorphism, whilst unaffected by the backwards one. On the other hand, {\color{orange}anti-eigenvalues} only see {\color{orange}backward} Stokes lines, not being affected by forward resurgent motions. As we shall show later in section~\ref{sec:ResurgenceInMatrixModels}, moving an eigenvalue to the non-physical sheet is equivalent to moving a ``hole eigenvalue'' into the physical sheet (\textit{i.e.}, an anti-eigenvalue). Removing one anti-eigenvalue from the perturbative cut hence exactly increases the number of eigenvalues in that same cut by one.}
\label{fig:eigenvaluesandholes}
\end{figure}
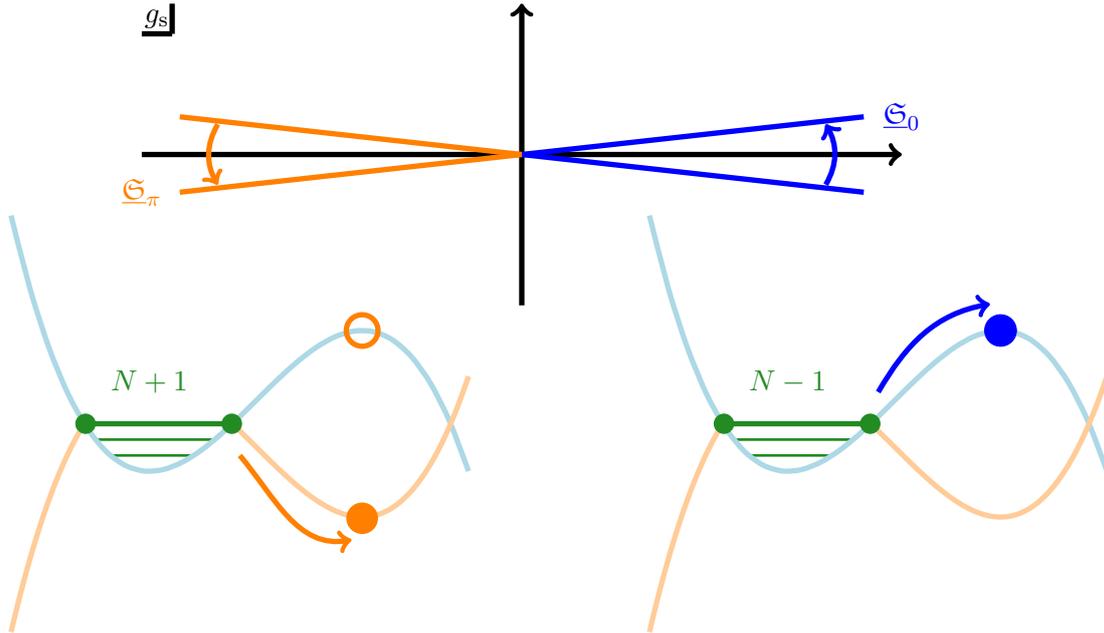

Better understanding the nature of anti-eigenvalues requires \eqref{eq:inversedeterminantrelation}, an expression we shall deduce later but which the reader may now glance at. What \eqref{eq:inversedeterminantrelation} shows is that moving an \textit{eigenvalue} out from the cut into the \textit{non-physical} sheet is equivalent to moving a \textit{``hole''} eigenvalue out of the cut and into the \textit{physical} sheet. This is somewhat reminiscent of the Dirac sea picture, hence the reason for dubbing these as \textit{``anti-eigenvalues''}. What \eqref{eq:inversedeterminantrelation} further implies is that tunneling of an anti-eigenvalue away from the cut starts with a ``creation'' of an eigenvalue--anti-eigenvalue pair, before having an anti-eigenvalue at our disposal which can be then be moved away---a process which leaves one extra eigenvalue on the perturbative cut. Furthermore, due to the nature of the Vandermonde interaction, eigenvalues and anti-eigenvalues feel an attractive force just like particles and anti-particles. This is also illustrated in figure~\ref{fig:eigenvaluesandholes}. The idea that ``different types'' of eigenvalues could be needed to account for resonant resurgence was first suggested in \cite{kmr10}. Anti-eigenvalues now make this precise.

\begin{figure}
\centering
	\begin{tikzpicture}[
	grayframe/.style={
		rectangle,
		draw=gray,
		text width=5em,
		align=center,
		rounded corners,
		minimum height=2em
	}, scale =1, line width=2
	]
	\node[grayframe] (a) at (0,0) {$Z^{(\ell,0)}(g_{\text{s}})$};
	\node[grayframe] (b) at (5, 0) {$Z^{(\ell+1,0)}(g_{\text{s}})$};
	\node[grayframe] (c) at (-5, 0) {$Z^{(\ell-1,0)}(g_{\text{s}})$};
	\node[grayframe] (d) at (0, 3) {$Z^{(\ell,1)}(g_{\text{s}})$};
	\node[grayframe] (e) at (-5, 3) {$Z^{(\ell-1,1)}(g_{\text{s}})$};
	\node[grayframe] (f) at (5, 3) {$Z^{(\ell+1,1)}(g_{\text{s}})$};
	\draw[brown, dashed] (c) -- (d);
	\draw[brown, dashed] (a) -- (f);
	\draw[brown, dashed] (e) -- (-5+0.2*5, 3+0.2*3);
	\draw[brown, dashed] (d) -- (0.2*5, 3+0.2*3);
	\draw[brown, dashed] (f) -- (5+0.2*5, 3+0.2*3);
	\draw[brown, dashed] (b) -- (5+0.2*5, 0+0.2*3);
	\draw[->, line width=2, blue] (a) -- (b);
	\draw[->, line width=2, orange] (a) -- (c);
	\draw[->, line width=2, orange] (a) -- (d);
	\node[orange] at (-2.4, 0.5) {$\mathsf{S}_{(\ell,0)\to(\ell-1,0)}$};
	\node[orange] at (-1.1, 1.4) {$\mathsf{S}_{(\ell,0)\to(\ell,1)}$};
	\node[blue] at (2.4, 0.5) {$\mathsf{S}_{(\ell,0)\to(\ell+1,0)}$};
	\begin{scope}[scale=0.5, shift={({-12},{-7})}]
	\draw[color=ForestGreen, line width=2pt] (-1.2,0.9) -- (1.55,0.9);
	\draw[color=ForestGreen, line width=1pt] (-1.03,0.6) -- (1.25,0.6);
	\draw[color=ForestGreen, line width=1pt] (-0.7,0.3) -- (0.85,0.3);
	\draw[color=Maroon, line width=1pt] (4-0.8,8/3-0.3) -- (4+0.7,8/3-0.3);
	 \draw[scale=2, domain=-1.3:3, smooth, variable=\x, LightBlue, line width=2pt] plot ({\x}, {\x*\x-2/6*\x*\x*\x});
	 \draw[scale=2, domain=1.55/2:3, smooth, variable=\x, deepsaffron, line width=2pt] plot ({\x}, {-\x*\x+2/6*\x*\x*\x+0.9});
	 \draw[scale=2, domain=-1.3:-0.6, smooth, variable=\x, deepsaffron, line width=2pt] plot ({\x}, {-\x*\x+2/6*\x*\x*\x+0.9});
	 \fill[orange, line width=2pt] (4,-0.89) circle (1.8ex);
	 \draw[orange, line width=2pt] (4,8/3) circle (1.8ex);
	 \draw[orange, line width=2pt, ->] (1.7, 0.3) to[out=310, in=190] (3.8, -1.3);
	\draw[ForestGreen, fill=ForestGreen] (-1.2,0.9) circle (1.1ex);
	 \draw[ForestGreen, fill=ForestGreen] (1.55,0.9) circle (1.1ex);
	 \node[ForestGreen] at (0,1.7) {$N-\ell+1$};
	 \node[Maroon] at (4,1.1) {$(\ell| 1)$};
	\end{scope}
	\begin{scope}[scale=0.5, shift={({-1},{-7})}]
	\draw[color=ForestGreen, line width=2pt] (-1.2,0.9) -- (1.55,0.9);
	\draw[color=ForestGreen, line width=1pt] (-1.03,0.6) -- (1.25,0.6);
	\draw[color=ForestGreen, line width=1pt] (-0.7,0.3) -- (0.85,0.3);
	\draw[color=Maroon, line width=1pt] (4-0.8,8/3-0.3) -- (4+0.7,8/3-0.3);
	 \draw[scale=2, domain=-1.3:3, smooth, variable=\x, LightBlue, line width=2pt] plot ({\x}, {\x*\x-2/6*\x*\x*\x});
	 \draw[scale=2, domain=1.55/2:3, smooth, variable=\x,deepsaffron, line width=2pt] plot ({\x}, {-\x*\x+2/6*\x*\x*\x+0.9});
	 \draw[scale=2, domain=-1.3:-0.6, smooth, variable=\x,deepsaffron, line width=2pt] plot ({\x}, {-\x*\x+2/6*\x*\x*\x+0.9});
	\draw[ForestGreen, fill=ForestGreen] (-1.2,0.9) circle (1.1ex);
	 \draw[ForestGreen, fill=ForestGreen] (1.55,0.9) circle (1.1ex);
	 \node[ForestGreen] at (0,1.7) {$N-\ell$};
	 \node[Maroon] at (4,1.1) {$(\ell| 0)$};
	\end{scope}
	\begin{scope}[scale=0.5, shift={({9},{-7})}]
	\draw[color=ForestGreen, line width=2pt] (-1.2,0.9) -- (1.55,0.9);
	\draw[color=ForestGreen, line width=1pt] (-1.03,0.6) -- (1.25,0.6);
	\draw[color=ForestGreen, line width=1pt] (-0.7,0.3) -- (0.85,0.3);
	\draw[color=Maroon, line width=1pt] (4-0.8,8/3-0.3) -- (4+0.7,8/3-0.3);
	 \draw[scale=2, domain=-1.3:3, smooth, variable=\x, LightBlue, line width=2pt] plot ({\x}, {\x*\x-2/6*\x*\x*\x});
	 \draw[scale=2, domain=1.55/2:3, smooth, variable=\x, deepsaffron, line width=2pt] plot ({\x}, {-\x*\x+2/6*\x*\x*\x+0.9});
	 \draw[scale=2, domain=-1.3:-0.6, smooth, variable=\x, deepsaffron, line width=2pt] plot ({\x}, {-\x*\x+2/6*\x*\x*\x+0.9});
	\draw[ForestGreen, fill=ForestGreen] (-1.2,0.9) circle (1.1ex);
	 \draw[ForestGreen, fill=ForestGreen] (1.55,0.9) circle (1.1ex);
	 \draw[blue, fill=blue] (4,8/3) circle (1.8ex);
	 \draw[blue, line width=2pt, ->] (1.7, 1.5) to[out=60, in=190] (3.8, 8/3+0.5);
	 \node[ForestGreen] at (0,1.7) {$N-\ell-1$};
	 \node[Maroon] at (3.9,1.1) {$(\ell+1| 0)$};
	\end{scope}
	\begin{scope}[scale=0.5,  shift={({-11},{-16})}]
	\fill[darktangerine,fill opacity=0.2] (-5, 0) -- (3, 0) -- (5,2) -- (-3,2) -- cycle;
	\draw[line width=1pt] (-3,2) -- (-5, 0) -- (3, 0) -- (5,2);
	\draw[line width=1pt] (-3, 2) -- (-2.5, 2);
	\draw[line width=1pt] (-1, 2) -- (1.95, 2);
	\draw[line width=1pt] (2.05, 2) -- (5, 2);
	\draw[fill=LightBlue,fill opacity=0.2,line width=1pt] (-5, 3) -- (3, 3) -- (5,5) -- (-3,5) -- cycle;
	\fill[darktangerine,fill opacity=0.2] (-2.5,2) to[out=90, in=270] (-2.5, 2.5) to[out=270, in=180] (-1.75,2.2) to[out=0, in=270] (-1,2.5) to[out=270, in=90] (-1,2) -- cycle; 
	\fill[LightBlue,fill opacity=0.2] (-2.5,3) to[out=270, in=90] (-2.5, 2.5) to[out=270, in=180] (-1.75,2.2) to[out=0, in=270] (-1,2.5) to[out=90, in=270] (-1,3) -- cycle;  
	\draw[dashed, line width=1pt] (-3, 4) to[out=0, in=100] (-2.5,3);
	\draw[dashed, line width=1pt] (-0.5, 4) to[out=180, in=80] (-1,3);
	\draw[line width=1pt] (-2.5, 3) to[out=272, in=90] (-2.5,2.5)        to[out=270, in=0] (-3,1);
	\draw[line width=1pt] (-1, 3) to[out=268, in=90] (-1,2.5) to[out=270, in=180] (-0.5,1);
	\draw[color=ForestGreen, line width=1.5pt] (-2.5, 2.5) to[out=270, in=180] (-1.75,2.2) to[out=0, in=270] (-1,2.5);
	\draw[dashed, color=ForestGreen, line width=1.5pt] (-2.5, 2.5) to[out=90, in=180] (-1.75,2.8) to[out=0, in=90] (-1,2.5);
	\draw[ForestGreen, fill=ForestGreen] (-2.5,2.5) circle (.5ex);
	\draw[ForestGreen, fill=ForestGreen] (-1,2.5) circle (.5ex);
	\fill[darktangerine,fill opacity=0.2] (2,2.5) -- (1.95, 2) -- (2.05,2) -- cycle;
	\fill[LightBlue,fill opacity=0.2] (2,2.5) -- (1.9, 3) -- (2.1,3) -- cycle;
	\draw[dashed, line width=1pt] (1, 4) to[out=0, in=110] (1.9,3);
	\draw[dashed, line width=1pt] (3, 4) to[out=180, in=70] (2.1,3);
	\draw[line width=1pt] (1.9, 3) to[out=290, in=90] (2,2.5)        to[out=270, in=0] (1,1);
	\draw[line width=1pt] (2.1, 3) to[out=250, in=90] (2,2.5) to[out=270, in=180] (3,1);
	\draw[Maroon, fill=Maroon] (2,2.5) circle (.7ex);
	\draw[orange, line width=1.5, ->] (-1, 2.4) to[out=240, in=180] (-0.5, 0.8) to[out=5, in=175] (1.5, 0.8)
	to[out=0, in=270] (2, 2.2);
	\draw[orange, line width=1.5] (2,2.5) circle (1.8ex);
	\end{scope}
	\begin{scope}[scale=0.5,  shift={({0},{-16})}]
	\fill[darktangerine,fill opacity=0.2] (-5, 0) -- (3, 0) -- (5,2) -- (-3,2) -- cycle;
	\draw[line width=1pt] (-3,2) -- (-5, 0) -- (3, 0) -- (5,2);
	\draw[line width=1pt] (-3, 2) -- (-2.5, 2);
	\draw[line width=1pt] (-1, 2) -- (1.95, 2);
	\draw[line width=1pt] (2.05, 2) -- (5, 2);
	\draw[fill=LightBlue,fill opacity=0.2,line width=1pt] (-5, 3) -- (3, 3) -- (5,5) -- (-3,5) -- cycle;
	\fill[darktangerine,fill opacity=0.2] (-2.5,2) to[out=90, in=270] (-2.5, 2.5) to[out=270, in=180] (-1.75,2.2) to[out=0, in=270] (-1,2.5) to[out=270, in=90] (-1,2) -- cycle; 
	\fill[LightBlue,fill opacity=0.2] (-2.5,3) to[out=270, in=90] (-2.5, 2.5) to[out=270, in=180] (-1.75,2.2) to[out=0, in=270] (-1,2.5) to[out=90, in=270] (-1,3) -- cycle;  
	\draw[dashed, line width=1pt] (-3, 4) to[out=0, in=100] (-2.5,3);
	\draw[dashed, line width=1pt] (-0.5, 4) to[out=180, in=80] (-1,3);
	\draw[line width=1pt] (-2.5, 3) to[out=272, in=90] (-2.5,2.5)        to[out=270, in=0] (-3,1);
	\draw[line width=1pt] (-1, 3) to[out=268, in=90] (-1,2.5) to[out=270, in=180] (-0.5,1);
	\draw[color=ForestGreen, line width=1.5pt] (-2.5, 2.5) to[out=270, in=180] (-1.75,2.2) to[out=0, in=270] (-1,2.5);
	\draw[dashed, color=ForestGreen, line width=1.5pt] (-2.5, 2.5) to[out=90, in=180] (-1.75,2.8) to[out=0, in=90] (-1,2.5);
	\draw[ForestGreen, fill=ForestGreen] (-2.5,2.5) circle (.5ex);
	\draw[ForestGreen, fill=ForestGreen] (-1,2.5) circle (.5ex);
	\fill[darktangerine,fill opacity=0.2] (2,2.5) -- (1.95, 2) -- (2.05,2) -- cycle;
	\fill[LightBlue,fill opacity=0.2] (2,2.5) -- (1.9, 3) -- (2.1,3) -- cycle;
	\draw[dashed, line width=1pt] (1, 4) to[out=0, in=110] (1.9,3);
	\draw[dashed, line width=1pt] (3, 4) to[out=180, in=70] (2.1,3);
	\draw[line width=1pt] (1.9, 3) to[out=290, in=90] (2,2.5)        to[out=270, in=0] (1,1);
	\draw[line width=1pt] (2.1, 3) to[out=250, in=90] (2,2.5) to[out=270, in=180] (3,1);
	\draw[Maroon, fill=Maroon] (2,2.5) circle (.7ex);
	\end{scope}
	\begin{scope}[scale=0.5,  shift={({11},{-16})}]
	\fill[darktangerine,fill opacity=0.2] (-5, 0) -- (3, 0) -- (5,2) -- (-3,2) -- cycle;
	\draw[line width=1pt] (-3,2) -- (-5, 0) -- (3, 0) -- (5,2);
	\draw[line width=1pt] (-3, 2) -- (-2.5, 2);
	\draw[line width=1pt] (-1, 2) -- (1.95, 2);
	\draw[line width=1pt] (2.05, 2) -- (5, 2);
	\draw[fill=LightBlue,fill opacity=0.2,line width=1pt] (-5, 3) -- (3, 3) -- (5,5) -- (-3,5) -- cycle;
	\fill[darktangerine,fill opacity=0.2] (-2.5,2) to[out=90, in=270] (-2.5, 2.5) to[out=270, in=180] (-1.75,2.2) to[out=0, in=270] (-1,2.5) to[out=270, in=90] (-1,2) -- cycle; 
	\fill[LightBlue,fill opacity=0.2] (-2.5,3) to[out=270, in=90] (-2.5, 2.5) to[out=270, in=180] (-1.75,2.2) to[out=0, in=270] (-1,2.5) to[out=90, in=270] (-1,3) -- cycle;  
	\draw[dashed, line width=1pt] (-3, 4) to[out=0, in=100] (-2.5,3);
	\draw[dashed, line width=1pt] (-0.5, 4) to[out=180, in=80] (-1,3);
	\draw[line width=1pt] (-2.5, 3) to[out=272, in=90] (-2.5,2.5)        to[out=270, in=0] (-3,1);
	\draw[line width=1pt] (-1, 3) to[out=268, in=90] (-1,2.5) to[out=270, in=180] (-0.5,1);
	\draw[color=ForestGreen, line width=1.5pt] (-2.5, 2.5) to[out=270, in=180] (-1.75,2.2) to[out=0, in=270] (-1,2.5);
	\draw[dashed, color=ForestGreen, line width=1.5pt] (-2.5, 2.5) to[out=90, in=180] (-1.75,2.8) to[out=0, in=90] (-1,2.5);
	\draw[ForestGreen, fill=ForestGreen] (-2.5,2.5) circle (.5ex);
	\draw[ForestGreen, fill=ForestGreen] (-1,2.5) circle (.5ex);
	\fill[darktangerine,fill opacity=0.2] (2,2.5) -- (1.95, 2) -- (2.05,2) -- cycle;
	\fill[LightBlue,fill opacity=0.2] (2,2.5) -- (1.9, 3) -- (2.1,3) -- cycle;
	\draw[dashed, line width=1pt] (1, 4) to[out=0, in=110] (1.9,3);
	\draw[dashed, line width=1pt] (3, 4) to[out=180, in=70] (2.1,3);
	\draw[line width=1pt] (1.9, 3) to[out=290, in=90] (2,2.5)        to[out=270, in=0] (1,1);
	\draw[line width=1pt] (2.1, 3) to[out=250, in=90] (2,2.5) to[out=270, in=180] (3,1);
	\draw[blue, fill=blue] (2,2.5) circle (1.8ex);
	\draw[blue, line width=1.5, ->] (-1, 2.6) to[out=110, in=170] (-0.5, 4.2) to[out=5, in=185] (1.5, 4.2)
	to[out=5, in=90] (2, 2.8);
	\end{scope}
	\end{tikzpicture}
\caption{Eigenvalue configurations describing the leading asymptotics of the $(\ell,0)$ transseries sector. First we depict the resurgent lattice of resonant sectors, illustrating resurgent contributions to the asymptotics from {\color{blue}forward} and {\color{orange}backward} transitions. Then we show what should be their associated (anti) eigenvalue configurations, and how tunneling anti-eigenvalues relates to the backward Stokes automorphism, whereas tunneling eigenvalues relates to the forward one. Finally we include the exact same illustration but for the spectral curve. It is important to note that (anti) eigenvalue configurations \textit{do not} strictly match transseries sectors, but rather \textit{match their resonant kernels} (represented by the brown dashed-lines of the topmost plot). These present plots build on figures first appearing in \cite{kmr10}.}
\label{fig:illustrationoftunnelingforresonantcase}
\end{figure}
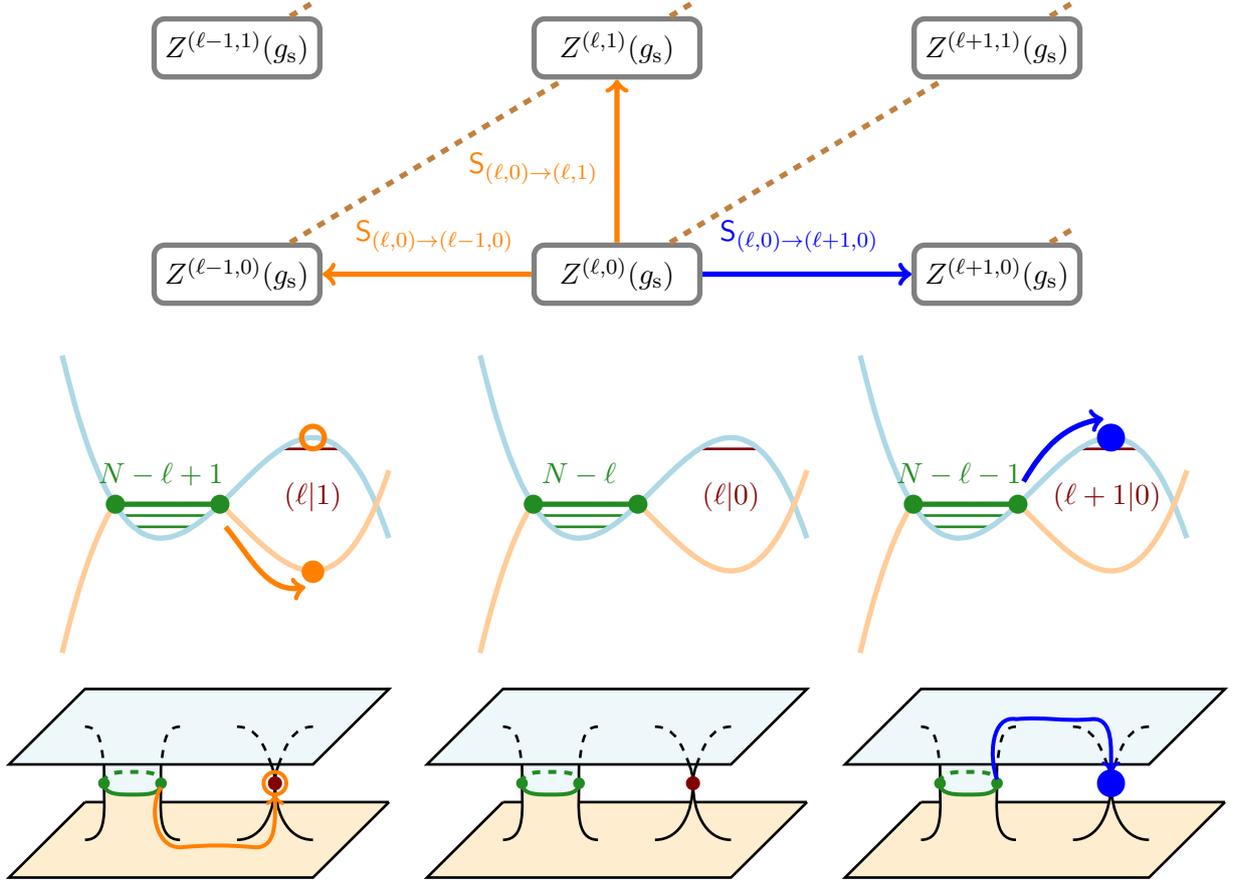

Finally, our simple picture of eigenvalue and anti-eigenvalue tunneling also gives clear intuition on how to access any sector on the full resonant transseries in figure~\ref{fig:walkingonesteppathsintro}. Consider a matrix-model configuration where $N-\ell$ eigenvalues sit in the perturbative cut and $\ell$ eigenvalues---and only eigenvalues---have tunneled away to the nonperturbative saddle. Let us denote this as the $(\ell|0)$ eigenvalue configuration, which we might at first expect would precisely match the $(\ell,0)$ transseries sector in \eqref{eq:transseriesexpansionGeneric} (note how we used \textit{different notation} to distinguish a $(n|\bar{m})$ (anti) eigenvalue configuration from a $(n,m)$ transseries sector as in \eqref{eq:transseriesexpansionGeneric-sectors}). This certainly seems to be the case going forward: from the transseries side the forward asymptotics is controlled by \eqref{eq:forwardStokesAutomorphism} with $(n,m)=(\ell,0)$, which at leading order resurges the $(\ell+1, 0)$ sector---and this is indeed known to correspond to tunneling a single eigenvalue to the $(\ell+1|0)$ configuration \cite{msw07}. However going backwards again leads to novelties. From the transseries side the asymptotics is now controlled by \eqref{eq:backwardStokesAutomorphism} with $(n,m)=(\ell,0)$, which at leading order now resurges \textit{both} $(\ell-1, 0)$ \textit{and} $(\ell, 1)$ sectors. It turns out that \textit{both} these contributions will arise from the \textit{single}\footnote{There is a subtlety which will be made clear later. Strictly speaking, this discussion only applies to free-energy transseries sectors, not partition-function transseries sectors. At the partition function level the $(\ell-1|0)$ eigenvalue configuration also contributes (albeit being absent for free energy). This is associated to the different natures of partition function and free energy transseries, and will be further discussed in appendix~\ref{app:resurgenceZvsF}. See \cite{abs18} as well.} $(\ell|1)$ configuration. In other words, we find \textit{two} transseries contributions being described by a \textit{single} eigenvalue configuration---a hallmark of \textit{resonance}. We illustrate this phenomenon in figure~\ref{fig:illustrationoftunnelingforresonantcase}.

Our two main points herein are the following. First, both Stokes transitions $\underline{\mathfrak{S}}_0$ and $\underline{\mathfrak{S}}_\pi$ may be accounted for once considering both eigenvalues and anti-eigenvalues. Second, the (anti) eigenvalues configurations to consider are not directly related to transseries sectors---but instead to their resonant combinations, already appearing in \eqref{eq:forwardStokesAutomorphism} and \eqref{eq:backwardStokesAutomorphism}. All these ideas will be made precise in the next section~\ref{sec:ResurgenceInMatrixModels}, and illustrated/checked with several examples in section~\ref{sec:ExamplesAndChecks}.

\subsection{Brief Comments on Supermatrix Models}\label{subsec:CommentOnSuperMatrixModel}

Let us end this section with some brief comments concerning supermatrix models. In fact the matrix integrals we will find in upcoming section~\ref{sec:ResurgenceInMatrixModels} have a  strong supermatrix model flavor \cite{am91, y91, mp09} (\textit{physical} supermatrix model, to be more precise) and it is interesting to first recall some of their generalities. That hermitian matrix models may have supermatrix reformulations is nothing new and arises for example when computing matrix-model correlation functions involving ratios of determinant insertions \cite{bh02, k14, bh18} (but see as well \cite{be09}), somewhat similar to what will happen for us (these are also useful tools in condensed matter and chaos, \textit{e.g.}, \cite{e96, hgk18}).

The supermatrix model partition-function is given by (compare with \eqref{eq:partitionfunctionhermitianmatrix})
\begin{equation}
\label{eq:supermatrixpartitionfunction}
\BZ_{N,\bar{N}} (g_{\text{s}}) = \int \text{d}\mathbb{M}\, \text{e}^{-\frac{1}{g_{\text{s}}}\,\text{Str}\, V(\mathbb{M})},
\end{equation}
\noindent
with $\mathbb{M}$ an hermitian $(N+\bar{N}) \times (N+\bar{N})$ supermatrix,
\begin{equation}
\label{eq:supermatrix}
\mathbb{M} = \begin{pmatrix}
A & \Psi \\
\Psi^{\dagger} & B
\end{pmatrix}.
\end{equation}
\noindent
Here $A$ and $B$ are, respectively, $N \times N$ and $\bar{N}\times \bar{N}$ hermitian matrices, and $\Psi$ is Grassmann odd with appropriate matrix size. There are two main types of supermatrix models, which essentially differ in how to perform the above integral: \textit{ordinary} and \textit{physical} supermatrix models \cite{am91, y91}.

\paragraph{Ordinary Supermatrix Models:}

In the ordinary supermatrix model \cite{am91, y91, de09}, the supermatrix partition function \eqref{eq:supermatrixpartitionfunction} is evaluated by considering matrices $A$ and $B$ with no ``soul'', \textit{i.e.}, matrices whose entries are pure-numbers not built out of combinations of Grassmann-odd contributions \cite{i15}. One may then integrate out the Grassmann contributions by integrating the $\Psi$. It turns out that, at perturbative level, such resulting supermatrix model reduces to the hermitian matrix model \eqref{eq:partitionfunctionhermitianmatrix} \cite{am91}. This type of supermatrix models plays no role in our analysis.

\paragraph{Physical Supermatrix Models:}

The physical supermatrix model proceeds along a distinct route \cite{y91}, where one instead focuses on immediately trying to diagonalize the supermatrix partition-function analogously to what was done in the hermitian case. This procedure, however, comes with a few caveats. Immediately, the volume of the super-unitary group can vanish in which case an hermitian-like reasoning cannot be followed (but see as well \cite{amu95}). But even if one just blindly diagonalizes and then simply drops this problematic volume factor, one is led to find the super-Vandermonde determinant (also known as the Berezinian) in the integrand,
\begin{equation}
\label{eq:supervandermondedeterminant}
\Delta_{N,\bar{N}} \left(\lambda,\bar{\lambda}\right) = \Delta_{N}(\lambda)\, \Delta_{\bar{N}}(\bar{\lambda})\, \prod\limits_{i=1}^{N} \prod\limits_{j=1}^{\bar{N}} \frac{1}{\lambda_i-\bar{\lambda}_j}.
\end{equation}
\noindent
This diverges whenever $\lambda_i=\bar{\lambda}_j$ hence turning the resulting eigenvalue integral ill-defined. One natural way to make sense of these resulting integrals is via a principal-value prescription. Consider two integration contours, $\CC$ and $\bar{\CC}$ parameterized by $x(\tau)$ and $\bar{x}(\bar{\tau})$, which meet at a point $x_0$ where the parameterization is chosen such that $x(0)=x_0=\bar{x}(0)$. Then, for $\epsilon>0$ and any integrand $f \left(x,\bar{x}\right)$, define\footnote{This definition includes a few technical assumptions. As usual, the parametrization of the contour integral must be generically piecewise-smooth; but we also ask it will be \textit{smooth inside the disk} which was removed around the pole. Further, this disk must be \textit{symmetric}, \textit{i.e.}, the same $\epsilon$ is chosen for both contour parametrizations.} the principle-value integral at $x_0$ by
\newcommand{\pvint}{\,\text{PV}\hspace{-4pt}\int}
\begin{equation}
\label{eq:principalvalueprescription}
\pvint_\CC \text{d}x\, \pvint_{\bar{\CC}} \text{d}\bar{x}\, f \left(x,\bar{x}\right) := \lim\limits_{\epsilon\to 0} \int\limits_{(-\infty,-\epsilon)\cup(\epsilon,+\infty)}\text{d}\tau \int\limits_{(-\infty,-\epsilon)\cup(\epsilon,+\infty)} \text{d}\bar{\tau}\,\, \dot{x}(\tau)\, \dot{\bar{x}}(\bar{\tau})\, f \left(x(\tau),\bar{x}(\bar{\tau})\right).
\end{equation}
\noindent
If contours intersect multiple times, this prescription still applies via implementation of piecewise parameterizations. In this way, one arrives at the definition of the physical supermatrix partition-function,
\begin{equation}
\label{eq:eigenvalueformualtionsuperpartitionfunction-super}
\BZ_{N,\bar{N}} (g_{\text{s}}) = \frac{1}{N! \bar{N}!}\, \pvint \prod\limits_{i=1}^{N} \frac{\text{d}\lambda_i}{2\pi}\, \prod\limits_{j=1}^{\bar{N}} \frac{\text{d}\bar{\lambda}_j}{2\pi}\, \Delta_{N,\bar{N}} (\lambda,\bar{\lambda})^2\, \text{e}^{-\frac{1}{g_{\text{s}}} \left( \sum\limits_{i=1}^{N} V(\lambda_i) - \sum\limits_{j=1}^{\bar{N}} V(\bar{\lambda}_j) \right)},
\end{equation}
\noindent
where we have further added the same combinatorial pre-factors as for the hermitian matrix models (compare with \eqref{eq:partitionfunctioneigenvalues}). The physical supermatrix nomenclature is the same as ours: the $\left\{ \lambda_i \right\}$ are eigenvalues and the $\left\{ \bar{\lambda}_j \right\}$ are anti-eigenvalues. When $\bar{N}=0$, this definition trivially reduces to the hermitian matrix model in \eqref{eq:partitionfunctioneigenvalues}. The idea that supermatrix models could be used to account for resonant resurgence was first suggested in \cite{mp09}. The upcoming match between non-physical sheet matrix integrals and physical supermatrix integrals now makes this precise.

The role of supergroups in the description of nonperturbative D-brane string physics has been advocated in \cite{v01, ot06, v14, dhjv16}, within the context of so-called ``ghost'' or ``negative'' D-branes. In fact a relation between supermatrix models and spectral-curve sheet-flipping was also suggested in \cite{v14}. In our context, this will be made fully explicit. As such, our set-up is rather reminiscent of some of those ideas, and it would be interesting to explore if our computational methods in the next section may further help in the quantitative analysis of these types of D-branes.

\section{Resurgent Instanton Calculus in Matrix Models}\label{sec:ResurgenceInMatrixModels}

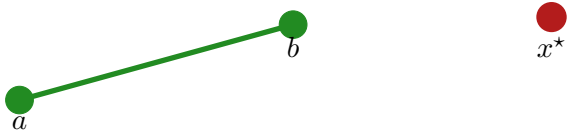
\begin{figure}
\centering
	\begin{tikzpicture}
		\draw[ForestGreen, line width=2pt] (-2, -1) -- (1.6, 0);
		\draw[ForestGreen, fill=ForestGreen] (-2,-1) circle (1.1ex);
		\node at (-2, -1.3) {$a$};
		\draw[ForestGreen, fill=ForestGreen] (1.6,0) circle (1.1ex);
		\node at (1.6, -.3) {$b$};
		\draw[cornellred, fill=cornellred, line width=1pt] (5,0.1) circle (1.1ex);
		\node at (5, -0.3) {$x^{\star}$};
	\end{tikzpicture}
\caption{Our schematic spectral curve. One single cut from $a$ to $b$ where the majority of the eigenvalues sit ({\color{ForestGreen} green}) and a nonperturbative saddle at $x^{\star}$ ({\color{cornellred} red}). In this paper we focus on the setting of a single non-trivial saddle-point. This basic plot will appear again in figures~\ref{fig:settingcutandsaddleStandardtunneling} and~\ref{fig:settingcutandsaddleGeneraltunneling} to help illustrate the required deformations of eigenvalue contours.}
\label{fig:settingcutandsaddle}
\end{figure}

Having outlined context and ideas on tunneling of eigenvalues and anti-eigenvalues, let us now make the previous heuristics fully precise. After reviewing the required matrix-model technology, we shall discuss our relevant (determinant) correlation functions---whose corresponding matrix-integrals must be carefully defined---and then, iteratively, obtain the different sectors appearing in the transseries lattice of figure~\ref{fig:walkingonesteppathsintro}. We focus on the one-cut spectral-curve setting depicted in figure~\ref{fig:settingcutandsaddle}, where, without loss of generality, we assume that the nonperturbative saddle-point $x^{\star}$ yields real action $A$ (such that the resulting resonant Stokes lines for $\underline{\mathfrak{S}}_0$ and $\underline{\mathfrak{S}}_{\pi}$ are basically those already illustrated in figures~\ref{fig:PictureTwoAutomorphisms} or~\ref{fig:eigenvaluesandholes}).

\definecolor{coralpink}{rgb}{0.97, 0.51, 0.47}

\subsection{Overview of the Matrix Model Set-Up}\label{subsec:matrixsetup}

Let us first collect several matrix model formulae which will play prominent roles throughout this section (we refer the reader to a vast literature; \textit{e.g.}, \cite{bipz78, ackm93, dgz93, m04, eo07a, msw07, ekr15}). We will be considering the hermitian matrix model \eqref{eq:partitionfunctioneigenvalues} around the one-cut perturbative saddle \eqref{eq:hermitianspectralcurve}, with generic gauge-invariant observables $\CO(M)$ computed via (already in diagonal gauge)
\begin{equation}
\label{eq:hermitianexpectationvalueeigenvalues}
\ev{\CO(M)} = \frac{1}{Z_{N}}\, \frac{1}{N!} \int \prod\limits_{i=1}^{N} \frac{\text{d}\lambda_i}{2\pi}\, \Delta_{N}(\lambda)^2\, \CO(\lambda)\, \text{e}^{-\frac{1}{g_{\text{s}}} \sum\limits_{i=1}^{N} V(\lambda_i)}.
\end{equation}
\noindent
A key such set of correlation functions are the main players of the topological recursion \cite{eo07a}
\begin{equation}
\label{eq:hermitiancorrelatorsgeneratinginteractions}
W_{h} \left( p_1, \ldots, p_h \right) = \ev{\text{Tr} \frac{1}{p_1-M} \cdots \text{Tr} \frac{1}{p_h-M}}_{(\text{c})},
\end{equation}
\noindent
with genus expansion in $g_{\text{s}}$ given by
\begin{equation}
\label{eq:wcorrelatorgsexpansionhermitian}
W_{h} \left(p_1, \ldots, p_h\right) \simeq \sum\limits_{g=0}^{\infty} W_{g, h} \left( p_1, \ldots, p_h; t \right) g_{\text{s}}^{2g+h-2}.
\end{equation}
\noindent
For the upcoming determinant-correlators we will have to compute, it is also useful to introduce integrations of the above multi-resolvents \cite{msw07},
\begin{align}
\label{eq:hermitianlogarithmiccorrelators}
A_{h} \left( x_1, \ldots, x_h \right) = \ev{\text{Tr} \log \left(x_1-M\right) \cdots \text{Tr} \log \left(x_h-M\right)}_{(\text{c})}
\end{align}
\noindent
computed via\footnote{The integration constants are fixed by the expansion into multi-trace correlators; see \cite{msw07}.}
\begin{align}
A_{h} \left(x_1, \ldots, x_h\right) = \int^{x_1} \text{d}p_1 \cdots \int^{x_h} \text{d}p_h\, W_{h} \left(p_1, \ldots, p_h\right).
\end{align}
\noindent
Clearly these integrated correlators inherit the $g_{\text{s}}$ expansion from \eqref{eq:wcorrelatorgsexpansionhermitian}. This translates to
\begin{align}
A_{h} \left(x_1, \ldots, x_h\right) \simeq \sum\limits_{g=0}^{\infty} A_{g; h} \left( x_1, \ldots, x_h; t \right) g_{\text{s}}^{2g+h-2}.
\end{align}

One of our main tasks going forward will be the computation of correlators involving ratios of determinant insertions (see \cite{be09} as well), and these may now be expressed in terms of the previous functions $A_{h_1, h_2, \dots, h_n}$. We find
\begin{align}
\label{eq:allorderratioofdeterminants}
\ev{\frac{\det \left(p_1-M\right)^2 \cdots \det \left(p_h-M\right)^2}{\det \left(q_{\bar{1}}-M\right)^2 \cdots \det \left(q_{\bar{h}}-M\right)^2}} &= \exp \Bigg\{ \underset{(n_1, \ldots, n_{h}, m_{\bar{1}}, \ldots, m_{\bar{h}}) \neq (0, \ldots, 0, 0, \ldots, 0)} {\sum\limits_{n_1=0}^{\infty} \cdots \sum\limits_{n_h=0}^{\infty}\, \sum\limits_{m_{\bar{1}}=0}^{\infty} \cdots \sum\limits_{m_{\bar{h}}=0}^{\infty}} \frac{2^{\sum_{i=1}^{h}n_i}}{n_1! \cdots n_h!}\,  \times \\
&
\hspace{-160pt}
\times\, \frac{(-2)^{\sum_{j=\bar{1}}^{\bar{h}}m_j}}{m_{\bar{1}}! \cdots m_{\bar{h}}!}\, A_{n_1 + \cdots + n_h + m_{\bar{1}} + \cdots m_{\bar{h}}} \big( \underbrace{p_1, \ldots, p_1}_{n_1},\, \ldots,\, \underbrace{p_h, \ldots, p_h}_{n_h},\, \underbrace{q_{\bar{1}}, \ldots, q_{\bar{1}}}_{m_{\bar{1}}},\, \ldots,\, \underbrace{q_{\bar{h}}, \ldots, q_{\bar{h}}}_{m_{\bar{h}}} \big) \Bigg\}. \nonumber
\end{align}
\noindent
Because we shall use them often later on, let us further spell-out the first few orders in the $g_{\text{s}}$ expansion of this correlator
\begin{align}
\label{eq:gsexpansionofdeterminants}
\ev{\frac{\det \left(p_1-M\right)^2 \cdots \det \left(p_h-M\right)^2}{\det \left(q_{\bar{1}}-M\right)^2 \cdots \det \left(q_{\bar{h}}-M\right)^2}} &= \exp \Bigg\{ \frac{1}{g_{\text{s}}} \left( \sum\limits_{i=1}^{h} 2 A_{0;1}(p_i) - \sum\limits_{j=\bar{1}}^{\bar{h}} 2 A_{0;1}(q_j) \right) + \nonumber \\
&
\hspace{-150pt}
+ \Bigg( \sum\limits_{i=1}^{h} 2 A_{0;2}(p_i,p_i) + \sum\limits_{j=\bar{1}}^{\bar{h}} 2 A_{0;2}(q_j,q_j) + \sum\limits_{i=1}^{h} \sum_{\ell=i+1}^{h} 4 A_{0;2}(p_i, p_\ell) + \sum\limits_{j=\bar{1}}^{\bar{h}} \sum_{k=j+1}^{\bar{h}} 4 A_{0;2}(q_j, q_k) - \nonumber \\
&
- \sum\limits_{i=1}^{h} \sum_{j=\bar{1}}^{\bar{h}} 4 A_{0;2}(p_i, q_j) \Bigg) + O (g_{\text{s}}) \Bigg\}.
\end{align}

Quite some more formulae are needed to tackle all calculations in this section, but which we will not explicitly use. In fact, some of the required steps throughout this section amount to revisiting and upgrading the calculations in \cite{msw07} and those will be omitted. The interested reader may consult \cite{ackm93, iy05, msw07} for a list of useful relations which are implicitly used in the following. Armed with all of these, let us next spell out the precise correlation functions we need to tackle.

\subsection{Determinant Correlators and Spectral Sheets in Matrix Models}\label{subsec:DeterminantCorrelators}

Eigenvalue tunneling on the physical sheet is very well established \cite{d92, hhikkmt04, st04, iy05, iky05, msw07, msw08}. In short, the $\ell$-th matrix-model instanton-sector has $\ell$ eigenvalues tunneled to the nonperturbative saddle at $x^{\star}$ and is given by\footnote{Recall the notation $(n|\bar{m})$ to denote (anti) eigenvalue configurations, as introduced in subsection~\ref{subsec:MatrixIntegralsWork}.}
\begin{equation}
\label{eq:integralkforwardcontributions}
\mathcal{Z}^{(\ell|0)} (t) = \frac{(N-\ell)!}{N!}\, \mathcal{Z}^{(0|0)} (t-\ell g_{\text{s}}) \int_{\CC^{\star}} \prod\limits_{i=1}^{\ell}\frac{\text{d}x_i}{2\pi}\, \Delta_{\ell}(x)^2\, \text{e}^{-\frac{1}{g_{\text{s}}} \sum\limits_{j=1}^{\ell}V(x_j)} \ev{\prod\limits_{k=1}^{\ell} \det \left(x_k-M\right)^{2}}_{N-\ell},
\end{equation}
\noindent
with $\mathcal{Z}^{(0| 0)}$ the perturbative partition function. Notice that, strictly as written, this expression is not fully precise on two accounts. On the one hand, we have not accounted for the required combinatorial factors arising from indistinguishability of eigenvalues \cite{hhikkmt04}. Such factors will appear naturally from the (anti) eigenvalue tunneling arguments in subsection~\ref{subsec:eigenvalueTunnelingContours}, in which case we opt for suppressing them throughout this subsection and only make them explicit later. On the other hand, specifying the contours $\CC^\star$ is absolutely crucial in defining what \eqref{eq:integralkforwardcontributions} means. In the present subsection we shall be focusing on the integrands and on their behavior as eigenvalues change sheets, in which case we opt for leaving the integration contours unspecified until the next subsection. Note how this also affects the left-hand side of the above equality: if the integration contour is \textit{not} to pass through a nonperturbative saddle, we are really \textit{not} computing any instanton sector. As such, all (integrand) formulae which now ensue are valid for arbitrary contours---albeit in the rest of the paper we are only interested in the nonperturbative ones.

We want to understand what happens when tunneling \textit{both}\footnote{We alert the reader not to be misled by notation, as generically $\ell$ and $\bar{\ell}$ are unrelated.} $\ell$ eigenvalues \textit{and} $\bar{\ell}$ anti-eigenvalues away from the cut. Our proposal for this matrix-model contribution is
\begin{align}
\label{eq:finalresultseparatedeigenvalues}
\mathcal{Z}^{(\ell|\bar{\ell})} (t) &= \frac{(N+\bar{\ell}-\ell)!}{(N+\bar{\ell})!\, \bar{\ell}!}\, \mathcal{Z}^{(0|0)} (t-\left(\ell-\bar{\ell}\right) g_{\text{s}}) \pvint \prod\limits_{i=1}^{\ell} \frac{\text{d}x_i}{2\pi}\, \pvint \prod\limits_{j=1}^{\bar{\ell}} \frac{\text{d}\bar{x}_j}{2\pi}\, \times \nonumber \\
&
\times\, \frac{\Delta_{\ell} (x)^2\, \Delta_{\bar{\ell}} (\bar{x})^2}{\prod\limits_{i=1}^{\ell} \prod\limits_{j=1}^{\bar{\ell}} \left(x_i-\bar{x}_j\right)^2}\, \rme^{-\frac{1}{g_{\text{s}}} \sum\limits_{i=1}^{\ell} V(x_i) + \frac{1}{g_{\text{s}}} \sum\limits_{j=1}^{\bar{\ell}} V(\bar{x}_j)} \ev{\frac{\prod\limits_{i=1}^{\ell} \det \left(x_i-M\right)^{2}}{\prod\limits_{j=1}^{\bar{\ell}} \det \left(\bar{x}_j-M\right)^{2}}}_{N+\bar{\ell}-\ell}.
\end{align}
\noindent
The principal-value prescription was defined in \eqref{eq:principalvalueprescription} and will be further discussed in the next subsection~\ref{subsec:eigenvalueTunnelingContours}. This proposal is valid in the 't~Hooft large $N$ limit, where it describes generic configurations with arbitrary number of eigenvalues and anti-eigenvalues tunneled away from the perturbative cut. The ``hole'' nature of anti-eigenvalues is also made clear by the signs in front of the $\bar{\ell}$s (more in the following). Let us also mention that similar types of matrix integrals have previously appeared in the literature, \textit{e.g.}, \cite{fs02a, fs02b, bds03, sss19, apshv22}.

Building upon the standard eigenvalue-tunneling expressions in \cite{hhikkmt04, msw07}, our proposal may be justified rather straightforwardly for the added inclusion of anti-eigenvalues. Let us consider in turns tunneling a single anti-eigenvalue, and then an arbitrary such number.

\paragraph{Tunneling One Anti-Eigenvalue:}

Even simpler, first just recall tunneling of a single eigenvalue \cite{msw07}, corresponding to $\ell=1$ in \eqref{eq:integralkforwardcontributions},
\begin{equation}
\label{eq:integral1tunneledeigenvalemsw07}
\mathcal{Z}^{(1|0)} (t) = \frac{1}{N}\, \mathcal{Z}^{(0|0)} (t-g_{\text{s}}) \int \frac{\text{d}x}{2\pi}\, \text{e}^{-\frac{1}{g_{\text{s}}} V(x)} \ev{\det \left(x-M\right)^{2}}_{N-1}.
\end{equation}

\begin{figure}
\centering
\begin{tikzpicture}[scale=1]
	\draw[fill=LightBlue,fill opacity=0.2, line width=1pt] (0,0)   to [out=90,in=95] (4,0)
	to [out=85,in=0] (1,2.1)
    to [out=180,in=90] (-2,0)
    to [out=270, in=180] (-1, -0.5)
    to [out=0, in=270] cycle;
    \draw[fill=darktangerine,fill opacity=0.2, line width=1pt] (-2,0)
    to [out=270,in=180] (1,-2.1)
    to [out=0,in=275] (4,0)
    to [out=265,in=270] (0,0)
    to [out=270, in=0] (-1, -0.5)
    to [out=180, in=270] cycle;
    \draw[color=ForestGreen, line width=2pt] (-2,0) to [out=270, in=180] (-1, -0.5)
    to [out=0, in=270] (0,0);
    \draw[dashed, color=ForestGreen, line width=2pt] (-2,0) to [out=90, in=180] (-1, 0.5)
    to [out=0, in=90] (0,0);
\draw[ForestGreen, fill=ForestGreen] (-2,0) circle (.7ex);
\draw[ForestGreen, fill=ForestGreen] (0,0) circle (.7ex);
\draw[cornellred, fill=cornellred] (4,0) circle (.7ex);
\node at (-2.3, 0) {$a$}; 
\node at (0.3, 0) {$b$};
\node at (4.4, 0) {$x^{\star}$}; 
\fill (-1.2, 1) circle (.6ex); 
\node at (-1.4, 1.2) {$\infty$};
\fill (-1.2, -1) circle (.6ex); 
\node at (-1.5, -1.2) {$\sigma(\infty)$};
\fill (1.5, 1.6) circle (.6ex);
\node at (1.8, 1.6) {$\zeta$};
\fill (1.5, -1.6) circle (.6ex);
\node at (2.1, -1.6) {$\sigma(\zeta)$};
\draw[line width=2pt, blue, ->] (1.3, -1.55) -- (-1.1, -1);
\draw[line width=2pt, blue, ->] (-1.2, -0.8) -- (-1.2, 0.9);
\draw[line width=2pt, blue, <-] (1.4, 1.6) -- (-1, 1.05);
	\end{tikzpicture}
	\caption{Illustration of the integration paths appearing in equations \eqref{eq:det-integration-contours-eggls-minus} and \eqref{eq:det-integration-contours-eggls-plus}.}
	\label{fig:integrationpathsFlippingDeterminants}
\end{figure}
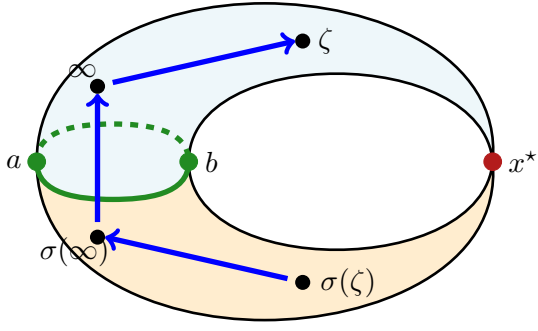

Now, in order to move to the non-physical sheet we need to understand how the determinant insertion behaves under the action of the involution map $\sigma(\upzeta)$ in \eqref{eq:SwitchingBetweenTwoSheets}. For this, it is instructive to first remember the well-known property of how a ratio of determinants is flipped when changing sheets,
\begin{equation}
\label{eq:ratioofdeterminantsflip}
\ev{\frac{\det \left( x(\sigma(\upzeta_1))-M \right)^2}{\det \left( x(\sigma(\upzeta_2))-M \right)^2}} = \ev{\frac{\det \left( x(\upzeta_2)-M\right)^2}{\det \left( x(\upzeta_1)-M \right)^2}}.
\end{equation}
\noindent
This is shown to all orders using the generic determinant-ratio \eqref{eq:allorderratioofdeterminants} and the flipping property of multi-resolvent correlators, $W_h \left( \sigma(\upzeta_1), \upzeta_2, \ldots, \upzeta_h \right) = - W_h \left( \upzeta_1, \upzeta_2, \ldots, \upzeta_h \right)$ \cite{bo18}. It further suggests a relation between a determinant-correlator and an inverse-determinant-correlator. Indeed, one can show the key relation
\begin{align}
\label{eq:inversedeterminantrelation}
\text{e}^{-\frac{1}{g_{\text{s}}} V(x(\sigma(\upzeta)))}\, \ev{\det \left( x(\sigma(\upzeta)) - M \right)^2}_{N-1}\, \mathcal{Z}^{(0|0)} (t-g_{\text{s}}) &= \\
&
\hspace{-85pt}
= \text{e}^{+\frac{1}{g_{\text{s}}} V(x(\upzeta))}\, \ev{\frac{1}{\det \left( x(\upzeta) - M \right)^{2}}}_{N+1}\, \mathcal{Z}^{(0|0)} (t+g_{\text{s}}). \nonumber
\end{align}
\noindent
To show this, follow \cite{eggls23} and consider determinant correlators as deformations of partition functions. Indexing the partition function by these deformations, via paths $\gamma$ that start at $\upzeta_{\text{i}}$ and end at $\upzeta_{\text{f}}$, we can write
\begin{equation}
\mathcal{Z} \left(\gamma_{\upzeta_{\text{i}} \to \upzeta_{\text{f}}}\right) = \exp \left\{ \sum\limits_{g\geq 0,n\geq 0} \frac{g_{\text{s}}^{2g-2+n}}{n!} \int_{\upzeta_{\text{i}}}^{\upzeta_{\text{f}}} \cdots \int_{\upzeta_{\text{i}}}^{\upzeta_{\text{f}}} W_{g,n} \left( x_1, \ldots, x_n \right) \frac{\partial x_1}{\partial \upzeta_1} \cdots \frac{\partial x_n}{\partial \upzeta_n}\, \text{d}\upzeta_1 \cdots \text{d}\upzeta_n \right\},
\end{equation}
\noindent
where the $x_i$ are related to the $\upzeta_i$ via \eqref{eq:uniformizationtransformation}. It then follows \cite{eggls23}
\begin{align}
\label{eq:det-integration-contours-eggls-minus}
\text{e}^{-\frac{1}{g_{\text{s}}} V(x(\upzeta))}\, \ev{\det \left( x(\upzeta) - M \right)^2}_{N-1}\, \mathcal{Z}^{(0|0)} (t-g_{\text{s}}) &= \mathcal{Z} \left( \gamma_{\sigma(\upzeta) \to \sigma(\infty)} + \gamma_{\sigma(\infty) \to \infty} + \gamma_{\infty \to \upzeta} \right) = \nonumber \\
&= \mathcal{Z} \left( \gamma_{\sigma(\upzeta) \to \upzeta} \right), \\
\label{eq:det-integration-contours-eggls-plus}
\text{e}^{\frac{1}{g_{\text{s}}} V(x(\upzeta))}\, \ev{\frac{1}{\det \left( x(\upzeta) - M \right)^{2}}}_{N+1}\, \mathcal{Z}^{(0|0)} (t+g_{\text{s}}) &= \mathcal{Z} \left( \gamma_{\upzeta \to \infty} + \gamma_{\infty \to \sigma(\infty)} + \gamma_{\sigma(\infty) \to \sigma(\upzeta)} \right) = \nonumber \\
&= \mathcal{Z} \left( \gamma_{\upzeta \to \sigma(\upzeta)} \right),
\end{align}
\noindent
where the integration path for the first relation is illustrated in figure~\ref{fig:integrationpathsFlippingDeterminants}. Under the action of the involution map $\sigma(\upzeta)$ the integration direction gets reversed, due to $W_h \left( \sigma(\upzeta_1), \upzeta_2, \ldots, \upzeta_h \right) = - W_h \left( \upzeta_1, \upzeta_2, \ldots, \upzeta_h \right)$ \cite{bo18}. This immediately implies that flipping sheets transforms the above \eqref{eq:det-integration-contours-eggls-minus} into \eqref{eq:det-integration-contours-eggls-plus}, thus yielding what we wanted to show: equation \eqref{eq:inversedeterminantrelation}.

As discussed in subsection~\ref{subsec:MatrixIntegralsWork} and illustrated in figures \ref{fig:eigenvaluesandholes} and \ref{fig:illustrationoftunnelingforresonantcase}, it is precisely relation \eqref{eq:inversedeterminantrelation} which gives rise to the anti-eigenvalue (or ``hole'' in the Dirac sea picture) interpretation. First, via the change in sign in front of the potential (``particle changes to anti-particle''); second, due to the change in the number of eigenvalues upon the cut (``particle--anti-particle pair creation'').

Making use of \eqref{eq:inversedeterminantrelation} back in the one-tunneled-eigenvalue partition function \eqref{eq:integral1tunneledeigenvalemsw07} leads us to the one-tunneled-anti-eigenvalue partition function. We label by $\bar{x}$ the anti-eigenvalue integration variable (to better keep track), and multiply the end result by $N$ in order to make up for the fact that we could have chosen\footnote{Albeit eigenvalue combinatorics is discussed in subsection~\ref{subsec:eigenvalueTunnelingContours}, we have no alternative but to discuss anti-eigenvalue combinatorics herein as we need to decide which of the eigenvalues in the cut is flipped.} any of the $N$ eigenvalues and changed its sheet. One immediately finds
\begin{equation}
\label{eq:separatedoneantieigenvalue}
\mathcal{Z}^{(0|1)} (t) = \mathcal{Z}^{(0|0)} (t+g_{\text{s}}) \int \frac{\text{d}\bar{x}}{2\pi}\, \text{e}^{\frac{1}{g_{\text{s}}} V(\bar{x})} \ev{\frac{1}{\det \left(\bar{x}-M\right)^{2}}}_{N+1}.
\end{equation}
\noindent
This is in agreement with our general proposal \eqref{eq:finalresultseparatedeigenvalues}.

Based on these simple considerations, one may next address a more intricate configuration still with one-tunneled-anti-eigenvalue but now with an arbitrary number of tunneled eigenvalues. The simplest way to do so is to extract these $\ell$ eigenvalues straight from the inverse determinant in the above \eqref{eq:separatedoneantieigenvalue}. First decompose its corresponding $(N+1)$-dimensional matrix-integral as
\begin{align}
\int \prod\limits_{i=1}^{N+1} \frac{\text{d}\lambda_i}{2\pi}\, \frac{\Delta_{N+1}(\lambda)^2}{\prod\limits_{j=1}^{N+1}\left(\bar{x}-\lambda_j\right)^2}\, \rme^{-\frac{1}{g_{\text{s}}} \sum\limits_{k=1}^{N+1}V(\lambda_k)} &= \int \prod\limits_{i=1}^{\ell} \frac{\text{d}x_{i}}{2\pi}\, \frac{\Delta_{\ell}(x)^2}{\prod\limits_{j=1}^{\ell}\left(\bar{x}-x_{j}\right)^2}\, \rme^{-\frac{1}{g_{\text{s}}} \sum\limits_{k=1}^{\ell}V(x_{k})}\, \times \\
&
\hspace{-60pt}
\times\, \int \prod\limits_{i=1}^{N+1-\ell} \frac{\text{d}\lambda_i}{2\pi}\, \frac{\Delta_{N+1-\ell}(\lambda)^2\, \prod\limits_{i=1}^{\ell} \prod\limits_{j=1}^{N+1-\ell} \left(x_{i}-\lambda_j\right)^2}{\prod\limits_{k=1}^{N+1-\ell} \left(\bar{x}-\lambda_k\right)^2}\, \rme^{-\frac{1}{g_{\text{s}}} \sum\limits_{k=1}^{N+1-\ell}V(\lambda_k)}, \nonumber
\end{align}
\noindent
where we further renamed the last $\ell$ eigenvalues as $x_i$, $i=1,\ldots,\ell$, for better readability. This factorization allows us to rewrite the inverse-determinant correlator in \eqref{eq:separatedoneantieigenvalue} as
\begin{align}
\mathcal{Z}^{(0|0)} (t+g_{\text{s}}) \ev{\frac{1}{\det \left(\bar{x}-M\right)^{2}}}_{N+1} &= \frac{(N+1-\ell)!}{(N+1)!}\, \mathcal{Z}^{(0|0)} (t+\left(1-\ell\right) g_{\text{s}})\, \times \\
&
\hspace{-150pt}
\times\, \int \prod\limits_{i=1}^{\ell} \frac{\text{d}x_{i}}{2\pi}\, \frac{\Delta_{\ell}(x)^2}{\prod\limits_{j=1}^{\ell} \left(\bar{x}-x_{j}\right)^2}\, \rme^{-\frac{1}{g_{\text{s}}} \sum\limits_{k=1}^{\ell} V(x_{k})} \ev{\frac{\prod\limits_{i=1}^{\ell} \det \left(x_{i}-M\right)^{2}}{\det \left(\bar{x}-M\right)^{2}}}_{N+1-\ell}, \nonumber
\end{align}
\noindent
which upon replacement back in \eqref{eq:separatedoneantieigenvalue} yields our desired result
\begin{align}
\label{eq:determinantformulak1sectors}
\mathcal{Z}^{(\ell|1)} (t) &= \frac{(N+1-\ell)!}{(N+1)!}\, \mathcal{Z}^{(0|0)} (t+\left(1-\ell\right) g_{\text{s}})\, \times \\
&\times\, \int \prod\limits_{i=1}^{\ell} \frac{\text{d}x_{i}}{2\pi} \int \frac{\text{d}\bar{x}}{2\pi}\, \frac{\Delta_{\ell}(x)^2}{\prod\limits_{i=1}^{\ell} \left(\bar{x}-x_{i}\right)^2}\, \rme^{-\frac{1}{g_{\text{s}}} \sum\limits_{i=1}^{\ell} V(x_{i}) + \frac{1}{g_{\text{s}}} V(\bar{x})} \ev{\frac{\prod\limits_{i=1}^{\ell} \det \left(x_{i}-M\right)^{2}}{\det \left(\bar{x}-M\right)^{2}}}_{N+1-\ell}. \nonumber
\end{align}
\noindent
Again, this is in agreement with our general proposal \eqref{eq:finalresultseparatedeigenvalues}. This integral (as our proposal) has double poles in the integrand whenever $x_{i} = \bar{x}$ and is hence ill-defined. Its definition is made via the principal-value prescription in \eqref{eq:principalvalueprescription} (already featured in \eqref{eq:finalresultseparatedeigenvalues}) which will come about with the discussion of integration contours in subsection~\ref{subsec:eigenvalueTunnelingContours}.

\paragraph{Tunneling Multiple Anti-Eigenvalues:}

The general case now follows almost straightforwardly. First, generalize \eqref{eq:inversedeterminantrelation} to multiple determinants, evaluated at multiple coordinates $x_i = x(\upzeta_i)$. Completely analogous calculations to the ones presented above now yield
\begin{align}
\label{eq:Multipleinversedeterminantrelation}
\text{e}^{-\frac{1}{g_{\text{s}}} \sum\limits_{i=1}^{\bar{\ell}} V(x(\sigma(\upzeta_i)))}\, \ev{\prod\limits_{j=1}^{\bar{\ell}} \det \left( x(\sigma(\upzeta_j)) - M \right)^{2}
}_{N-\bar{\ell}}\, \mathcal{Z}^{(0|0)} (t-\bar{\ell} g_{\text{s}}) &= \\
&
\hspace{-185pt}
= \text{e}^{+\frac{1}{g_{\text{s}}} \sum\limits_{i=1}^{\bar{\ell}} V(x(\upzeta_i))}\, \ev{\frac{1}{\prod\limits_{j=1}^{\bar{\ell}} \det \left( x(\upzeta_j) - M \right)^{2}}
}_{N+\bar{\ell}}\, \mathcal{Z}^{(0|0)} (t+\bar{\ell} g_{\text{s}}). \nonumber
\end{align}
\noindent
This relation may again be derived via the deformed partition-function argument in \cite{eggls23}.

Given this relation, the generic $(\ell|\bar{\ell})$ configuration follows. In the same way we used \eqref{eq:inversedeterminantrelation} in the one-tunneled-eigenvalue partition function \eqref{eq:integral1tunneledeigenvalemsw07} to obtain $\mathcal{Z}^{(0|1)}$ in \eqref{eq:separatedoneantieigenvalue}, we may now use the above \eqref{eq:Multipleinversedeterminantrelation} in the multi-eigenvalue partition function \eqref{eq:integralkforwardcontributions} to obtain an expression for $\mathcal{Z}^{(0|\bar{\ell})}$---which the reader may check is precisely the one from our proposal \eqref{eq:finalresultseparatedeigenvalues}, once we factor-in the extra symmetry factor ${N}\choose{\bar{\ell}}$ accounting for the fact that out of $N$ indistinguishable eigenvalues, exactly $\bar{\ell}$ have flipped sheet. Similarly, in the same way we factored-out $\ell$ eigenvalues from the matrix-integral associated to the inverse-determinant correlator in \eqref{eq:separatedoneantieigenvalue} to obtain $\mathcal{Z}^{(\ell|1)}$ in \eqref{eq:determinantformulak1sectors}, we may now factor-out this same $\ell$ eigenvalues from the product of inverse-determinants correlator in \eqref{eq:Multipleinversedeterminantrelation} to finally obtain \eqref{eq:finalresultseparatedeigenvalues} in full generality. All that is now missing is properly understanding the appropriate integration contours associated to (anti) eigenvalue tunneling.

To end this subsection, let us make the connection to the supermatrix discussion in subsection~\ref{subsec:CommentOnSuperMatrixModel}. Head back to our proposal for $\mathcal{Z}^{(\ell|\bar{\ell})}$ in \eqref{eq:finalresultseparatedeigenvalues} and assume that \textit{all contours are perturbative} (which we have left unspecified but were thinking ``nonperturbative'' in the back of our minds). There is then no reason to single-out the $x_i$ or $\bar{x}_j$ integrations, and we may rather include them all with the eigenvalue $\lambda_i$ and anti-eigenvalue $\bar{\lambda}_j$ integrations, inside the ratio-of-determinants correlator. In fact that correlator is just denoting
\begin{align}
\mathcal{Z}^{(0|0)} (t-\left(\ell-\bar{\ell}\right) g_{\text{s}}) \ev{\frac{\prod\limits_{i=1}^{\ell} \det \left(x_i-M\right)^{2}}{\prod\limits_{j=1}^{\bar{\ell}} \det \left(\bar{x}_j-M\right)^{2}}}_{N+\bar{\ell}-\ell} &= \\
&
\hspace{-200pt}
= \frac{1}{(N+\bar{\ell}-\ell)!} \int \prod\limits_{i=1}^{N+\bar{\ell}-\ell} \frac{\text{d}\lambda_i}{2\pi}\, \Delta_{N+\bar{\ell}-\ell}^2(\lambda)\, \frac{\prod\limits_{i=1}^{\ell} \prod\limits_{j=1}^{N+\bar{\ell}-\ell} \left(x_{i} - \lambda_j\right)^{2}}{\prod\limits_{k=1}^{\bar{\ell}} \prod\limits_{j=1}^{N+\bar{\ell}-\ell} \left(\bar{x}_k - \lambda_j\right)^{2}}\, \rme^{-\frac{1}{g_{\text{s}}} \sum\limits_{i=1}^{N+\bar{\ell}-\ell} V(\lambda_{i})}. \nonumber
\end{align}
\noindent
Plugging this expression back in \eqref{eq:finalresultseparatedeigenvalues} and renaming $x_i \mapsto \lambda_{i+N+\bar{\ell}-\ell}$ and $\bar{x}_j \mapsto \bar{\lambda}_j$ therein as mentioned above, then \eqref{eq:finalresultseparatedeigenvalues} will exactly match \eqref{eq:eigenvalueformualtionsuperpartitionfunction-super} with the supermatrix assignments:
\be
\BZ_{N+\bar{\ell},\bar{\ell}} \left(g_{\text{s}}\right).
\ee
\noindent
Interestingly, we now explicitly see how the eigenvalue--anti-eigenvalue partition function may be written as a physical supermatrix integral, which we had already mentioned in subsection~\ref{subsec:CommentOnSuperMatrixModel}.

\subsection{Eigenvalue Tunneling, Resurgence, and Integration Contours}\label{subsec:eigenvalueTunnelingContours}

Having understood the matrix integrals (and integrands) we wish to evaluate, let us now make their (anti) eigenvalue-tunneling integration-contours precise. In particular, we want to make sure that the $\mathcal{Z}^{(\ell|\bar{\ell})}$-amplitudes from the previous subsection \textit{are indeed} instanton-amplitudes, \textit{i.e.}, that all tunneled (anti) eigenvalue contours pass through \textit{nonperturbative} saddles. For the $(\ell|0)$ multi-eigenvalue configurations this is pretty standard and nonperturbative contours are steepest-descent contours; \textit{e.g.}, \cite{msw07}. But this \textit{will not} be the case for more general sectors, where residue contours will also appear. This is a subtle discussion in which we now engage.

In order to properly understand how to formulate these integration contours we will begin by directly relating eigenvalue tunneling with the resurgence relations of our problem. This immediately comes about once considering the discontinuity in the asymptotic evaluation of the matrix-model partition-function when crossing a Stokes line. This discontinuity \eqref{eq:discontinutiystokesautomorphism} may be evaluated in two equivalent ways:
\begin{itemize}
\item Directly within the resurgence framework, by using the Stokes automorphisms \eqref{eq:forwardStokesAutomorphism}-\eqref{eq:backwardStokesAutomorphism}.
\item Directly for the eigenvalue matrix-integral, where it acts as:
\begin{align}
\label{eq:discontinuityongeneralfunctionexpression}
\text{Disc}_{\theta}\, \mathcal{Z} (g_s) = \mathcal{Z} (\abs{g_{\text{s}}}\, \rme^{\rmi \theta^{+}}) - \mathcal{Z} (\abs{g_{\text{s}}}\, \rme^{\rmi \theta^{-}}), \qquad \theta = \arg g_{\text{s}}.
\end{align}
\noindent
On the physical sheet, this directly translates to standard eigenvalue tunneling\footnote{Note how even though in principle the partition-function matrix-integral results in an actual function, its evaluation via saddle-point approximation does not---it produces an asymptotic series. Hence when comparing matrix-integral results with resurgence results, one is really consistently comparing two asymptotic series.}. However, when considering both sheets of the spectral curve, we will uncover additional \textit{residue} contributions which need to be taken into account.
\end{itemize} 

\paragraph{Eigenvalue Tunneling on the Physical Sheet:}

\begin{figure}
\centering
	\begin{tikzpicture}[
	 scale =1, line width=2
	]
	\draw[->] (-5,0) -- (5, 0);
	\draw[->] (0,-2) -- (0, 2);
	\draw (-5, 1.6) -- (-4.6, 1.6);
	\draw (-4.6, 1.6) -- (-4.6, 2);
	\node at (-4.8, 1.8) {$g_{\text{s}}$};
	\draw[blue] (0,0) -- (4.5, -0.5);
	\draw[blue] (0,0) -- (4.5, 0.5);
	\draw[->, blue] (4, -0.4) to[bend right =30] (4, 0.4);
	\node[blue] at (5, 0.5) {$\underline{\mathfrak{S}}_0$};
	\end{tikzpicture}\\
	\begin{tikzpicture}[
	grayframe/.style={
		rectangle,
		draw=gray,
		text width=4em,
		align=center,
		rounded corners,
		minimum height=2em
	}, scale =1, line width=2
	]
	\foreach \n in {0,...,3}{
	\node[grayframe] (\n_0) at (4*\n,0) {$Z^{(\n, 0)}(g_{\text{s}})$};
	}
	\draw[blue, ->] (0_0) -- (1_0) node[midway, above]{$\mathsf{S}_{(0,0)\to(1,0)}$};
	\draw[color=blue, ->] (0_0) to [out=315, in=225, looseness=0.6] (2_0);
	\node[color=blue] at (4, -0.8){$\mathsf{S}_{(0,0)\to(2,0)}$};
	\draw[color=blue, ->] (0_0) to [out=315, in=225, looseness=0.6] (3_0);
	\node[color=blue] at (7, -2.15){$\mathsf{S}_{(0,0)\to(3,0)}$};
	\end{tikzpicture}
\caption{The forward Stokes automorphism ({\color{blue}blue}). When crossing the positive real axis starting from the perturbative sector we pick-up new resurgent sectors. The Stokes ``lattice'' is however one-dimensional as we are only considering the forward discontinuity.}
\label{fig:walkingonedimresurgentlattice}
\end{figure}
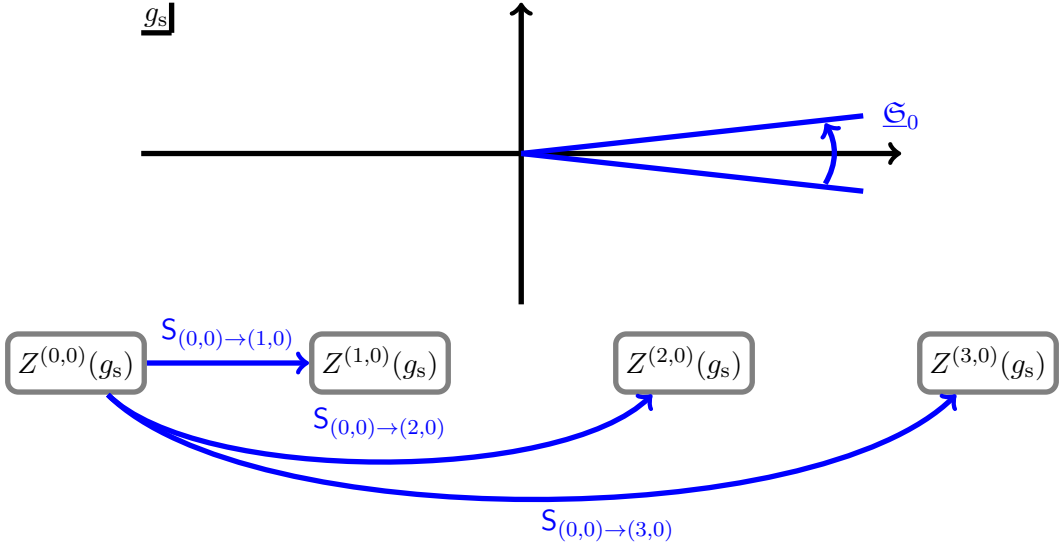 

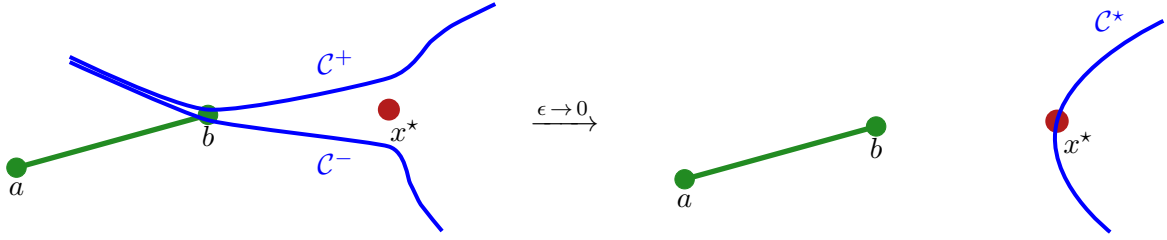
\begin{figure}
\centering
    \begin{minipage}{.5\textwidth}
    \centering
	\begin{tikzpicture}[scale=0.7]
		\draw[ForestGreen, line width=2pt] (-2, -1) -- (1.6, 0);
		\draw[ForestGreen, fill=ForestGreen] (-2,-1) circle (1.1ex);
		\node at (-2, -1.4) {$a$};
		\draw[ForestGreen, fill=ForestGreen] (1.6,0) circle (1.1ex);
		\node at (1.6, -.4) {$b$};
		\draw[cornellred, fill=cornellred, line width=1pt] (5,0.1) circle (1.1ex);
		\node at (5.3, -0.3) {$x^{\star}$};
		\draw[blue, line width=1.5pt] plot [smooth, tension=0.4] coordinates{(-1, 1.1)(1.6, 0.1)(5,0.7)(5.8,1.4)(6.2,1.7)(7, 2.1)};
		\draw[blue, line width=1.5pt] plot [smooth, tension=0.3] coordinates{(-1, 1)(1.6,-0.1)(5,-0.6)(5.4, -1.3)(5.6, -1.7)(6, -2.2)};
		\node at (8.3,0) {$\xrightarrow{\epsilon\,\rightarrow\,0}$};
		\node[blue] at (4, 1) {$\CC^{+}$};
		\node[blue] at (4, -0.9) {$\CC^{-}$};
	\end{tikzpicture}
	\end{minipage}%
    \begin{minipage}{0.5\textwidth}
    \centering
		\begin{tikzpicture}[scale=0.7]
		\draw[ForestGreen, line width=2pt] (-2, -1) -- (1.6, 0);
		\draw[ForestGreen, fill=ForestGreen] (-2,-1) circle (1.1ex);
		\node at (-2, -1.4) {$a$};
		\draw[ForestGreen, fill=ForestGreen] (1.6,0) circle (1.1ex);
		\node at (1.6, -.4) {$b$};
		\draw[cornellred, fill=cornellred, line width=1.5pt] (5,0.1) circle (1.1ex);
		\node at (5.4, -0.3) {$x^{\star}$};
		\draw[blue, line width=1.5pt] plot [smooth, tension=1] coordinates{(6, -2)(5,0.1)(7, 2)};
		\node[blue] at (6, 2) {$\CC^{\star}$};
	\end{tikzpicture}
	\end{minipage}
\caption{Integration contours associated with eigenvalue tunneling (drawn on top of figure~\ref{fig:settingcutandsaddle}). We have two perturbative contours $\CC^{+}$ and $\CC^{-}$ which, for $\epsilon\rightarrow 0$, get deformed into the nonperturbative saddle-point contour $\CC^{\star}$ (throughout the figures of this subsection, $\epsilon \equiv \arg g_{\text{s}}$). This picture will change in the presence of anti-eigenvalues (see upcoming figure~\ref{fig:settingcutandsaddleGeneraltunneling}).}
\label{fig:settingcutandsaddleStandardtunneling}
\end{figure}

Start by considering the perturbative sector and undergo a forward Stokes transition, as shown in figure~\ref{fig:walkingonedimresurgentlattice}. From resurgence, the discontinuity reads
\begin{equation}
\label{eq:resurgentforwarddiscontinuity}
\text{Disc}_{0}\, Z^{(0,0)} (t) = \sum_{\ell=1}^{\infty} \rme^{-\ell \frac{A}{g_{\text{s}}}}\,  \mathsf{S}_{(0,0)\to(\ell,0)}\, Z^{(\ell,0)} (t).
\end{equation}
\noindent
This should reproduce standard eigenvalue tunneling---but to evaluate the discontinuity on the matrix integral side, we first need to be fully specific concerning eigenvalue integration-contours. Keeping track of contours in the following is easier if introducing added subscripts into \eqref{eq:partitionfunctioneigenvalues}, as:
\begin{align}
\label{eq:partitionfunctionwithintegrationcontours}
\mathcal{Z}_{N} (g_{\text{s}}) \equiv \mathcal{Z}_{\lbrace \CC_1, \ldots, \CC_N \rbrace} (g_{\text{s}}) = \frac{1}{N!} \int_{\CC_1} \cdots \int_{\CC_N} \prod\limits_{i=1}^{N} \frac{\text{d}\lambda_i}{2\pi}\, \Delta_{N} (\lambda)^2\, \text{e}^{-\frac{1}{g_{\text{s}}} \sum\limits_{i=1}^{N} V(\lambda_i)}.
\end{align}
\noindent
Given the different saddle-points of the potential, $x_{\text{sp}}$, contours $\CC_i$ are taken as the corresponding paths of steepest-descent, dictated by
\begin{align}
\label{eq:steepestdescentconditionV}
\Im \left( \frac{V(x)-V(x_{\text{sp}})}{g_{\text{s}}} \right) = 0.
\end{align}
\noindent
Start with $N=1$ and evaluate the discontinuity \eqref{eq:discontinuityongeneralfunctionexpression} of the perturbative partition-function across $\theta=0$. It follows
\begin{align}
\label{eq:tunnelingoneeigenvalueone}
\text{Disc}_{0}\, \mathcal{Z}_{1}^{(0|0)} (g_s) &= \mathcal{Z}_1^{(0|0)} (\abs{g_{\text{s}}}\, \text{e}^{\rmi 0^{+}}) - \mathcal{Z}_1^{(0|0)} (\abs{g_{\text{s}}}\, \text{e}^{\rmi 0^{-}}) \equiv \mathcal{Z}_{\lbrace \CC^{+} \rbrace} (\abs{g_{\text{s}}}\, \text{e}^{\rmi 0^{+}}) - \mathcal{Z}_{\lbrace \CC^{-} \rbrace} (\abs{g_{\text{s}}}\, \text{e}^{\rmi 0^{-}}) = \nonumber \\
&= \mathcal{Z}_{\lbrace \CC^{+}-\CC^{-} \rbrace} (\abs{g_{\text{s}}}\, \text{e}^{\rmi 0^{-}}) = \mathcal{Z}_{\lbrace \CC^{\star} \rbrace} (\abs{g_{\text{s}}}\, \text{e}^{\rmi 0^{-}}),
\end{align}
\noindent
where, as usual \cite{msw07}, in the last step we have deformed the difference of perturbative contours into the nonperturbative-saddle contour $\CC^{\star}$. This general setting is  illustrated in figure~\ref{fig:settingcutandsaddleStandardtunneling}.

We may be more explicit in the concrete example of the quartic matrix model (for which we have already shown steepest-descent contours in \eqref{eq:rotationperturbativecontour}-\eqref{eq:quarticexampleeigenvaluetunnelingscetch}, and which we will use as an explicit example throughout this subsection). In fact, one may immediately rewrite  \eqref{eq:quarticexampleeigenvaluetunnelingscetch} in our present discontinuity language, where it simply becomes
\begin{equation}
\label{eq:quarticexampleeigenvaluetunneling}
\CC^{+}_{\text{quartic}} - \CC^{-}_{\text{quartic}} =\,\ZPertpe\,-\,\ZPertme\,=\,\ZSuperSaddles\,\xrightarrow{\epsilon\,\rightarrow\,0}\,\ZSadpzero\,= \CC^{\star}_{\text{quartic}}.
\end{equation}

Analogous calculations for higher values of $N$ go through by simple iteration of the discontinuity \eqref{eq:tunnelingoneeigenvalueone} and using linearity of the integral. One finds
\begin{align}
\label{eq:anyNforwardcontours}
\text{Disc}_{0}\, \mathcal{Z}_{N}^{(0|0)} (g_s) &= \mathcal{Z}_N^{(0|0)} (\abs{g_{\text{s}}}\, \text{e}^{\rmi 0^{+}}) - \mathcal{Z}_N^{(0|0)} (\abs{g_{\text{s}}}\, \text{e}^{\rmi 0^{-}}) = {{N}\choose{1}}\, \mathcal{Z}_{\lbrace \CC^{\star}, \CC^{-}, \CC^{-}, \ldots \rbrace} (\abs{g_{\text{s}}}\, \text{e}^{\rmi 0^{-}}) + \nonumber \\
&
+ {{N}\choose{2}}\, \mathcal{Z}_{\lbrace \CC^{\star}, \CC^{\star}, \CC^{-}, \ldots \rbrace} (\abs{g_{\text{s}}}\, \text{e}^{\rmi 0^{-}}) + \cdots + {{N}\choose{\ell}}\, \mathcal{Z}_{\lbrace \underbrace{\scriptstyle{\CC^{\star}, \ldots, \CC^{\star}}}_{\ell}, \underbrace{\scriptstyle{\CC^{-}, \ldots, \CC^{-}}}_{N-\ell} \rbrace} (\abs{g_{\text{s}}}\, \text{e}^{\rmi 0^{-}}) + \nonumber \\
&
+ \cdots + \mathcal{Z}_{\lbrace \CC^{\star}, \CC^{\star}, \CC^{\star}, \ldots \rbrace} (\abs{g_{\text{s}}}\, \text{e}^{\rmi 0^{-}}).
\end{align}
\noindent
Herein the combinatorial factors ${N\choose{\ell}}$ alluded to in subsection~\ref{subsec:DeterminantCorrelators} naturally appear, and this hence further allows us to finally make \eqref{eq:integralkforwardcontributions} fully precise (in fact, as anticipated therein). Recall we were missing the combinatorics of indistinguishable eigenvalues, and the specification of the $\CC^{\star}$ integration contours through the nonperturbative saddles. Both accounts are now solved. The $\ell$-instanton contribution $\mathcal{Z}^{(\ell|0)}$ is the contribution with $\ell$ tunneled eigenvalues in \eqref{eq:anyNforwardcontours}, \textit{i.e.}, the missing specifications in the matrix integral \eqref{eq:integralkforwardcontributions} are\footnote{From here onwards we only list the ``missing'' integration contours we are interested in, as opposed to all. The perturbative  contours $\CC^{-}$ which are left implicit are taken at $g_{\text{s}} = \abs{g_{\text{s}}}\, \rme^{\rmi 0^{-}}$. In principle one should always keep track of this phase albeit for readability we will most times just write $g_{\text{s}}$, with the phase clear form context.} just:
\begin{equation}
\label{eq:forwardcontributionstunneled}
\mathcal{Z}^{(\ell|0)} \equiv {{N}\choose{\ell}}\, \mathcal{Z}_{\lbrace \underbrace{\scriptstyle{\CC^{\star}, \ldots, \CC^{\star}}}_{\ell} \rbrace}.
\end{equation}

All that is left to do, is to now relate the resurgence discontinuity \eqref{eq:resurgentforwarddiscontinuity} with the matrix-integral discontinuity \eqref{eq:anyNforwardcontours}. This also makes precise how eigenvalue tunneling essentially \textit{yields} resurgence. Recalling we are comparing asymptotic series, we first equate\footnote{We have normalized both sides of this asymptotic equality by the perturbative sector, for two reasons. First, this normalization is convenient in order to compare matrix-integral results for the free energy instead of the partition function \cite{msw07}. Second, in this way we do not have to worry about neither the Gaussian normalization nor the global partition-function pre-factor; both earlier mentioned in subsection~\ref{subsec:resurgenttranssriesandresonance}.} both discontinuities as:
\begin{equation}
\label{eq:comparisondiscontinuities}
\frac{\text{Disc}_{0}\, Z^{(0,0)}(t, g_{\text{s}})}{Z^{(0,0)}(t, g_{\text{s}})} \simeq \frac{\text{Disc}_{0}\, \mathcal{Z}^{(0|0)}(t, g_{\text{s}})}{\mathcal{Z}^{(0|0)}(t, g_{\text{s}})}.
\end{equation}
\noindent
Note how on both sides above all terms are listed by their instanton number. It is then immediate to further identify them one-by-one as:
\begin{equation}
\label{eq:forwardresurgentEVtunneling}
\text{e}^{-\ell\frac{A}{g_{\text{s}}}}\, \mathsf{S}_{(0,0)\rightarrow (\ell,0)}\, \frac{Z^{(\ell,0)}(t, g_{\text{s}})}{Z^{(0,0)}(t, g_{\text{s}})} \simeq \frac{\mathcal{Z}^{(\ell|0)}(t, g_{\text{s}})}{\mathcal{Z}^{(0| 0)}(t, g_{\text{s}})}.
\end{equation}
\noindent
Let us comment on this result. First, it establishes the \textit{bridge} between resurgent-transseries quantities on its left-hand side, and matrix-integrals on its right-hand side. Second, via explicit matrix integrals---with the appropriate eigenvalue configurations---this equality \textit{directly computes} Borel residues; which is to say, directly computes the Stokes data of the resurgent transseries. Third, although we have started-out focused upon the partition-function, at the end-of-the-day we are really facing its \textit{ratios}; \textit{i.e.}, what we are really computing is the \textit{free energy} transseries.

\paragraph{Generic Eigenvalue Tunneling:}

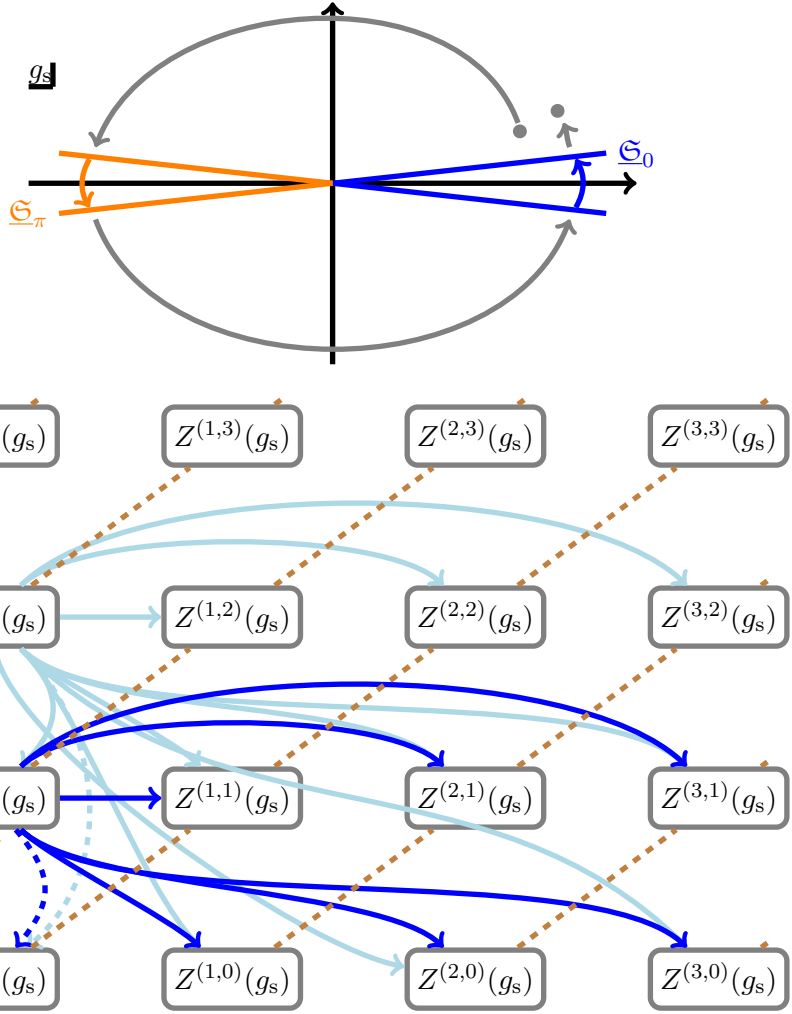
\begin{figure}
\centering
\begin{tikzpicture}[
	 scale =0.8, line width=2
	]
	\draw[->] (-5,0) -- (5, 0);
	\draw[->] (0,-3) -- (0, 3);
	\draw (-5, 1.6) -- (-4.6, 1.6);
	\draw (-4.6, 1.6) -- (-4.6, 2);
	\node at (-4.8, 1.8) {$g_{\text{s}}$};
	\draw[blue] (0,0) -- (4.5, -0.5);
	\draw[blue] (0,0) -- (4.5, 0.5);
	\draw[->, blue] (4, -0.4) to[bend right =30] (4, 0.4);
	\node[blue] at (5, 0.5) {$\underline{\mathfrak{S}}_0$};
	\draw[color=orange] (0,0) -- (-4.5, -0.5);
	\draw[color=orange] (0,0) -- (-4.5, 0.5);
	\draw[->, color=orange] (-4, 0.4) to[bend right =30] (-4, -0.4);
	\node[color=orange] at (-5, -0.5) {$\underline{\mathfrak{S}}_{\pi}$};
	\draw[gray, ->] (3.05, 1) to[out=110, in=70] (-3.9, 0.6);
	\draw[gray, ->] (3.9, 0.6) to[out=110, in=290] (3.8, 1);
	\draw[gray, ->] (-3.9, -0.6) to[out=290, in=250] (3.9, -0.6);
	\filldraw[gray] (3.7, 1.2) circle (2pt);
	\filldraw[gray] (3.08, 0.86) circle (2pt);
	\end{tikzpicture}\\
	\begin{tikzpicture}[
	grayframe/.style={
		rectangle,
		draw=gray,
		text width=4em,
		align=center,
		rounded corners,
		minimum height=2em
	}, scale =0.8, line width=2
	]
	\foreach \n in {0,...,3}{
	  \foreach \m in {0,..., 3}{
	\node[grayframe] (\n_\m) at (4*\n,3*\m) {$Z^{(\n, \m)}(g_{\text{s}})$};
	}}
		\draw[color=orange, ->] (0_0) -- (0_1);
	\draw[color=orange, ->] (0_0) to [out=135, in=225, looseness=0.6] (0_2);
	\draw[color=orange, ->] (0_0) to [out=135, in=225, looseness=0.6] (0_3);
		\draw[color=LightBlue, ->] (0_2) -- (1_2);
	\draw[color=LightBlue, ->] (0_2) to [out=315, in=45, looseness=1.3] (0_1);
		\draw[color=LightBlue, ->] (0_2) to [out=45, in=135, looseness=0.5] (2_2);
		\draw[color=LightBlue, ->] (0_2) to [out=315, in=135, looseness=0.5] (1_1);
		\draw[color=LightBlue, ->] (0_2) to [out=45, in=135, looseness=0.6] (3_2);
		\draw[color=LightBlue, ->] (0_2) to [out=315, in=135, looseness=0.6] (2_1);
				\draw[color=LightBlue, ->] (0_2) to [out=315, in=135, looseness=0.5] (3_1);	
		\draw[color=LightBlue, ->] (0_2) to [out=315, in=135, looseness=0.6] (1_0);
		\draw[color=LightBlue, ->] (0_2) to [out=280, in=170, looseness=0.5] (2_0);
				\draw[color=LightBlue, ->] (0_2) to [out=315, in=135, looseness=1] (3_0);	
				\draw[color=LightBlue, ->, dashed] (0_2) to [out=315, in=40, looseness=1] (0_0);
			\draw[color=blue, ->] (0_1) -- (1_1);
	\draw[color=blue, ->, dashed] (0_1) to [out=310, in=50, looseness=1.3] (0_0);
				\draw[color=blue, ->] (0_1) to [out=45, in=135, looseness=0.5] (2_1);
		\draw[color=blue, ->] (0_1) to [out=315, in=135, looseness=0.5] (1_0);
		\draw[color=blue, ->] (0_1) to [out=45, in=135, looseness=0.6] (3_1);
		\draw[color=blue, ->] (0_1) to [out=315, in=135, looseness=0.5] (2_0);
				\draw[color=blue, ->] (0_1) to [out=315, in=135, looseness=0.5] (3_0);				
		\draw[dashed, brown] (0_0) -- (1_1);
		\draw[dashed, brown] (1_1) -- (2_2);
		\draw[dashed, brown] (2_2) -- (3_3);
		\draw[dashed, brown] (3_3) -- (12.8, 9.6);
		\draw[dashed, brown] (1_0) -- (2_1);
		\draw[dashed, brown] (2_1) -- (3_2);
		\draw[dashed, brown] (3_2) -- (12.8, 6.6);	
		\draw[dashed, brown] (2_0) -- (3_1);
		\draw[dashed, brown] (3_1) -- (12.8, 3.6);
		\draw[dashed, brown] (3_0) -- (12.8, 0.6);
		\draw[dashed, brown] (0_1) -- (1_2);
		\draw[dashed, brown] (1_2) -- (2_3);
		\draw[dashed, brown] (2_3) -- (8.8, 9.6);
		\draw[dashed, brown] (0_2) -- (1_3);
		\draw[dashed, brown] (1_3) -- (4.8, 9.6);
		\draw[dashed, brown] (0_3) -- (0.8, 9.6);
	\end{tikzpicture}
	\caption{The intricacies of resurgence. Take the {\color{gray}gray} path depicted on the top image, where we first undergo a {\color{orange}backward} Stokes automorphism, then followed by a {\color{blue}forward} Stokes automorphism. The (intricate) resurgence relations which emerge from such a sequence of transitions are depicted upon the transseries lattice of the bottom image. In dashed {\color{brown}brown} we represent the directions of the resonant kernel. These resurgence relations literally provide access to the \textit{full} transseries.}
\label{fig:resonantresurgentlatticeTwoAutomorphisms}
\end{figure}

Having established the forward Stokes automorphism, with its resurgent sectors as in figure~\ref{fig:walkingonedimresurgentlattice}, let us next turn to arbitrary ``resurgence motions'' and try to access the full transseries as shown in figure~\ref{fig:resonantresurgentlatticeTwoAutomorphisms}. As illustrated, one way in which this is achievable is by starting instead with the backward Stokes automorphism.

Let us first underline that the perturbative eigenvalues on the physical sheet are not affected by the backward discontinuity. This is because while on the physical sheet nothing really happens at $\theta = \pi$---there is no room for a perturbative-contour deformation of the type shown in figure~\ref{fig:settingcutandsaddleStandardtunneling} or in \eqref{eq:quarticexampleeigenvaluetunneling}. In fact, recall that the rotation of the perturbative quartic contour was already shown in \eqref{eq:rotationperturbativecontour}, where it was rather explicit how Stokes transitions only occurred at $\theta=0$ and $\theta=2\pi$. And these were exactly the transition now illustrated in \eqref{eq:quarticexampleeigenvaluetunneling}. On the other hand, when crossing $\theta=\pi$ no such deformation occurred. 

Start again by considering the perturbative sector but now undergo a backward Stokes transition, as shown in figure~\ref{fig:resonantresurgentlatticeTwoAutomorphisms}, where we will need to include anti-eigenvalues. From resurgence, the discontinuity reads (compare with \eqref{eq:resurgentforwarddiscontinuity})
\begin{equation}
\label{eq:purelybackwardsdiscontinuity}
\text{Disc}_{\pi}\, Z^{(0,0)} (t) = \sum_{\ell=1}^{\infty} \rme^{\ell \frac{A}{g_{\text{s}}}}\, \mathsf{S}_{(0,0)\to(0,\ell)}\, Z^{(0,\ell)}(t).
\end{equation}
\noindent
As the reader already knows, our goal is to reproduce this formula via anti-eigenvalue tunneling.

Akin to what we did earlier for eigenvalues, it is rather instructive to first evaluate the discontinuity \eqref{eq:discontinuityongeneralfunctionexpression} starting with a single \textit{anti-eigenvalue}. In this case, let us focus on trying to make sense of
\begin{equation}
\text{Disc}_{\pi}\, \mathcal{Z}^{(0|0)} = \mathcal{Z}^{(0|1)} + \cdots.
\end{equation}
\noindent
We will deal with multiple anti-eigenvalues later; for now let us just figure out a single anti-eigenvalue in the first transition of figure~\ref{fig:resonantresurgentlatticeTwoAutomorphisms}. This we shall do by resorting to \eqref{eq:separatedoneantieigenvalue} \textit{but with $\bar{x}$ integrated over a perturbative contour} $\bar{\CC}$, as\footnote{It is now important to keep track of both \textit{super}scripts and \textit{sub}scripts to properly define each quantity.}
\begin{equation}
\label{eq:separatedoneantieigenvalueWithCOntour}
\mathcal{Z}^{(0|1)}_{\lbrace \bar{\CC} \rbrace} (t) = \mathcal{Z}^{(0|0)} (t+g_{\text{s}}) \int_{\bar{\CC}} \frac{\text{d}\bar{x}}{2\pi}\, \text{e}^{\frac{1}{g_{\text{s}}} V(\bar{x})} \ev{\frac{1}{\det \left(\bar{x}-M\right)^{2}}}_{N+1}.
\end{equation}
\noindent
The discontinuity \eqref{eq:discontinuityongeneralfunctionexpression} of this integral across $\theta=\pi$ follows. The backward Stokes transition occurs at $-g_{\text{s}}$, where anti-eigenvalues at $\theta=\pi$ behave\footnote{As the minus sign in $g_{\text{s}}$ cancels the one from switching sheets (recall the corresponding discussion of Stokes automorphisms in subsection~\ref{subsec:MatrixIntegralsWork}).} just like eigenvalues at $\theta=0$, in which case we find (compare with \eqref{eq:tunnelingoneeigenvalueone})
\begin{equation}
\label{eq:01correctedcontour}
\text{Disc}_{\pi}\, \mathcal{Z}_{\lbrace \bar{\CC} \rbrace}^{(0|1)} (g_s) = \mathcal{Z}^{(0|1)}_{\lbrace \bar{\CC}^{+} \rbrace} (\abs{g_{\text{s}}}\, \text{e}^{\rmi \pi^{+}}) - \mathcal{Z}^{(0|1)}_{\lbrace \bar{\CC}^{-} \rbrace} (\abs{g_{\text{s}}}\, \text{e}^{\rmi \pi^{-}}) = \mathcal{Z}^{(0|1)}_{\lbrace \bar{\CC}^{\star} \rbrace} (\abs{g_{\text{s}}}\, \text{e}^{\rmi \pi^{-}}).
\end{equation}
\noindent
Herein we are considering the same saddle-point $x^{\star}$ as above, and we have accordingly denoted the corresponding tunneled-anti-eigenvalue contour by $\bar{\CC}^{\star}$. But we still need to continue the path outlined in figure~\ref{fig:resonantresurgentlatticeTwoAutomorphisms} to reach positive $g_{\text{s}}$. As that happens $\bar{\CC}^{\star}$ also rotates; in particular, on the physical sheet, $\bar{\CC}^{\star}$ stops being steepest-ascent and becomes steepest-descent again---once more reflecting the minus-sign which appears when switching sheets in the spectral curve. Anti-eigenvalues have hence tunneled to the nonperturbative saddle in the same way that eigenvalues do, only they do so via the backward automorphism (and unaffected by the forward one).

This may be illustrated in our familiar quartic example, similarly to what happened in \eqref{eq:quarticexampleeigenvaluetunneling}. In this case, taking into account all aforementioned rotations, the above transition looks like
\begin{equation}
\label{eq:antieigenvaluetunnelingpictorial}
\ZPertBarpipe\,-\,\ZPertBarpime\,\xrightarrow{\epsilon\rightarrow0}\,\ZSadBarpipe\,\xrightarrow{-g_{\text{s}} \rightarrow g_{\text{s}}}\,\ZSadBarpe\,,
\end{equation}
\noindent
where we see anti-eigenvalue tunneling at $\theta=\pi$ before rotating back to $\theta=0$; and where one can explicitly see how anti-eigenvalue contours are eigenvalue contours rotated\footnote{Notice how under the rotation by $\pi$ the contours seem to rotate only by $\frac{\pi}{4}$. This is just due to the fact that for this illustrative example we are using the steepest-descent \eqref{eq:steepestdescentconditionV} with a \textit{quartic} potential \eqref{eq:quarticPotential}.} by $\pi$.

\begin{figure}
\centering
	\begin{tikzpicture}[scale=0.7]
		\begin{scope}[shift={({-6},{0})}]
		\draw[gray, line width=1.5pt, ->] (-2,0) -- (5,0);
		\draw[gray, line width=1.5pt, ->] (0,-4) -- (0,4);
		\draw[ForestGreen, fill=ForestGreen] (-1,0) circle (1.1ex);
		\draw[ForestGreen, fill=ForestGreen] (1,0) circle (1.1ex);
		\draw[ForestGreen, line width=2pt] (-1,0) -- (1,0);
		\draw[Maroon, fill=Maroon] (4,0) circle (1.1ex);
		\draw[blue, line width=1.5pt, ->] (-2,-0.4) to[out=5, in=185] (4,0.7) to[out=80, in=230] (5, 3.1);
		\draw[orange, line width=1.5pt, <-] (-1, -4) to[out=65, in=240] (1, -0.5) to[out=5, in=185] (5, -0.2);
		\node at (0, 5) {$\text{arg}(g_{\text{s}})=0^{+}$};
		\node[blue] at (4,2.5) {$\CC^{+}$};
		\node[orange] at (4, -1) {$\bar{\CC}^{\star,+}$};
		\end{scope}
		\begin{scope}[shift={({6},{0})}]
		\draw[gray, line width=1.5pt, ->] (-2,0) -- (5,0);
		\draw[gray, line width=1.5pt, ->] (0,-4) -- (0,4);
		\draw[ForestGreen, fill=ForestGreen] (-1,0) circle (1.1ex);
		\draw[ForestGreen, fill=ForestGreen] (1,0) circle (1.1ex);
		\draw[ForestGreen, line width=2pt] (-1,0) -- (1,0);
		\draw[Maroon, fill=Maroon] (4,0) circle (1.1ex);
		\draw[blue, line width=1.5pt, ->] (-2,0.4) to[out=355, in=175] (4,-0.7) to[out=280, in=130] (5, -3.1);
		\draw[orange, line width=1.5pt, <-] (-1, 4) to[out=295, in=120] (1, 0.5) to[out=355, in=175] (5, 0.2);
		\node at (0, 5) {$\text{arg}(g_{\text{s}})=0^{-}$};
		\node[blue] at (4,-2.5) {$\CC^{-}$};
		\node[orange] at (4, 1) {$\bar{\CC}^{\star,-}$};
		\end{scope}
	\end{tikzpicture}
\caption{Integration contours in the cubic matrix model (to be discussed in subsection~\ref{subsec:CubicMatrixModelChecks}) drawn on top of its figure~\ref{fig:settingcutandsaddle}, for both (perturbative) {\color{blue}eigenvalues} and (nonperturbative) {\color{orange}anti-eigenvalues}. At a forward Stokes transition both contours undergo a Stokes jump. Further, the eigenvalue perturbative contour crosses the anti-eigenvalue nonperturbative contour somewhere in-between the endpoint of the cut and the nonperturbative saddle-point. For our double-pole integrals as in \eqref{eq:oneeigenvalueoneantieigenvalueforwardtunneling} this leads to a residue contribution as generically illustrated in figure~\ref{fig:settingcutandsaddleGeneraltunneling}.}
\label{fig:cubicexampletransition}
\end{figure}
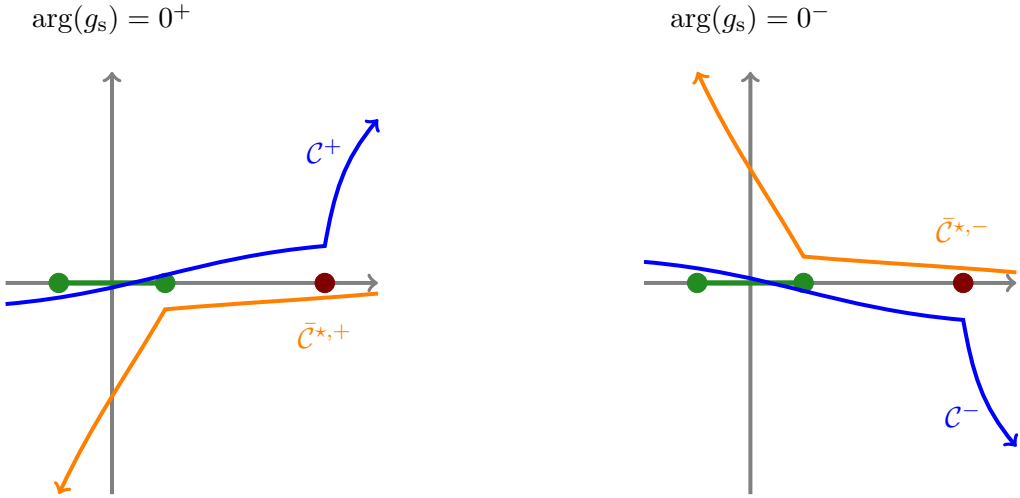

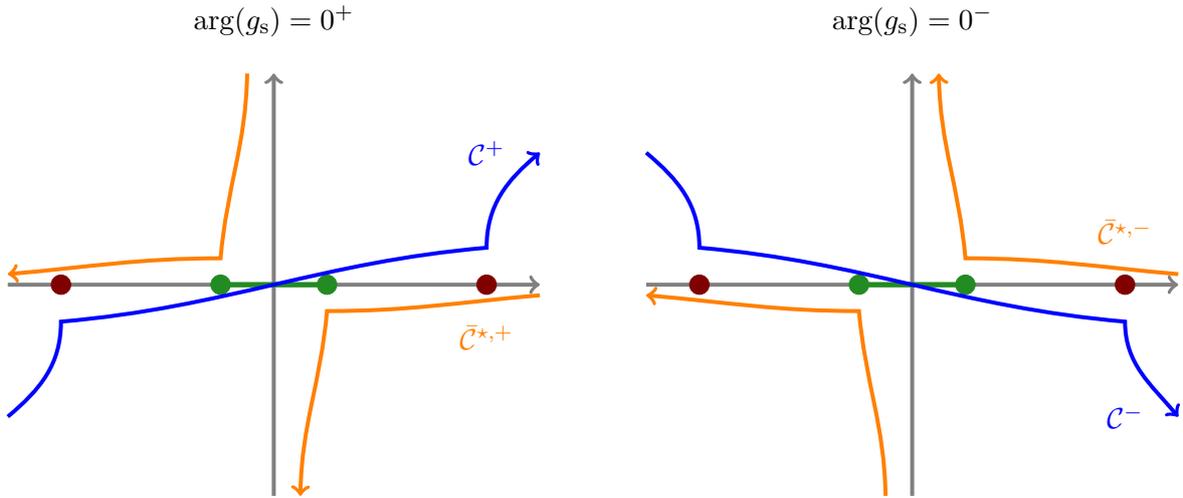
\begin{figure}
\centering
	\begin{tikzpicture}[scale=0.7]
		\begin{scope}[shift={({-6},{0})}]
		\draw[gray, line width=1.5pt, ->] (-5,0) -- (5,0);
		\draw[gray, line width=1.5pt, ->] (0,-4) -- (0,4);
		\draw[ForestGreen, fill=ForestGreen] (-1,0) circle (1.1ex);
		\draw[ForestGreen, fill=ForestGreen] (1,0) circle (1.1ex);
		\draw[ForestGreen, line width=2pt] (-1,0) -- (1,0);
		\draw[Maroon, fill=Maroon] (4,0) circle (1.1ex);
		\draw[Maroon, fill=Maroon] (-4,0) circle (1.1ex);
		\draw[blue, line width=1.5pt, ->] (-5, -2.5) to[out=40, in=270] (-4,-0.7) to[out=5, in=185] (4,0.7) to[out=90, in=220] (5, 2.5);
		\draw[orange, line width=1.5pt, ->] (-0.5, 4) to[out=269, in=85] (-1, 0.5) to[out=180, in=5] (-5, 0.2);
		\draw[orange, line width=1.5pt, <-] (0.5, -4) to[out=89, in=265] (1, -0.5) to[out=0, in=185] (5, -0.2);
		\node at (0, 5) {$\text{arg}(g_{\text{s}})=0^{+}$};
		\node[blue] at (4,2.5) {$\CC^{+}$};
		\node[orange] at (4, -1) {$\bar{\CC}^{\star,+}$};
		\end{scope}
		\begin{scope}[shift={({6},{0})}]
		\draw[gray, line width=1.5pt, ->] (-5,0) -- (5,0);
		\draw[gray, line width=1.5pt, ->] (0,-4) -- (0,4);
		\draw[ForestGreen, fill=ForestGreen] (-1,0) circle (1.1ex);
		\draw[ForestGreen, fill=ForestGreen] (1,0) circle (1.1ex);
		\draw[ForestGreen, line width=2pt] (-1,0) -- (1,0);
		\draw[Maroon, fill=Maroon] (4,0) circle (1.1ex);
		\draw[Maroon, fill=Maroon] (-4,0) circle (1.1ex);
		\draw[blue, line width=1.5pt, ->] (-5, 2.5) to[out=320, in=90] (-4,0.7) to[out=355, in=175] (4,-0.7) to[out=270, in=130] (5, -2.5);
		\draw[orange, line width=1.5pt, ->] (-0.5, -4) to[out=91, in=275] (-1, -0.5) to[out=180, in=355] (-5, -0.2);
		\draw[orange, line width=1.5pt, <-] (0.5, 4) to[out=271, in=95] (1, 0.5) to[out=0, in=175] (5, 0.2);
		\node at (0, 5) {$\text{arg}(g_{\text{s}})=0^{-}$};
		\node[blue] at (4,-2.5) {$\CC^{-}$};
		\node[orange] at (4, 1) {$\bar{\CC}^{\star,-}$};
		\end{scope}
	\end{tikzpicture}
\caption{Integration contours in the quartic matrix model (to be discussed in subsection~\ref{subsec:quarticexample}) drawn on top of its figure~\ref{fig:settingcutandsaddle} (which now has two symmetric nonperturbative saddles), for both (perturbative) {\color{blue}eigenvalues} and (nonperturbative) {\color{orange}anti-eigenvalues}. For this example, eigenvalue perturbative-contours were already shown in \eqref{eq:rotationperturbativecontour}, and anti-eigenvalue nonperturbative-contours in \eqref{eq:antieigenvaluetunnelingpictorial}. At a forward Stokes transition, both contours jump and eigenvalue/anti-eigenvalue contours cross, leading to residue contributions for \eqref{eq:oneeigenvalueoneantieigenvalueforwardtunneling} as illustrated in figure~\ref{fig:settingcutandsaddleGeneraltunneling}.}
\label{fig:quarticexampletransition}
\end{figure}

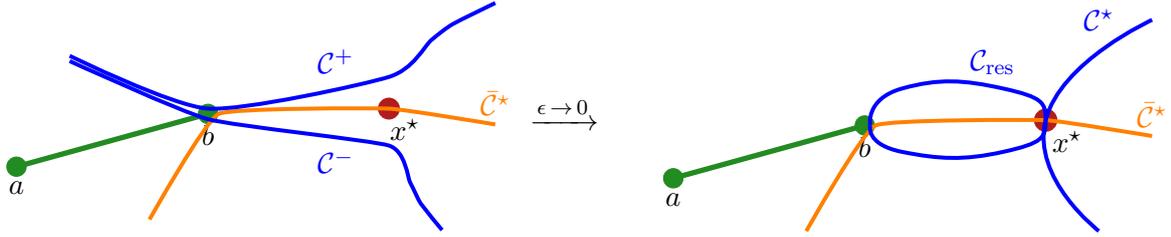
\begin{figure}
\centering
    \begin{minipage}{.5\textwidth}
    \centering
	\begin{tikzpicture}[scale=0.7]
		\draw[ForestGreen, line width=2pt] (-2, -1) -- (1.6, 0);
		\draw[ForestGreen, fill=ForestGreen] (-2,-1) circle (1.1ex);
		\node at (-2, -1.4) {$a$};
		\draw[ForestGreen, fill=ForestGreen] (1.6,0) circle (1.1ex);
		\node at (1.6, -.4) {$b$};
		\draw[cornellred, fill=cornellred, line width=1pt] (5,0.1) circle (1.1ex);
		\node at (5.3, -0.3) {$x^{\star}$};
		\draw[orange, line width=1.5pt] plot [smooth, tension=0.2] coordinates{(7, -0.2)(5, 0.1)(1.8,0)(0.5,-2)};
		\draw[blue, line width=1.5pt] plot [smooth, tension=0.4] coordinates{(-1, 1.1)(1.6, 0.1)(5,0.7)(5.8,1.4)(6.2,1.7)(7, 2.1)};
		\draw[blue, line width=1.5pt] plot [smooth, tension=0.3] coordinates{(-1, 1)(1.6,-0.1)(5,-0.6)(5.4, -1.3)(5.6, -1.7)(6, -2.2)};
		\node at (8.8,0) {$\xrightarrow{\epsilon\,\rightarrow\,0}$};
		\node[blue] at (4, 1) {$\CC^{+}$};
		\node[blue] at (4, -0.9) {$\CC^{-}$};
		\node[orange] at (7, 0.2) {$\bar{\CC}^{\star}$};
	\end{tikzpicture}
	\end{minipage}%
    \begin{minipage}{0.5\textwidth}
    \centering
		\begin{tikzpicture}[scale=0.7]
		\draw[ForestGreen, line width=2pt] (-2, -1) -- (1.6, 0);
		\draw[ForestGreen, fill=ForestGreen] (-2,-1) circle (1.1ex);
		\node at (-2, -1.4) {$a$};
		\draw[ForestGreen, fill=ForestGreen] (1.6,0) circle (1.1ex);
		\node at (1.6, -.4) {$b$};
		\draw[cornellred, fill=cornellred, line width=1.5pt] (5,0.1) circle (1.1ex);
		\node at (5.4, -0.3) {$x^{\star}$};
		\draw[orange, line width=1.5pt] plot [smooth, tension=0.2] coordinates{(7, -0.2)(5, 0.1)(1.8,0)(0.5,-2)};
		\draw[blue, line width=1.5pt] plot [smooth, tension=1] coordinates{(6, -2)(5,0.1)(7, 2)};
		\node[blue] at (6, 2) {$\CC^{\star}$};
		\node[blue] at (4, 1.2) {$\CC_{\text{res}}$};
		\node[orange] at (7, 0.2) {$\bar{\CC}^{\star}$};
		\draw[blue, line width=1.5pt] plot [smooth, tension=1] coordinates{(1.7, 0)(2.3,0.7)(4.2, 0.7)(5,0.1)(4.2, -0.5)(2.3, -0.5)(1.7,0)};
	\end{tikzpicture}
	\end{minipage}
\caption{Integration contours associated with anti-eigenvalue tunneling (drawn on top of figure~\ref{fig:settingcutandsaddle}). We have two perturbative contours $\CC^{+}$ and $\CC^{-}$ which, for $\epsilon\rightarrow 0$, get deformed into the nonperturbative saddle-point contour $\CC^{\star}$. But in the presence of an anti-eigenvalue a novelty appears: we pick-up an additional residue contribution $\CC_{\text{res}}$ when deforming the contours $\CC^{+}$ and $\CC^{-}$ past the anti-eigenvalue contour, due to the poles in the integral \eqref{eq:oneeigenvalueoneantieigenvalueforwardtunneling}.}
\label{fig:settingcutandsaddleGeneraltunneling}
\end{figure}

Let us move towards the general case in steps. First consider the one tunneled-anti-eigenvalue configuration in \eqref{eq:01correctedcontour}-\eqref{eq:separatedoneantieigenvalueWithCOntour}, but now with one added eigenvalue which we would like to tunnel to the nonperturbative saddle. This is then described by the matrix integral \eqref{eq:determinantformulak1sectors}, where we initially consider an eigenvalue \textit{perturbative} contour,
\begin{equation}
\label{eq:oneeigenvalueoneantieigenvalueforwardtunneling}
\mathcal{Z}^{(1|1)}_{\lbrace \CC, \bar{\CC}^{\star} \rbrace} (t) = \frac{1}{N+1}\, \mathcal{Z}^{(0|0)} (t) \pvint_{\CC} \frac{\text{d}x}{2\pi} \pvint_{\bar{\CC}^{\star}} \frac{\text{d}\bar{x}}{2\pi}\, \frac{1}{\left(\bar{x}-x\right)^2}\, \rme^{-\frac{1}{g_{\text{s}}} V(x) + \frac{1}{g_{\text{s}}} V(\bar{x})} \ev{\frac{\det \left(x-M\right)^{2}}{\det \left(\bar{x}-M\right)^{2}}}_{N}.
\end{equation}
\noindent
Consider the above integration contours close to a forward Stokes transition. This is illustrated for the example of the cubic matrix model in figure~\ref{fig:cubicexampletransition}, for the example of the quartic matrix model in figure~\ref{fig:quarticexampletransition}, and generically in figure~\ref{fig:settingcutandsaddleGeneraltunneling}. The eigenvalue perturbative contour has been previously studied. The anti-eigenvalue contour $\bar{\CC}^{\star}$ through the nonperturbative saddle $x^{\star}$ is more subtle: contrary to anti-eigenvalue \textit{perturbative} contours, anti-eigenvalue \textit{nonperturbative} contours do undergo a Stokes jump at $\theta=0$ (this is quite evident for the cubic in figure~\ref{fig:cubicexampletransition} and for the quartic in figure~\ref{fig:quarticexampletransition}). Focusing on our favorite quartic example, the jump of the anti-eigenvalue nonperturbative contour from $\theta=0^+$ to $\theta=0^-$ is\footnote{The factor of $2$ arises from the fact that the quartic example has $2$ symmetric saddle points. In the setting of a single nonperturbative saddle such factor would be absent, and, as such, we shall drop it in the ensuing discussion. But, depending on the example in question, one may have to carefully take such symmetry factors into account.} (compare with \eqref{eq:quarticexampleeigenvaluetunneling})
\begin{equation}
\label{eq:antieigenvaluesaddles}
\bar{\CC}^{\star,+}_{\text{quartic}} - \bar{\CC}^{\star,-}_{\text{quartic}} = \ZSadBarpe\,-\,\ZSadBarme\,\xrightarrow{\epsilon\,\rightarrow\,0}\,-2\,\ZPertBarme\,= - 2\, \bar{\CC}_{\text{quartic}}.
\end{equation}
\noindent
In other words, nonperturbative anti-eigenvalues can ``tunnel back'' to the perturbative saddle at the positive Stokes ray. But, it turns out, any such contributions may be always absorbed by the constant component in the free-energy transseries (which always exists as this transseries is not quite purely two-parameters; see \cite{abs18} and the discussion in appendix~\ref{app:resurgenceZvsF}). As purely-additive free-energy corrections, we shall ignore them in the following for the sake of simplicity.

In light of this added subtlety, let us then evaluate the discontinuity \eqref{eq:discontinuityongeneralfunctionexpression} of the above integral \eqref{eq:oneeigenvalueoneantieigenvalueforwardtunneling} across $\theta=0$; which is
\begin{align}
\label{eq:disc0Z11CstarCbarstar}
\text{Disc}_{0}\, \mathcal{Z}^{(1|1)}_{\lbrace \CC, \bar{\CC}^{\star} \rbrace} (g_s) &= \mathcal{Z}^{(1|1)}_{\lbrace \CC^{+}, \bar{\CC}^{\star,+} \rbrace} (\abs{g_{\text{s}}}\, \rme^{\rmi 0^{+}}) - \mathcal{Z}^{(1|1)}_{\lbrace \CC^{-}, \bar{\CC}^{\star,-} \rbrace} (\abs{g_{\text{s}}}\, \rme^{\rmi 0^{-}}) = \nonumber \\
&= \mathcal{Z}^{(1|1)}_{\lbrace \CC^{+}-\CC^{-}, \bar{\CC}^{\star,-} \rbrace} (\abs{g_{\text{s}}}\, \rme^{\rmi 0^{-}}) + \mathcal{Z}^{(1|1)}_{\lbrace \CC^{+}-\CC^{-}, - \bar{\CC}^{-} \rbrace} (\abs{g_{\text{s}}}\, \rme^{\rmi 0^{-}}) + \mathcal{Z}^{(1|1)}_{\lbrace \CC^{-}, -\bar{\CC}^{-} \rbrace} (\abs{g_{\text{s}}}\, \rme^{\rmi 0^{-}}) = \nonumber \\ 
&= \mathcal{Z}^{(1|1)}_{\lbrace \CC^{+}-\CC^{-}, \bar{\CC}^{\star,-} \rbrace} (\abs{g_{\text{s}}}\, \rme^{\rmi 0^{-}}) + \cdots.
\end{align}
\noindent
In the last line we dropped the aforementioned terms associated to additive free-energy constants---and we refer the reader to appendix~\ref{app:resurgenceZvsF} for further details on this.

In principle we would now like to repeat the deformation of contours in \eqref{eq:tunnelingoneeigenvalueone} and write $\CC^{+}-\CC^{-} = \CC^{\star}$. But due to the pole in \eqref{eq:oneeigenvalueoneantieigenvalueforwardtunneling} when eigenvalues and anti-eigenvalues meet, this alone cannot be the case---clearly, if we are to make sense of this integral the double pole in the integrand must be tackled. Instead, on top of the nonperturbative contour $\CC^{\star}$, we also find an additional closed contour $\CC_{\text{res}}$ which is picking-up pole contributions in-between the endpoint of the cut and the nonperturbative saddle. This crossing of the eigenvalue contour $\CC^{+}$ through the anti-eigenvalue contour $\bar{\CC}^{\star}$ at a forward Stokes transition is also illustrated in figures~\ref{fig:cubicexampletransition}, \ref{fig:quarticexampletransition}, and~\ref{fig:settingcutandsaddleGeneraltunneling}. In our favorite example, this deformation of contours finally amounts to
\begin{equation}
\ZPertpe\,\ZSadBarpe\,-\,\ZPertme\,\ZSadBarme\,=\,\ZSuperSaddles\,\ZSadBarpe\,-\,\cdots,
\end{equation}
\noindent
where the dots again stand for the dropped additive contributions---and because we henceforth only focus on physical contributions, we also drop such dots from now on. The final contour is then exactly the contour shown in figure~\ref{fig:settingcutandsaddleGeneraltunneling}. We may then finally ``abbreviate'' \eqref{eq:disc0Z11CstarCbarstar} as
\begin{equation}
\text{Disc}_{0}\, \mathcal{Z}^{(1|1)}_{\lbrace \CC, \bar{\CC}^{\star} \rbrace} (g_s) = \mathcal{Z}^{(1|1)}_{\lbrace \CC^{+}, \bar{\CC}^{\star} \rbrace} (\abs{g_{\text{s}}}\, \rme^{\rmi 0^{+}}) - \mathcal{Z}^{(1|1)}_{\lbrace \CC^{-}, \bar{\CC}^{\star} \rbrace} (\abs{g_{\text{s}}}\, \rme^{\rmi 0^{-}}) = \mathcal{Z}^{(1|1)}_{\lbrace \CC_{\text{res}}+\CC^{\star}, \bar{\CC}^{\star} \rbrace} (\abs{g_{\text{s}}}\, \rme^{\rmi 0^{-}}).
\end{equation}
\noindent 
More details on these types of integration contours may be found in appendices~\ref{app:detailsintegrationcontours} and~\ref{app:resurgenceZvsF}.

The exact same contour deformations go through for any number of eigenvalues in the presence of anti-eigenvalues. In this way, as long as one carefully keeps track of all additional residue closed contours, eigenvalue-tunneling calculations go through. Consider the $(\ell|1)$ sectors appearing in
\begin{equation}
\text{Disc}_0\, \mathcal{Z}^{(0|1)} (g_{\text{s}}) = \sum\limits_{\ell=1}^{\infty} \mathcal{Z}^{(\ell|1)} (g_{\text{s}}),
\end{equation}
\noindent
whose matrix integral representation is in \eqref{eq:determinantformulak1sectors}, but which still lacked full specification of the relevant integration contours. The fully precise matrix integral \eqref{eq:determinantformulak1sectors} is now just: 
\begin{equation}
\mathcal{Z}^{(\ell|1)} \equiv {{N+1}\choose{\ell}}\, \mathcal{Z}^{(\ell|1)}_{\lbrace \underbrace{\scriptstyle{\CC_{\text{res}} + \CC^{\star}, \ldots, \CC_{\text{res}} + C^{\star}}}_{\ell}, \bar{\CC}^{\star} \rbrace}.
\end{equation}
\noindent
For illustration, these integration contours in the quartic matrix model example are
\begin{equation}
\underbrace{\ZSuperSaddles\,\dots\,\ZSuperSaddles}_{\ell}\,\underbrace{\ZPertpe\,\dots\,\ZPertpe}_{N-\ell+1}\,\,\ZSadBarpe\,,
\end{equation}
\noindent
alongside the required residue-contour deformations of figure~\ref{fig:settingcutandsaddleGeneraltunneling}. 

Let us go back to figure~\ref{fig:resonantresurgentlatticeTwoAutomorphisms} and address the full rotation depicted therein. This implies applying a forward transition at $\theta=0$ on top of the backwards one at $\theta=\pi$. Within the resurgence framework, this translates to (iteration of \eqref{eq:resurgentforwarddiscontinuity} and \eqref{eq:purelybackwardsdiscontinuity})
\begin{align}
\label{eq:TwoAutomorphismsAllSectors}
\text{Disc}_{0}\, \text{Disc}_{\pi}\, Z^{(0,0)} (t) = \sum_{\ell=1}^{\infty}\sum_{\bar{\ell}=1}^{\infty} \rme^{\left(\bar{\ell}-\ell\right) \frac{A}{g_{\text{s}}}} \sum_{p=0}^{\min(\ell,\bar{\ell})} \mathsf{S}_{(0,0)\to(0,\bar{\ell})}\, \mathsf{S}_{(0,\bar{\ell})\to(\ell-p,\bar{\ell}-p)}\, Z^{(\ell-p,\bar{\ell}-p)} (t).
\end{align}
\noindent
Proceeding step-by-step as usual, first focus on the $\bar{\ell}=1$ contribution. If our previous matrix-integral arguments hold, this should now establish a relation between resurgence and matrix-integral discontinuities, \textit{i.e.}, we should be able to identify their respective terms one-by-one as (compare with \eqref{eq:forwardresurgentEVtunneling})
\begin{align}
\label{eq:generalexpressionforeigenvaluetunnelingtwoparameters-ell1}
\rme^{-\left(\ell-1\right) \frac{A}{g_{\text{s}}}} \sum_{p=0}^{\min(\ell,1)} \mathsf{S}_{(0,0)\to(0,1)}\, \mathsf{S}_{(0,1)\to(\ell-p,1-p)}\, \frac{Z^{(\ell-p,1-p)} (t, g_{\text{s}})}{Z^{(0,0)} (t, g_{\text{s}})} \simeq \frac{\mathcal{Z}^{(\ell|1)}(t, g_{\text{s}})}{\mathcal{Z}^{(0|0)}(t, g_{\text{s}})}.
\end{align}
\noindent
This makes precise how \textit{anti-eigenvalue} tunneling now yields the resurgence of the \textit{full} transseries. In fact, this expression is taking us \textit{inside} the resurgence lattice, as in figure~\ref{fig:resonantresurgentlatticeTwoAutomorphisms}. There is, however, a price to pay: the additional $p$-sum over the resonant kernel (originating from the resurgence discontinuity \eqref{eq:TwoAutomorphismsAllSectors}). This implies our (anti) eigenvalue configurations no longer are in one-to-one correspondence with resurgent transseries sectors---but only with their resonant kernels. This is exactly the phenomenon which was anticipated in figure~\ref{fig:illustrationoftunnelingforresonantcase}. Finally, as in \eqref{eq:forwardresurgentEVtunneling}, this expression yields a comparison of partition-function \textit{ratios}, \textit{i.e.}, we are matching \textit{free energies}.

All we have left to do is to generalize the above formulae for any (anti) eigenvalue configuration. The procedure is exactly the same as described above. Pure anti-eigenvalue tunneling is described by our generic matrix integral \eqref{eq:finalresultseparatedeigenvalues} specialized to the $(0|\bar{\ell})$ backward sector and where we still need\footnote{Note how in this pure anti-eigenvalue case the correct symmetry factor is already included in \eqref{eq:finalresultseparatedeigenvalues}.} to specify the anti-eigenvalue integration contours. These go through the nonperturbative saddle $x^{\star}$; so that the fully precise $(0|\bar{\ell})$ matrix integral \eqref{eq:finalresultseparatedeigenvalues} is now:
\begin{equation}
\label{eq:backwardsectoreigenvaluetunneling}
\mathcal{Z}^{(0|\bar{\ell})} \equiv \mathcal{Z}^{(0|\bar{\ell})}_{\lbrace \underbrace{\scriptstyle{\bar{\CC}^{\star}, \ldots, \bar{\CC}^{\star}}}_{\bar{\ell}} \rbrace}.
\end{equation}
\noindent
In complete analogy with the pure eigenvalue-tunneling case in \eqref{eq:forwardresurgentEVtunneling}, we now find the bridge between resurgence and matrix-integrals to be
\begin{equation}
\label{eq:backwardresurgentEVtunneling}
\text{e}^{\bar{\ell}\frac{A}{g_{\text{s}}}}\, \mathsf{S}_{(0,0)\rightarrow (0,\bar{\ell})}\, \frac{Z^{(0,\bar{\ell})}(t, g_{\text{s}})}{Z^{(0,0)}(t, g_{\text{s}})} \simeq \frac{\mathcal{Z}^{(0|\bar{\ell})}(t, g_{\text{s}})}{\mathcal{Z}^{(0| 0)}(t, g_{\text{s}})}.
\end{equation}
\noindent
Enter the bulk of the resurgence lattice following the full motion in figure~\ref{fig:resonantresurgentlatticeTwoAutomorphisms} and undergoing the remaining forward Stokes transition. Comparing exponential orders we arrive at the general result (compare with \eqref{eq:generalexpressionforeigenvaluetunnelingtwoparameters-ell1} above)
\begin{align}
\label{eq:generalexpressionforeigenvaluetunnelingtwoparameters}
\rme^{-(\ell-\bar{\ell}) \frac{A}{g_{\text{s}}}} \sum_{p=0}^{\min(\ell,\bar{\ell})} \mathsf{S}_{(0,0)\to(0,\bar{\ell})}\, \mathsf{S}_{(0,\bar{\ell})\to(\ell-p,\bar{\ell}-p)}\, \frac{Z^{(\ell-p,\bar{\ell}-p)} (t, g_{\text{s}})}{Z^{(0,0)} (t, g_{\text{s}})} \simeq \frac{\mathcal{Z}^{(\ell|\bar{\ell})}(t, g_{\text{s}})}{\mathcal{Z}^{(0|0)}(t, g_{\text{s}})}, \qquad \ell,\bar{\ell} \geq 1,
\end{align}
\noindent
where the explicit matrix integral on the right-hand side is given by the precise specification of integration contours and combinatorial factors required in \eqref{eq:finalresultseparatedeigenvalues}, which finally becomes:
\begin{equation}
\label{eq:generalsectorfromeigenvaluetunneling}
\mathcal{Z}^{(\ell|\bar{\ell})} \equiv {{N+\bar{\ell}}\choose{\ell}}\, \mathcal{Z}^{(\ell|\bar{\ell})}_{\lbrace \underbrace{\scriptstyle{\CC_{\text{res}} + \CC^{\star}, \ldots, \CC_{\text{res}} + \CC^{\star}}}_{\ell}, \underbrace{\scriptstyle{\bar{\CC}^{\star}, \ldots, \bar{\CC}^{\star}}}_{\bar{\ell}} \rbrace}.
\end{equation}

Some final comments are in order. As in \eqref{eq:generalexpressionforeigenvaluetunnelingtwoparameters-ell1}, also \eqref{eq:generalexpressionforeigenvaluetunnelingtwoparameters} makes clear that (anti) eigenvalue configurations are related to resonant sums of transseries sectors. Our formulae directly compute Borel residues (hence, Stokes data) for these sectors. But an additional subtlety emerges: depending on the sequence of Stokes transitions one carries through in these iterative processes, different Borel residues (Stokes data) may appear associated to the very same transseries sectors. This is associated to the non-trivial nature of Borel plane singularities, and is captured by consistency conditions on higher Stokes data which we discuss in appendix~\ref{app:HigherOrderStokes}. Having set-up our matrix integrals in the previous subsection~\ref{subsec:DeterminantCorrelators}, and made precise which contours to integrate them over in the present subsection, we may now move towards explicit computations.

\subsection{On the Boundary of a Resonant Transseries}\label{subsec:edgesofresurgentlattice}

The natural place to begin our explicit calculations is on the boundary of the resurgence lattice illustrated back in figure~\ref{fig:walkingonesteppathsintro}. In other words, we begin by focusing on the simplest (anti) eigenvalue configurations, $(\ell|0)$ and $(0|\bar{\ell})$. Both computations are immediate (given the literature).

Computing the ``forward boundary'' of the resurgence lattice has been done in many places via saddle-point methods, \cite{msw07, msw08, emms22b}. Such saddle-point evaluation of \eqref{eq:generalsectorfromeigenvaluetunneling}-\eqref{eq:finalresultseparatedeigenvalues} for the $(\ell|0)$ configuration yields, \textit{to first order} in $g_{\text{s}}$,
\begin{equation}
\label{eq:resultedgeforward}
\frac{\mathcal{Z}^{(\ell|0)}(t, g_{\text{s}})}{\mathcal{Z}^{(0|0)}(t, g_{\text{s}})} = \frac{G_{2}\left(\ell+1\right)}{(2\pi)^{\ell/2}}\, \rme^{-\frac{\ell}{g_{\text{s}}} \left( V_{\text{h;eff}}(x^{\star})-V_{\text{h;eff}}(b) \right)} \left(\frac{g_{\text{s}}\left(b-a\right)^2}{16 V_{\text{h;eff}}^{\prime\prime}(x^{\star})}\right)^{\frac{\ell^2}{2}} \frac{1}{\left( \left(x^{\star}-a\right) \left(x^{\star}-b\right) \right)^{\ell^2}}.
\end{equation}
\noindent
Hopefully by now, it should be clear to the reader that the computation for the ``backward boundary'' is completely analogous to the one above (and we shall henceforth give no details). Consider again \eqref{eq:generalsectorfromeigenvaluetunneling}-\eqref{eq:finalresultseparatedeigenvalues} but now for the $(0|\bar{\ell})$ configuration, and perform its saddle-point evaluation \textit{to first order} in $g_{\text{s}}$. We find
\begin{equation}
\label{eq:resultedgebackward}
\frac{\mathcal{Z}^{(0|\bar{\ell})}(t, g_{\text{s}})}{\mathcal{Z}^{(0|0)}(t, g_{\text{s}})} = \frac{G_{2}\left(\bar{\ell}+1\right)}{(2\pi)^{\bar{\ell}/2}}\, \rme^{\frac{\bar{\ell}}{g_{\text{s}}} \left( V_{\text{h;eff}}(x^{\star})-V_{\text{h;eff}}(b) \right)} \left( - \frac{g_{\text{s}}\left(b-a\right)^2}{16 V_{\text{h;eff}}^{\prime\prime}(x^{\star})}\right)^{\frac{\bar{\ell}^2}{2}} \frac{1}{\left( \left(x^{\star}-a\right) \left(x^{\star}-b\right) \right)^{\bar{\ell}^2}}.
\end{equation}
\noindent
Unsurprisingly, forward and backward results are extremely similar. Up to signs they are in fact related, which is a common feature occurring in resonant problems and is sometimes dubbed a backward-forward relation \cite{bssv22}. The instanton actions in the arguments of the transmonomials have changed sign, just as expected in accordance with \eqref{eq:forwardresurgentEVtunneling} and \eqref{eq:backwardresurgentEVtunneling}. These formulae further imply that the above expressions should be computing Stokes data, and in fact they are. They immediately compute the forward and backward Borel residues in \eqref{eq:forwardbackwardBorelResidues}, which is to say they yield analytical expressions for the two Stokes vectors
\be
\boldsymbol{S}_{(1,0)} \quad \text{ and } \quad \boldsymbol{S}_{(0,1)}.
\ee
\noindent
The first one is the usually well-known ``canonical'' Stokes coefficient; for instance computed from matrix models for Painlev\'e~I in, \textit{e.g.}, \cite{d91, msw07}. The second one is its backward sibling.

\subsection{In the Bulk of a Resonant Transseries}\label{subsec:bulkoneonesector}

As we finally venture inside the resurgence lattice (recall figure~\ref{fig:walkingonesteppathsintro}) the difficulty of calculations quickly escalates. In this subsection we focus on the first such non-trivial ``bulk'' sector one needs to face, which is the $(1|1)$ sector. As we have already discussed at length, in this sector one has to deal with residue contours \eqref{eq:generalsectorfromeigenvaluetunneling}---which prevent us from performing a direct saddle-point evaluation. We will find a way to circumvent this problem by considering derivatives with respect to $g_{\text{s}}$ of the integrals in question. Before diving into such direct calculation, it is useful to take a step back and consider a toy example.

\paragraph{A Toy Example:}

Let us consider the following toy integral
\begin{equation}
\label{eq:toymodelexample11}
\CI = \pvint_{\CC} \frac{\text{d}x}{2\pi} \pvint_{\bar{\CC}} \frac{\text{d}\bar{x}}{2\pi}\, \frac{1}{(x-\bar{x})^2}\, \rme^{-\frac{1}{g_{\text{s}}} \left(V(x)-V(\bar{x})\right)},
\end{equation}
\noindent
where $V(x)$ is a polynomial potential with a saddle-point $x^{\star}$, $\CC$ its associated steepest-descent, and $\bar{\CC}$ its associated steepest-ascent contours. The principle-value prescription is as always defined in \eqref{eq:principalvalueprescription}, where we recall that $\CC$ and $\bar{\CC}$ are parameterized by $x(\tau)$ and $\bar{x}(\bar{\tau})$, respectively, such that $x(0)=x^{\star}=\bar{x}(0)$. To better understand this toy-integral let us try to perform a na\"\i ve saddle-point evaluation, and see how it fails. Set $(\tau,  \bar{\tau}) = (\sqrt{g_{\text{s}}}\, s, \sqrt{g_{\text{s}}}\, \bar{s})$ and expand its integrand in $g_{\text{s}}$. For the singular term, we have the expansion (recall $x(0)-\bar{x}(0)=0$)
\begin{equation}
\label{eq:expansionofpolethatdestroysgsexpansion}
\frac{1}{\left( x(\sqrt{g_{\text{s}}}\, s) - \bar{x}(\sqrt{g_{\text{s}}}\, \bar{s}) \right)^2} = \frac{1}{g_{\text{s}} \left( x^{\prime}(0)\, s - \bar{x}^{\prime}(0)\, \bar{s} \right)^2} - \frac{x^{\prime\prime}(0)\, s^2 - \bar{x}^{\prime\prime}(0)\, \bar{s}^2}{\sqrt{g_{\text{s}}} \left( x^{\prime}(0)\, s - \bar{x}^{\prime}(0)\, \bar{s} \right)^3} + \frac{(\cdots)}{\left( x^{\prime}(0)\, s - \bar{x}^{\prime}(0)\, \bar{s}\right)^4} + \cdots,
\end{equation}
\noindent
which, for $s=\bar{s}$, produces ever higher-order poles for higher $g_{\text{s}}$ orders. As such, blind saddle-point does not result in a sensible $g_{\text{s}}$-expansion, and we must resort to another procedure. 

It is not too hard to realize that taking adequate $g_{\text{s}}$-derivatives of $\CI \equiv \CI (g_{\text{s}})$ can kill the singularity in the integrand. In particular, the integral\footnote{In ordinary supermatrix models, a similar observation evaluates integrals over $(1,1)$ supermatrices \cite{hgk18}.} $\partial_{1/g_{\text{s}}}^2 \CI$ does not contain poles. We find
\begin{equation}
\partial_{1/g_{\text{s}}}^2 \CI = \pvint_{\CC} \frac{\text{d}x}{2\pi} \pvint_{\bar{\CC}} \frac{\text{d}\bar{x}}{2\pi} \left(\frac{V(x)-V(\bar{x})}{x-\bar{x}}\right)^2 \rme^{-\frac{1}{g_{\text{s}}}\left(V(x)-V(\bar{x})\right)}.
\end{equation} 
\noindent
Now that the integrand is no longer singular, standard saddle-point evaluation follows as usual. The one thing left to do, however, is to revert the $g_{\text{s}}$-derivatives with their appropriate primitives, in order to find the final answer. But it first seems there is no way to fix the resulting integration constants. This conundrum may be solved by recasting the above derivative-procedure as an exchange in the order of integrations. This is done as follows. First write
\begin{equation}
\label{eq:integraltermtodefinederivativetrick}
\rme^{-\frac{1}{g_{\text{s}}} \left(V(x)-V(\bar{x})\right)} = \int_{1}^{+\infty} \text{d}u \int_{u}^{+\infty} \text{d}v\, \frac{1}{g_{\text{s}}^2} \left(V(x)-V(\bar{x})\right)^2 \rme^{-\frac{v}{g_{\text{s}}} \left(V(x)-V(\bar{x})\right)}.
\end{equation}
\noindent
These integrals are well defined along $x\in \CC$ and $\bar{x}\in\bar{\CC}$, except for the point $x=\bar{x}=x^{\star}$---where the principal-value prescription \eqref{eq:principalvalueprescription} comes to our rescue. Then we can write
\begin{equation}
\CI = \pvint_{\CC} \frac{\text{d}x}{2\pi} \pvint_{\bar{\CC}} \frac{\text{d}\bar{x}}{2\pi} \int_{1}^{+\infty} \text{d}u \int_{u}^{+\infty} \text{d}v\, \frac{1}{g_{\text{s}}^2} \left(\frac{V(x)-V(\bar{x})}{x-\bar{x}}\right)^2 \rme^{-\frac{v}{g_{\text{s}}} \left(V(x)-V(\bar{x})\right)}.
\end{equation}
\noindent
In this way the above derivative-procedure comes down to an exchange of integrations: the $x$ and $\bar{x}$ principal-value integrals with the $u$ and $v$ ordinary integrals. This is by no means a straightforward process. For instance, spelling out the principal-value prescription in \eqref{eq:principalvalueprescription} we see that on top of exchanging ordinary integrals we also need to exchange the integrals in \eqref{eq:integraltermtodefinederivativetrick} with the $\epsilon\to 0$ limit---which unsurprisingly requires a notion of uniform convergence of the integrand. A detailed discussion on this subtlety may be found in appendix~\ref{app:GaussianToyModel}. After changing the order of integrations, one is left with the evaluation of
\begin{equation}
\CI = \int_{1}^{+\infty} \text{d}u \int_{u}^{+\infty} \text{d}v \int_{\CC} \frac{\text{d}x}{2\pi} \int_{\bar{\CC}} \frac{\text{d}\bar{x}}{2\pi}\, \frac{1}{g_{\text{s}}^2} \left(\frac{V(x)-V(\bar{x})}{x-\bar{x}}\right)^2 \rme^{-\frac{v}{g_{\text{s}}} \left(V(x)-V(\bar{x})\right)},
\end{equation}
\noindent
where we no longer need to use the principle-value prescription. Then, using the example of a quartic potential $V(x) = \frac{1}{2}x^2 - \frac{\lambda}{24}x^4$ and making use of standard saddle-point, we find to leading order in $g_{\text{s}}$
\begin{equation}
\CI = \frac{1}{(2\pi)^2} \int_{1}^{+\infty} \text{d}u \int_{u}^{+\infty} \text{d}v \left( \frac{\rmi\pi\lambda}{12 v^3}\, g_{\text{s}} + O (g_{\text{s}}^2) \right).
\end{equation}
\noindent
Performing the final double-integration yields the \textit{unambiguous} result
\begin{equation}
\CI = \frac{\rmi\lambda}{96\pi}\, g_{\text{s}} + O (g_{\text{s}}^2).
\end{equation}
\noindent
Further details alongside a numerical exposition of this integral may be found in appendix~\ref{app:11QuarticToyModel}. Let us next use this derivative-method to evaluate our matrix-integral $(1|1)$ eigenvalue configuration.

\paragraph{The $(1|1)$ Eigenvalue Sector:}

Using our general proposal \eqref{eq:generalsectorfromeigenvaluetunneling}-\eqref{eq:finalresultseparatedeigenvalues} with $(\ell|\bar{\ell})=(1|1)$ leads to the following integral:
\begin{align}
\label{eq:explicitintegraloneone}
\frac{\mathcal{Z}^{(1|1)} (t, g_{\text{s}})}{\mathcal{Z}^{(0|0)} (t, g_{\text{s}})} = \pvint_{\CC_{\text{res}} + \CC^{\star}} \frac{\text{d}x}{2\pi}\, \pvint_{\bar{\CC}^{\star}} \frac{\text{d}\bar{x}}{2\pi}\, \frac{1}{\left(x-\bar{x}\right)^2}\, \rme^{-\frac{1}{g_{\text{s}}} \left( V(x) - V(\bar{x}) \right)} \ev{\frac{\det \left(x-M\right)^{2}}{\det \left(\bar{x}-M\right)^{2}}}_{N},
\end{align}
\noindent
where the integration contours $\CC_{\text{res}}$, $\CC^{\star}$, $\bar{\CC}^{\star}$ are the familiar ones in figure~\ref{fig:settingcutandsaddleGeneraltunneling}. What our ``bridge'' between matrix integrals and transseries sectors in \eqref{eq:generalexpressionforeigenvaluetunnelingtwoparameters} says is that the above integral is computing the following free-energy transseries sectors\footnote{The reader may want to recall that $\mathsf{S}_{(0,1)\rightarrow (0,0)} = 0$ \cite{abs18, bssv22}.}
\begin{align}
\label{eq:oneonesectortransseries}
\mathsf{S}_{(0,0)\rightarrow (0,1)}\, \mathsf{S}_{(0,1)\rightarrow (1,1)}\, \frac{Z^{(1,1)} (t, g_{\text{s}})}{Z^{(0,0)} (t, g_{\text{s}})} &= \\
&
\hspace{-50pt}
= \mathsf{S}_{(0,0)\rightarrow (0,1)}\, \mathsf{S}_{(0,1)\rightarrow (1,1)} \left\{ F^{(1,1)} (t, g_{\text{s}}) + F^{(1,0)} (t, g_{\text{s}})\, F^{(0,1)} (t, g_{\text{s}}) \right\}. \nonumber
\end{align}

Let us evaluate this integral, \eqref{eq:explicitintegraloneone}. First, spell out the first few $g_{\text{s}}$-orders of the determinant-correlator in the integrand
\begin{equation}
\rme^{-\frac{1}{g_{\text{s}}} \left( V(x) - V(\bar{x}) \right)} \ev{\frac{\det \left(x-M\right)^{2}}{\det \left(\bar{x}-M\right)^{2}}}_{N} = \rme^{-\frac{1}{g_{\text{s}}} \left( V_{\text{h;eff}}(x) - V_{\text{h;eff}} (\bar{x}) \right)} \exp \Big\{ \CA_0 (x,\bar{x}) + g_{\text{s}}\, \CA_1 (x,\bar{x}) + \cdots \Big\},
\end{equation}
\noindent
where $\CA_0(x,\bar{x})$ and $\CA_1(x,\bar{x})$ are given by the adequate combinations which follow in comparison to \eqref{eq:gsexpansionofdeterminants} and were explicitly given in \cite{msw07}. We start with the residue contribution, before addressing the saddle-point evaluation (following the previous example \eqref{eq:toymodelexample11}). There only is such a residue contribution when $\bar{x}$ is inside the closed contour $\CC_{\text{res}}$; in which case we can drop the principal-value prescription and calculate\footnote{Recall that correlators and their integrals vanish on the cut.}
\begin{align}
&
\int_{\CC_{\text{res}}} \frac{\text{d}x}{2\pi} \int_{\bar{\CC}^{\star}} \frac{\text{d}\bar{x}}{2\pi}\, \frac{1}{\left(x-\bar{x}\right)^2}\, \rme^{-\frac{1}{g_{\text{s}}} \left( V(x) - V(\bar{x}) \right)} \ev{\frac{\det \left(x-M\right)^{2}}{\det \left(\bar{x}-M\right)^{2}}}_{N} = \nonumber \\
&=
\int_{b}^{x^{\star}} \frac{\text{d}\bar{x}}{2\pi} \int_{\CC_{\text{res}}} \frac{\text{d}x}{2\pi}\, \frac{1}{\left(x-\bar{x}\right)^2}\, \rme^{-\frac{1}{g_{\text{s}}} \left( V_{\text{h;eff}}(x) - V_{\text{h;eff}}(\bar{x}) \right)}\, \rme^{\CA_0 (x,\bar{x})}\,\Big( 1 + g_{\text{s}}\, \CA_1 (x,\bar{x}) + \cdots \Big) = \nonumber \\
&=
\frac{1}{(2\pi)^2} \int_{b}^{x^{\star}} \text{d}\bar{x}\, 2\pi\rmi \left( \frac{1}{g_{\text{s}}}\, \partial_{\bar{x}} V_{\text{h;eff}}(\bar{x}) + 0 - g_{\text{s}} \left. \partial_x \CA_1 (x,\bar{x}) \right|_{x=\bar{x}} + \cdots \right) = \nonumber \\
&=
\frac{1}{g_{\text{s}}}\, \frac{\rmi}{2\pi}\, \Big(V_{\text{h;eff}}(x^{\star}) - V_{\text{h;eff}}(b) \Big) - g_{\text{s}}\, \frac{\rmi}{2\pi} \int_{b}^{x^{\star}} \text{d}\bar{x} \left. \partial_x \CA_1 (x,\bar{x}) \right|_{x=\bar{x}} + \cdots.
\end{align}
\noindent
Having established the residue contribution, let us move to the saddle-point contour. In order to use the derivative-method outlined above it is useful to consider the integral
\begin{equation}
\label{eq:parameterintegraloneonesector}
\int_{\CC^{\star}} \frac{\text{d}x}{2\pi} \int_{\bar{\CC}^{\star}} \frac{\text{d}\bar{x}}{2\pi}\, \frac{1}{\left(x-\bar{x}\right)^2}\, \rme^{-\frac{v}{g_{\text{s}}} \left( V_{\text{h;eff}}(x) - V_{\text{h;eff}}(\bar{x}) \right)}\, \rme^{\CA_0(x,\bar{x})}\, \Big( 1 + g_{\text{s}}\, \CA_1 (x,\bar{x}) + \cdots \Big)
\end{equation}
\noindent
where we have introduced the additional parameter $v$ which places us in the earlier framework of our toy example. When $v=1$ we recover our saddle-point integral. Now, applying the derivative-method in $v$ is straightforward, essentially following the steps in the toy-example, and yields to leading order in $g_{\text{s}}$
\begin{equation}
-g_{\text{s}}\, \frac{\rmi}{48\pi}\, \frac{1}{V_{\text{h;eff}}^{\prime\prime} (x^{\star})^3} \left( \left( V_{\text{h;eff}}^{(3)} (x^{\star}) \right)^2 - V_{\text{h;eff}}^{\prime\prime} (x^{\star})\, V_{\text{h;eff}}^{(4)} (x^{\star}) + 3 V_{\text{h;eff}}^{\prime\prime} (x^{\star})^2\, \frac{\left(b-a\right)^2}{\left(a-x^{\star}\right)^2 \left(b-x^{\star}\right)^2} \right),
\end{equation}
\noindent
where we have already set $v=1$. Assembling the above residue and saddle-point contributions ensemble yields the general formula for the lowest three $g_{\text{s}}$-orders of the $(1|1)$ (anti) eigenvalue configuration at the $x^{\star}$ nonperturbative saddle-point. This is purely written in terms of spectral-curve data as
\begin{align}
\label{eq:generalfinalresultoneonesector}
\frac{\mathcal{Z}^{(1|1)} (t, g_{\text{s}})}{\mathcal{Z}^{(0|0)} (t, g_{\text{s}})} &\simeq \frac{1}{g_{\text{s}}}\, \frac{\rmi}{2\pi}\, \Big(V_{\text{h;eff}}(x^{\star}) - V_{\text{h;eff}}(b) \Big) - g_{\text{s}} \left\{ \frac{\rmi}{2\pi} \int_{b}^{x^{\star}} \text{d}\bar{x} \left. \partial_x \CA_1 (x,\bar{x}) \right|_{x=\bar{x}} + \frac{\rmi}{48\pi}\, \frac{1}{V_{\text{h;eff}}^{\prime\prime} (x^{\star})^3} \times \right. \nonumber \\
&
\left. \times \left( \left( V_{\text{h;eff}}^{(3)} (x^{\star}) \right)^2 - V_{\text{h;eff}}^{\prime\prime} (x^{\star})\, V_{\text{h;eff}}^{(4)} (x^{\star}) + 3 V_{\text{h;eff}}^{\prime\prime} (x^{\star})^2\, \frac{\left(b-a\right)^2}{\left(a-x^{\star}\right)^2 \left(b-x^{\star}\right)^2} \right)
\right\} + O (g_{\text{s}}^2).
\end{align} 

This result computes the first few $g_{\text{s}}$-orders of the free-energy resurgent-transseries, according to \eqref{eq:oneonesectortransseries}, for arbitrary one-cut matrix models. It predicts a negative starting-power in $g_{\text{s}}$, which is a known feature in the $(1,1)$ sector of matrix-model transseries \cite{asv11}. It also makes very explicit how it is possible to calculate nonperturbative contributions in the ``bulk'' of the resurgence lattice using our methods. Herein we focused on the first three terms in the $g_{\text{s}}$-expansion. Of course our methods work to arbitrary order, albeit computations quickly become very intricate.

\subsection{The Interplay Between Resonance and Logarithms}\label{subsec:resonanceandlogs}

Having established an accessible ``bulk'' contribution, it is now time to move to a more complicated one. One of the main features in resonant transseries for matrix models and their double-scaling limits is that they contain logarithms (recall \eqref{eq:transseriesexpansionGeneric} and \eqref{eq:transseriesexpansionGeneric-sectors}); see \cite{gikm10, asv11, sv13, as13, bssv22}. Let us now see how to compute such logarithmic contributions with our methods. The first such contribution will appear with the $(2|1)$ (anti) eigenvalue configuration. Motivated by our ``bridge'' \eqref{eq:generalexpressionforeigenvaluetunnelingtwoparameters-ell1}, this configuration is described by\footnote{There is actually a subtlety herein as we are not directly using \eqref{eq:generalexpressionforeigenvaluetunnelingtwoparameters-ell1}; rather just implementing the relation it conveys. There are ``three-steps'' resurgence motions on the left-hand side and a factor of $2$ on the right-hand side. This is explained in appendix~\ref{app:HigherOrderStokes}, in particular formula \eqref{eq:resurgenteigenvaluetunnleing21onestepscopy}, and they arise from the different possible paths on the resurgence lattice and the symmetry factors associated to (anti) eigenvalue tunneling.}
\begin{equation}
\label{eq:resurgenceeigenvaluetunneling21onesteps}
\rme^{-\frac{A}{g_{\text{s}}}}\, \mathsf{S}_{(0,0)\rightarrow (0,1)}\, \mathsf{S}_{(0,1)\rightarrow (1,1)} \left( \mathsf{S}_{(1,1)\to(2,1)}\, \frac{Z^{(2,1)}(t, g_{\text{s}})}{Z^{(0,0)}(t, g_{\text{s}})} + \mathsf{S}_{(1,1)\to(1,0)}\, \frac{Z^{(1,0)}(t, g_{\text{s}})}{Z^{(0,0)}(t, g_{\text{s}})} \right) \simeq 2\, \frac{\mathcal{Z}^{(2|1)}(t, g_{\text{s}})}{\mathcal{Z}^{(0|0)}(t, g_{\text{s}})},
\end{equation}
\noindent
where we have considered the case where we walked in single-steps on the resurgence lattice of transseries sectors (see appendix~\ref{app:HigherOrderStokes}). On top of the resonant sum, another novelty appears in this formula: this is the Borel residue $\mathsf{S}_{(1,1)\to(1,0)}$ which is associated to a rather non-trivial Stokes coefficient \cite{bssv22}. Note how it comes attached to the $(1,0)$ transseries sector, not $(2,1)$ as one could have expected---it sits \textit{inside} the resonant sum. This resonant contribution is in fact the one of greatest\footnote{The non-resonant $(2,1)[0]$ contribution (the reader may want to recall \eqref{eq:transseriesexpansionGeneric-sectors}), up to $\sqrt{g_{\text{s}}}$ order, turns out to be just the product of the previously calculated $(1,0)$ and $(1,1)$ contributions. In this sense, whilst it is a check on the validity of the transseries structure, it contains no real new information.} interest: not only it comes with this non-trivial Borel residue but it further fixes the logarithmic contributions in the transseries as in \eqref{eq:transseriesexpansionGeneric-sectors}. Let us try to understand this better.

First, we need to extract the contribution of the $(2,1)[0]$ sector from \eqref{eq:resurgenceeigenvaluetunneling21onesteps}, in order to isolate the resonant contribution. Following \eqref{eq:transseriesexpansionGeneric-sectors}, the full $(2,1)$ sector is given by 
\begin{align}
\label{eq:formof21sector-1}
Z^{(2,1)} (t, g_{\text{s}}) &= Z^{(2,1)[0]} (t, g_{\text{s}}) + Z^{(2,1)[1]} (t, g_{\text{s}})\, \log \frac{f(t)}{g_{\text{s}}^2} = \\
\label{eq:formof21sector-2}
&= Z^{(2,1)[0]} (t, g_{\text{s}}) + \alpha\, Z^{(1,0)} (t, g_{\text{s}})\, \log \frac{f(t)}{g_{\text{s}}^2},
\end{align}
\noindent
where from the first to the second line we used \eqref{eq:Phinmk=alphaPhinm0}. Furthermore up to order $\sqrt{g_{\text{s}}}$, the $(2,1)[0]$-sector in the partition function factorizes into $(1,1)$ and $(1,0)$ sectors as
\begin{equation}
\label{eq:factorization21sector}
Z^{(2,1)[0]} (t, g_{\text{s}}) = Z^{(1,0)} (t, g_{\text{s}})\, Z^{(1,1)} (t, g_{\text{s}}) + O (g_{\text{s}}^{3/2}).
\end{equation}
\noindent
Then, using this decomposition in \eqref{eq:formof21sector-2}, and further using \eqref{eq:forwardresurgentEVtunneling} and \eqref{eq:generalexpressionforeigenvaluetunnelingtwoparameters}, it is possible to finally isolate the resonant contribution of \eqref{eq:resurgenceeigenvaluetunneling21onesteps} in the new ``bridge'' relation
\begin{align}
\label{eq:resurgenceeigenvaluetunneling21simplified}
\rme^{-\frac{A}{g_{\text{s}}}}\, \mathsf{S}_{(0,0)\rightarrow (0,1)}\, \mathsf{S}_{(0,1)\rightarrow (1,1)} \left( \mathsf{S}_{(1,1)\to(1,0)} + \mathsf{S}_{(1,1)\to(2,1)}\, \alpha\, \log \frac{f(t)}{g_{\text{s}}^2} \right) \frac{Z^{(1,0)} (t, g_{\text{s}})}{Z^{(0,0)} (t, g_{\text{s}})} + O (g_{\text{s}}^{3/2}) &\simeq \nonumber \\
&
\hspace{-300pt}
\simeq 2\, \frac{\mathcal{Z}^{(2|1)} (t, g_{\text{s}})}{\mathcal{Z}^{(0|0)} (t, g_{\text{s}})} - 2\, \frac{\mathcal{Z}^{(1|0)} (t, g_{\text{s}})}{\mathcal{Z}^{(0|0)} (t, g_{\text{s}})}\, \frac{\mathcal{Z}^{(1|1)} (t, g_{\text{s}})}{\mathcal{Z}^{(0|0)} (t, g_{\text{s}})}.
\end{align}
\noindent
Notice how $\mathcal{Z}^{(1|0)} (t, g_{\text{s}})$ and $\mathcal{Z}^{(1|1)} (t, g_{\text{s}})$ appear in the right-hand side canceling the not-so-interesting contributions associated to $Z^{(2,1)[0]}(t, g_{\text{s}})$. These are also known expressions by now, as they were calculated in subsections~\ref{subsec:edgesofresurgentlattice} and~\ref{subsec:bulkoneonesector}. Let us tackle relation \eqref{eq:resurgenceeigenvaluetunneling21simplified}.

To obtain a direct matrix-integral computation of \eqref{eq:resurgenceeigenvaluetunneling21simplified} our first concern is with the evaluation of $\mathcal{Z}^{(2|1)} (t, g_{\text{s}})$. Using our general proposal \eqref{eq:generalsectorfromeigenvaluetunneling}-\eqref{eq:finalresultseparatedeigenvalues} with $(\ell|\bar{\ell})=(2|1)$ leads to
\begin{align}
\label{eq:integral21sectornontrivialstokes}
\mathcal{Z}^{(2|1)} (t, g_{\text{s}}) &= \frac{1}{2}\, \mathcal{Z}^{(0|0)} (t-g_{\text{s}}, g_{\text{s}})\, \pvint_{\CC_{\text{res}} + \CC^{\star}} \frac{\text{d}x_1}{2\pi}\, \pvint_{\CC_{\text{res}} + \CC^{\star}} \frac{\text{d}x_2}{2\pi}\, \pvint_{\bar{\CC}^{\star}} \frac{\text{d}\bar{x}}{2\pi}\, \times \\
&
\times\, \frac{\left(x_1-x_2\right)^2}{\left(x_1-\bar{x}\right)^2 \left(x_2-\bar{x}\right)^2}\, \rme^{- \frac{1}{g_{\text{s}}} \left( V(x_1) + V(x_2) - V(\bar{x}) \right)} \ev{\frac{\det \left(x_1-M\right)^{2} \det \left(x_2-M\right)^{2}}{\det \left(\bar{x}-M\right)^{2}}}_{N-1}. \nonumber
\end{align}
\noindent
The actual evaluation of this integral is quite lengthy but we shall nevertheless give a few details. Start by using partial-fraction decomposition to simplify the poles involved; via
\begin{equation}
\label{eq:partialfractiondecomposition21}
\frac{\left(x_1-x_2\right)^2}{\left(x_1-\bar{x}\right)^2 \left(x_2-\bar{x}\right)^2} = \frac{1}{\left(x_1-\bar{x}\right)^2} + \frac{1}{\left(x_2-\bar{x}\right)^2} - \frac{2}{\left(x_1-\bar{x}\right) \left(x_2-\bar{x}\right)}.
\end{equation}
\noindent
By linearity this splits \eqref{eq:integral21sectornontrivialstokes} into three parts. The first two are simply related by symmetry, and may moreover be directly evaluated using the precise same methods as those outlined in the previous subsection~\ref{subsec:bulkoneonesector}. We shall henceforth merely state the result (where it is convenient to also include the additional terms associated with $\mathcal{Z}^{(1|0)} (t, g_{\text{s}})$ and $\mathcal{Z}^{(1|1)} (t, g_{\text{s}})$). Up to and including $\sqrt{g_{\text{s}}}$-order we find
\begin{align}
\label{eq:integral21sectornontrivialstokes-new}
& -\frac{\mathcal{Z}^{(1|0)} (t, g_{\text{s}})}{\mathcal{Z}^{(0|0)} (t, g_{\text{s}})}\, \frac{\mathcal{Z}^{(1|1)} (t, g_{\text{s}})}{\mathcal{Z}^{(0|0)} (t, g_{\text{s}})} + \frac{1}{2}\, \frac{\mathcal{Z}^{(0|0)} (t-g_{\text{s}}, g_{\text{s}})}{\mathcal{Z}^{(0|0)} (t, g_{\text{s}})}\, \pvint_{\CC_{\text{res}} + \CC^{\star}} \frac{\text{d}x_1}{2\pi}\, \pvint_{\CC_{\text{res}} + \CC^{\star}} \frac{\text{d}x_2}{2\pi}\, \pvint_{\bar{\CC}^{\star}} \frac{\text{d}\bar{x}}{2\pi}\, \times \nonumber \\
&
\hspace{60pt}
\times \frac{2}{\left(x_1-\bar{x}\right)^2}\, \rme^{- \frac{1}{g_{\text{s}}} \left( V(x_1) + V(x_2) - V(\bar{x}) \right)} \ev{\frac{\det \left(x_1-M\right)^{2} \det \left(x_2-M\right)^{2}}{\det \left(\bar{x}-M\right)^{2}}}_{N-1} \simeq \nonumber \\
&
\hspace{60pt}
\simeq \frac{\rmi}{8\pi^2} \left(b-a\right) \sqrt{\frac{2\pi g_{\text{s}}}{V^{\prime\prime}_{\text{h;eff}}(x^{\star})}}\, \rme^{-\frac{1}{g_{\text{s}}} V_{\text{h;eff}}(x^{\star})}\, \frac{\log\left(\frac{4 \left(a-x^{\star}\right)}{a-b}\right)}{\left(x^{\star}-a\right) \left(x^{\star}-b\right)} + O (g_{\text{s}}^{3/2}).
\end{align}
\noindent
Having dealt with the first two terms let us now turn our attention to the last term of \eqref{eq:partialfractiondecomposition21}. Specifically, we want to compute \eqref{eq:integral21sectornontrivialstokes} with the following integrand replacement
\begin{equation}
\frac{\left(x_1-x_2\right)^2}{\left(x_1-\bar{x}\right)^2 \left(x_2-\bar{x}\right)^2} \quad \mapsto \quad - \frac{2}{\left(x_1-\bar{x}\right) \left(x_2-\bar{x}\right)}.
\end{equation}
\noindent
It makes sense to use linearity of the integral and separately consider the different combinations of integration contours. There are the following three distinct contributions:
\begin{itemize}
\item Start with the configuration $x_1 \in \CC_{\text{res}}$, $x_2 \in \CC^{\star}$, $\bar{x} \in \bar{\CC}^{\star}$ (which comes with an attached symmetry factor of $2$). The $x_1$ residue contribution is straightforward to evaluate. The ensuing $\bar{x}$ integral is then direct, and one arrives at
\begin{equation}
\frac{\rmi}{2\pi^2} \int_{\CC^{\star}} \text{d}x_2\, \frac{\log\left(\frac{x_2-x^{\star}}{x_2-b}\right)}{\left(x_2-a\right) \left(x_2-b\right)}\, \rme^{-\frac{1}{g_{\text{s}}} V_{\text{h;eff}}(x_2)}.
\end{equation}
\noindent
After an adequate change of variables, $x_2 = x^{\star} + \sqrt{\frac{2g_{\text{s}}}{V^{\prime\prime}_{\text{h;eff}}(x^{\star})}} \tau$, this becomes a simple saddle-point evaluation with a good $g_{\text{s}}$ expansion after integration. The final result reads
\begin{equation}
\frac{\rmi}{2\pi^2} \sqrt{\frac{2\pi g_{\text{s}}}{V^{\prime\prime}_{\text{h;eff}} (x^{\star})}}\, \frac{\rme^{-\frac{1}{g_{\text{s}}}V_{\text{h;eff}} (x^{\star})}}{\left(x^{\star}-a\right)\left(x^{\star}-b\right)} \left\{ \gamma_{\text{E}} + \log \left( - \frac{2 V^{\prime\prime}_{\text{h;eff}} (x^{\star}) \left(x^{\star}-b\right)^2}{g_{\text{s}}} \right) \right\} + O (g_{\text{s}}^{3/2}),
\end{equation}
\noindent
where $\gamma_{\text{E}}$ denotes the Euler--Mascheroni constant $\gamma_{\text{E}} \approx 0.5772156649...$ (which played a featuring role in the Stokes data of \cite{bssv22}).
\item Next, there is a pure saddle-point configuration $x_1 \in \CC^{\star}$, $x_2 \in \CC^{\star}$, $\bar{x} \in \bar{\CC}^{\star}$. This contribution is dealt with using the same derivative-method as in subsection~\ref{subsec:bulkoneonesector}, except for an additional subtlety we must tackle: instead of a single pole, we now have to deal with \textit{two} poles, located at $x_1=\bar{x}$ and $x_2=\bar{x}$. To understand how to deal with these, first recall that we introduced the derivative-method as the na\"\i ve expansion of a pole-factor (arising from an eigenvalue--anti-eigenvalue ``collision'') in the integrand was producing ever higher order poles and thus destroying the $g_{\text{s}}$-expansion; recall \eqref{eq:expansionofpolethatdestroysgsexpansion}. What we now observe is that a similar expansion in-between two eigenvalues or two anti-eigenvalues does not suffer from this problem (again, compare with \eqref{eq:expansionofpolethatdestroysgsexpansion}):
\begin{align}
\frac{1}{\left( x(\sqrt{g_{\text{s}}}\, s_1) - x(\sqrt{g_{\text{s}}}\, s_2) \right)^2} &= \frac{1}{g_{\text{s}}\, x^{\prime}(0)^2 \left(s_1-s_2\right)^2} - \frac{x^{\prime\prime}(0)\, s_1^2 + x^{\prime\prime}(0)\, s_2^2}{\sqrt{g_{\text{s}}}\, x^{\prime}(0)^3 \left(s_1-s_2\right)^2} + \nonumber \\
&+ \frac{(\cdots)}{x^{\prime}(0)^4 \left(s_1-s_2\right)^2} + \cdots.
\end{align}
\noindent
Indeed we are not finding ever higher order poles. In addition we observe the identity
\begin{equation}
\label{eq:reductionproblematicpoles}
\frac{1}{\left(x_1-\bar{x}\right) \left(x_2-\bar{x}\right)} = \frac{1}{x_1-x_2} \left( \frac{1}{x_2-\bar{x}} - \frac{1}{x_1-\bar{x}} \right),
\end{equation}
\noindent
which essentially reduces the number of would-be problematic eigenvalue--anti-eigenvalue poles of the form \eqref{eq:expansionofpolethatdestroysgsexpansion} from two to one. Using \eqref{eq:reductionproblematicpoles} and introducing a $v$ parameter as in subsection~\ref{subsec:bulkoneonesector} in order to perform the derivative-method which cancels the last problematic pole, we simply need to evaluate
\begin{align}
\pvint_{\CC^{\star}} \frac{\text{d}x_1}{2\pi}\, \pvint_{\CC^{\star}} \frac{\text{d}x_2}{2\pi}\, \pvint_{\bar{\CC}^{\star}} \frac{\text{d}\bar{x}}{2\pi}\, \frac{-4}{\left(x_1-x_2\right) \left(x_2-\bar{x}\right)} &\times \\
&
\hspace{-200pt}
\times \exp \Bigg\{- \frac{1}{g_{\text{s}}} \Big( V_{\text{h;eff}}(x_1) + v \big\{ V_{\text{h;eff}}(x_2) - V_{\text{h;eff}}(\bar{x}) \big\} \Big) +\nonumber\\
& 
\hspace{-200pt}
+ \CA_0 (x_1, x_2, \bar{x}) + \partial_t V_{\text{h;eff}}(x_1)  + \partial_t V_{\text{h;eff}}(x_2) - \partial_t V_{\text{h;eff}}(\bar{x}) + \cdots \Bigg\}. \nonumber
\end{align}
\noindent
Following the procedure in subsection~\ref{subsec:bulkoneonesector}, we arrive at the final result for this configuration
\begin{equation}
-\frac{\rmi}{2\pi^2} \sqrt{\frac{2\pi g_{\text{s}}}{V^{\prime\prime}_{\text{h;eff}} (x^{\star})}}\, \rme^{-\frac{1}{g_{\text{s}}} V_{\text{h;eff}} (x^{\star})}\, \frac{\log 2}{\left(x^{\star}-a\right) \left(x^{\star}-b\right)}.
\end{equation}
\item Finally, there is the ``residue dominated'' configuration $x_1 \in \CC_{\text{res}}$, $x_2 \in \CC_{\text{res}}$, $\bar{x} \in \bar{\CC}^{\star}$. The two residue calculations are straightforward, the remaining $\bar{x}$ integral as well, and we find
\begin{equation}
\frac{2}{(2\pi)^3} \left(2\pi\rmi\right)^2 \sqrt{\frac{2\pi g_{\text{s}}}{V^{\prime\prime}_{\text{h;eff}}(x^{\star})}}\, \rme^{-\frac{1}{g_{\text{s}}} V_{\text{h;eff}}(x^{\star})}\, \frac{1}{\left(x^{\star}-a\right)\left(x^{\star}-b\right)}.
\end{equation}
\noindent
Interestingly, this term just rotates logarithms in the terms acquired above.
\end{itemize}

Having computed all integrations, one still needs to take into account the contribution arising from the pre-factor $\mathcal{Z}^{(0,0)}\left(t-g_{\text{s}}, g_{\text{s}}\right)$ in \eqref{eq:integral21sectornontrivialstokes}, which further requires a straightforward $g_{\text{s}}$ power-series expansion as explained in \cite{msw07}. Once we gather every factor, the final result for the right-hand side of the ``bridge'' relation \eqref{eq:resurgenceeigenvaluetunneling21simplified} is adequately written as 
\begin{align}
\label{eq:generlaformula21}
\frac{\mathcal{Z}^{(2|1)} (t, g_{\text{s}})}{\mathcal{Z}^{(0|0)} (t, g_{\text{s}})} - \frac{\mathcal{Z}^{(1|0)} (t, g_{\text{s}})}{\mathcal{Z}^{(0|0)} (t, g_{\text{s}})}\, \frac{\mathcal{Z}^{(1|1)} (t, g_{\text{s}})}{\mathcal{Z}^{(0|0)} (t, g_{\text{s}})} &\simeq \\
&
\hspace{-80pt}
\simeq \frac{\rmi}{2 \left(2\pi\right)^2} \sqrt{\frac{2\pi g_{\text{s}}}{V^{\prime\prime}_{\text{h;eff}}(x^{\star})}}\, \frac{1}{\left(x^{\star}-a\right) \left(x^{\star}-b\right)}\, \frac{\left(b-a\right)}{4}\, \rme^{-\frac{1}{g_{\text{s}}} \left(V_{\text{h;eff}}(x^{\star}) - V_{\text{h;eff}}(b)\right)} \times \nonumber \\
&
\hspace{-80pt}
\times \left\{ 2 \gamma_{\text{E}} + \log \left( \frac{2^8}{g_{\text{s}}^2}\, \frac{V^{\prime\prime}_{\text{h;eff}}(x^{\star})^2 \left(x^{\star}-a\right)^4 \left(x^{\star}-b\right)^4}{\left(a-b\right)^4} \right) \right\} + O (g_{\text{s}}^{3/2}). \nonumber
\end{align}
\noindent
Plugging this result into \eqref{eq:resurgenceeigenvaluetunneling21simplified} we then find
\begin{align}
\label{eq:finalresult21}
&\rme^{-\frac{A}{g_{\text{s}}}}\, \mathsf{S}_{(0,0)\rightarrow (0,1)}\, \mathsf{S}_{(0,1)\rightarrow (1,1)} \left( \mathsf{S}_{(1,1)\to(1,0)} + \mathsf{S}_{(1,1)\to(2,1)}\, \alpha\, \log \frac{f(t)}{g_{\text{s}}^2} \right) \frac{Z^{(1,0)} (t, g_{\text{s}})}{Z^{(0,0)} (t, g_{\text{s}})} = \nonumber \\
&
\hspace{40pt}
= \rme^{-\frac{A}{g_{\text{s}}}}\, \mathsf{S}_{(0,0)\rightarrow (0,1)}\, \mathsf{S}_{(0,1)\rightarrow (1,1)} \left( \mathsf{S}_{(1,1)\to(1,0)} + \mathsf{S}_{(1,1)\to(2,1)}\, \alpha\, \log \frac{f(t)}{g_{\text{s}}^2} \right) F^{(1,0)} (t, g_{\text{s}}) \simeq \nonumber \\
&
\hspace{40pt}
\simeq \frac{\rmi}{\left(2\pi\right)^2} \sqrt{\frac{2\pi g_{\text{s}}}{V^{\prime\prime}_{\text{h;eff}}(x^{\star})}}\, \frac{1}{\left(x^{\star}-a\right) \left(x^{\star}-b\right)}\, \frac{\left(b-a\right)}{4}\, \rme^{-\frac{1}{g_{\text{s}}} \left(V_{\text{h;eff}}(x^{\star}) - V_{\text{h;eff}}(b)\right)} \times \nonumber \\
&
\hspace{40pt}
\times \left\{ 2 \gamma_{\text{E}} + \log \left( \frac{2^8}{g_{\text{s}}^2}\, \frac{V^{\prime\prime}_{\text{h;eff}}(x^{\star})^2 \left(x^{\star}-a\right)^4 \left(x^{\star}-b\right)^4}{\left(a-b\right)^4} \right) \right\} + O (g_{\text{s}}^{3/2}).
\end{align}

Our final result is again written purely in terms of spectral-curve data, hence applicable to any one-cut hermitian matrix model. As such, it shows some \textit{universal features}. One may always expect the appearance of the Euler--Mascheroni constant and logarithm contributions (which is certainly consistent with all resonant examples worked-out so far \cite{gikm10, asv11, sv13, as13, bssv22}). In particular, the Euler--Mascheroni constant seems to be a general feature of the $\mathsf{S}_{(1,1)\to(1,0)}$ Borel residue, implying some degree of universality in the transcendental number content of matrix model Stokes data. In this way, one may wonder if more of the ``Stokes numerology'' unveiled in \cite{bssv22} will be manifest across different one-cut (and double-scaled) examples.

\subsection{Resurgence, Transseries, and Their Stokes Data}\label{subsec:partitionfunction}

Let us recapitulate what have we achieved so far. Using both eigenvalues and anti-eigenvalues, tunneled via their appropriate contours, we established clear bridges between transseries sectors, their Stokes data, and matrix integrals (subsections~\ref{subsec:DeterminantCorrelators} and~\ref{subsec:eigenvalueTunnelingContours}); which we then carefully illustrated by evaluating the matrix integrals associated to the examples of the $(\ell|0)$, $(0,\bar{\ell})$, $(1|1)$ and $(2|1)$ (anti) eigenvalue configurations (subsections~\ref{subsec:edgesofresurgentlattice},~\ref{subsec:bulkoneonesector} and~\ref{subsec:resonanceandlogs}). How can we obtain generic $(n,m)$-sectors of a resonant resurgent transseries of the form \eqref{eq:transseriesexpansionGeneric} using this approach?

\definecolor{amethyst}{rgb}{0.6, 0.4, 0.8}
\definecolor{dartmouthgreen}{rgb}{0.05, 0.5, 0.06}
\definecolor{darkbrown}{rgb}{0.4, 0.26, 0.13}
\begin{figure}
\centering
\begin{tikzpicture}[
	blueframe/.style={
		rectangle,
		draw=blue,
		fill=blue!20,
		text width=4em,
		align=center,
		rounded corners,
		minimum height=2em
	},
	redframe/.style={
		rectangle,
		draw=orange,
		fill=orange!20,
		text width=4em,
		align=center,
		rounded corners,
		minimum height=2em
	},
	grayframe/.style={
		rectangle,
		draw=gray,
		text width=4em,
		align=center,
		rounded corners,
		minimum height=2em
	}, scale =0.6, line width=2
	]
	\foreach \n in {0,...,2}{
	  \foreach \m in {0,..., 2}{
	\node[grayframe] (\n_\m) at (4*\n-17,3*\m-3) {$Z^{(\n, \m)}(g_{\text{s}})$};
	}}
	\draw[gray, ->] (0, -4.5) -- (0, 4.5);
	\draw[gray, ->] (-6.5, 0) -- (6.5, 0);
\node[blueframe] (a1) at (4,0) {$\boldsymbol{S}_{(1, 0)}$};
\node[blueframe] (a2) at (4,-3) {$\boldsymbol{S}_{(1, -1)}$};
\node[blueframe] (a3) at (0,-3) {$\boldsymbol{S}_{(0, -1)}$};
\node[redframe] (a4) at (0,3) {$\boldsymbol{S}_{(0, 1)}$};
\node[redframe] (a5) at (-4,3) {$\boldsymbol{S}_{(-1, 1)}$};
\node[redframe] (a6) at (-4,0) {$\boldsymbol{S}_{(-1, 0)}$};
\draw[rounded corners, amethyst] (-18.8, -1.3) rectangle (-15.2, 4.3);
\draw[rounded corners, dartmouthgreen] (-14.8, -1.3) rectangle (-11.2, 1.3);
\draw[rounded corners, red] (-10.8, -1.3) rectangle (-7.2, 1.3);
\draw[rounded corners, darkbrown] (-14.8, -4.3) rectangle (-7.2, -1.7);
\draw[black, line width=1.5pt] (-6.8, -8) -- (-6.8,10);
\draw[black, line width=1.5pt] (-18, 9) -- (6,9);
\node[black] at (-13, 9.5) {Lattice of Transseries Sectors};
\node[black] at (0, 9.5) {Lattice of Stokes Vectors};
\draw[rounded corners, amethyst] (-1.8, 1.7) rectangle (1.8, 4.3);
\draw[rounded corners, red] (-1.8, -4.3) rectangle (5.8, -1.7);
\draw[rounded corners, darkbrown] (2.2, -1.3) rectangle (5.8, 1.3);
\node[amethyst] (a) at (-10, 8) {Purely backwards transseries sectors};
\draw[amethyst, line width=1.5, ->] (a) to[out=280, in=120] (-1, 4.8);
\draw[amethyst, line width=1.5, ->] (a) to[out=260, in=40] (-16, 4.8);
\node[dartmouthgreen] (b) at (0, 7) {The $(1|1)$ contribution};
\draw[dartmouthgreen, line width=1.5, ->] (b) to[out=270, in=70] (-12, 1.8);
\node[darkbrown] (c) at (-12, -7) {Purely forwards transseries sectors};
\draw[darkbrown, line width=1.5, ->] (c) to[out=120, in=260] (-12, -4.8);
\draw[darkbrown, line width=1.5, ->] (c) to[out=20, in=180] (2, -0.4);
\node[red] (d) at (0, -8) {The $(2|1)$ contribution};
\draw[red, line width=1.5, ->] (d) to[out=120, in=300] (-6.9, -1.6);
\draw[red, line width=1.5, ->] (d) to[out=80, in=270] (2, -4.7);
\end{tikzpicture}
\caption{Transseries contributions which we have calculated using formulae obtained throughout section~\ref{sec:ResurgenceInMatrixModels}. On the left-column we plot the lattice of transseries sectors (as in figure~\ref{fig:walkingonesteppathsintro}) and in the right-column we plot the lattice of Stokes vectors (as in figure~\ref{fig:stokeslattice}). The contributions we have explicitly computed are highlighted (albeit for Stokes data we highlight sectors in which they \textit{first} appear, \textit{i.e.}, the $(1|1)$ contribution contains the already-known ``trivial'' Stokes but we have not specifically highlighted this).}
\label{fig:CalculatedContributions}
\end{figure}
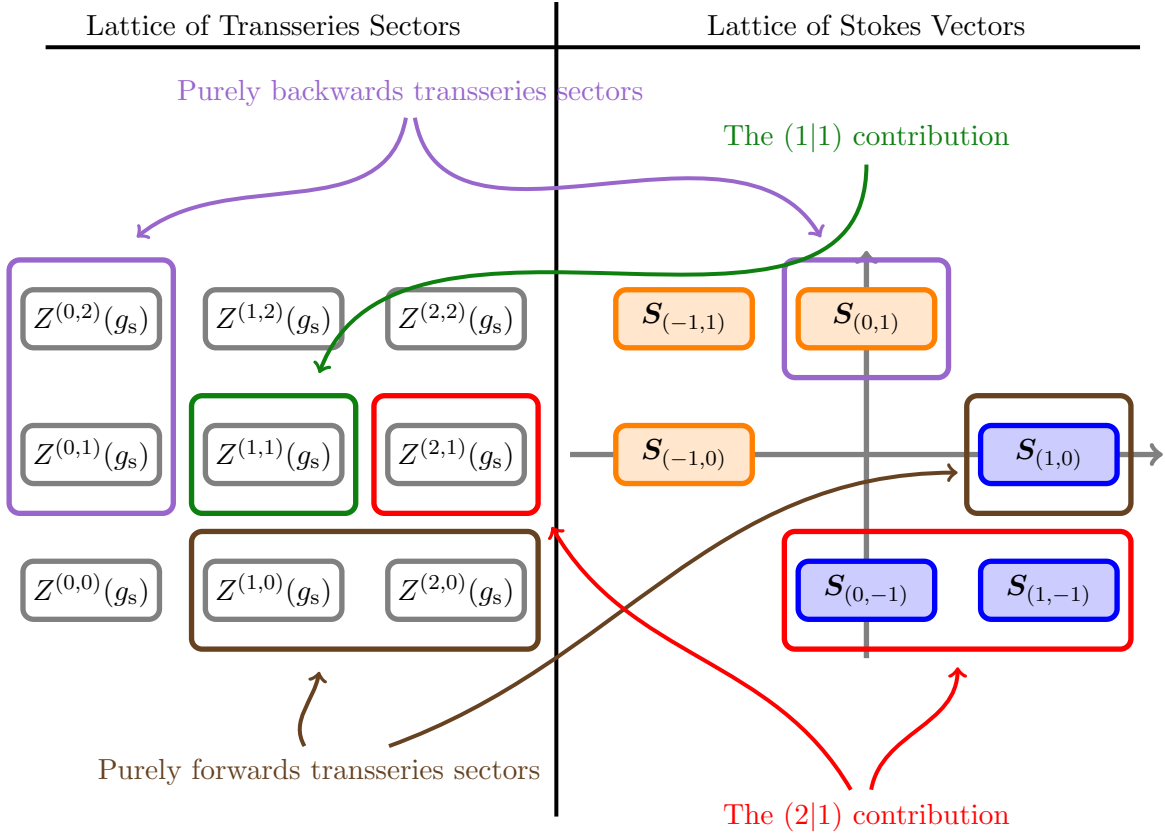

Let us list our explicit results (summarized in figure~\ref{fig:CalculatedContributions}). They read: 
\begin{itemize}
\item We repeated the calculation in \cite{msw07, msw08, emms22b} of the leading-$g_{\text{s}}$ term in all purely forward contributions $(\ell|0)$, resulting in formula \eqref{eq:resultedgeforward}. This calculation further confirms the already analytically-known Stokes vector $\boldsymbol{S}_{(1,0)}$.
\item We have extended this result to all (leading-$g_{\text{s}}$) purely backward contributions $(0|\bar{\ell})$ in equation \eqref{eq:resultedgebackward}, which yields the (also already analytically-known) Stokes vector $\boldsymbol{S}_{(0,1)}$.
\item Moving into the ``bulk'' of the resurgence lattice, we have computed the first $3$ terms in the $g_{\text{s}}$-expansion of the $(1|1)$ configuration in formula \eqref{eq:generalfinalresultoneonesector}. This both illustrates how we are not restricted to purely eigenvalue or anti-eigenvalue configurations, and how our methods are not restricted to leading $g_{\text{s}}$ order.
\item We calculated the $(2|1)$ configuration up to $\sqrt{g_{\text{s}}}$ order in formula \eqref{eq:finalresult21}. From this result we are able to extract the Stokes vector $\boldsymbol{S}_{(0,-1)}$, with non-trivial number content. In addition, this result reveals the nature of the logarithmic structures in the underlying transseries.
\item We can also compute a few other Stokes vectors, purely from consistency of eigenvalue tunneling with resurgence relations. This is illustrated in formula \eqref{eq:higherorderStokes}.
\end{itemize}
\noindent
Note that computing higher $g_{\text{s}}$ orders quickly becomes quite cumbersome, and so does disentangling the free-energy transseries out from these higher $g_{\text{s}}$ orders in the partition-function transseries. The contributions we listed above were however both feasible and at the same time containing a lot of non-trivial information.

Nonetheless, carrying through our calculations for any (anti) eigenvalue configuration is algorithmical, no matter how convoluted, and there are no new ingredients on top of what we have already used. Let us quickly lay-down the general strategy to compute integrals of the form \eqref{eq:generalsectorfromeigenvaluetunneling}-\eqref{eq:finalresultseparatedeigenvalues}. The procedure is as follows.
\begin{itemize}
\item First evaluate residue integrals. When there are combinations of saddle-point and residue contours, first evaluate the residue integrals and then do the saddle-point calculations.
\item Saddle-point calculations require that one first deals with the multitude of poles arising from the ratios of Vandermonde determinants in \eqref{eq:finalresultseparatedeigenvalues}. One iteratively uses fractional factorizations as \eqref{eq:reductionproblematicpoles} until the problem has been reduced to only one ``problematic'' pole.
\item The final step requires handling the one last ``problematic'' pole. This can be done by making use of the derivative-method introduced in subsection~\ref{subsec:bulkoneonesector}, and then just performing the saddle-point evaluation.
\end{itemize}
\noindent
Following this procedure, it should be possible---albeit increasingly cumbersome---to access all resonant nonperturbative contributions of our matrix-model resurgent transseries.

\section{Examples and Checks on the Proposal}\label{sec:ExamplesAndChecks}

Let us finally address a few very explicit examples of resurgent transseries, in the well-known cases of the cubic \cite{msw08, kmr10} and quartic \cite{m08, asv11, csv15} matrix models, as well as in the case of the Painlev\'e~I equation \cite{msw07, gikm10, kmr10, asv11, bssv22}. These examples will also serve as checks on our proposal, in particular by both matching against their resurgent-transseries data discussed in appendix~\ref{app:TransseriesData} and against the non-trivial Stokes data recently obtained in \cite{bssv22}.

\subsection{The Cubic Matrix Model}\label{subsec:CubicMatrixModelChecks}

Let us begin with the cubic hermitian matrix model, which is defined by \eqref{eq:partitionfunctionhermitianmatrix} with potential
\be
\label{eq:cubicPotential}
V_{\text{cubic}}(x) = \frac{1}{2} x^2 - \frac{\lambda}{6}x^3.
\ee
\noindent
This is the sole example in the present subsection, in which case we will drop ``cubic'' subscripts henceforth. The orthogonal-polynomial recursion-coefficients $\left\{ r_{n} \right\}$ for the cubic potential \eqref{eq:cubicPotential} satisfy the string equation \cite{biz80, m04}
\begin{align}
\label{eq:stringequationcubicmatrixmodel}
\frac{1}{2} r_{n} \left( \sqrt{1 - \lambda^2\, r_{n} - \lambda^2\, r_{n-1}} + \sqrt{1 - \lambda^2\, r_{n} - \lambda^2\, r_{n+1}} \right) = n g_{\text{s}}.
\end{align}
\noindent
In the large $N$ 't~Hooft limit, the ``continuous'' recursion coefficients \eqref{eq:rn-to-Rtgs} $r_n \mapsto R(t)$ satisfy the corresponding finite-difference equation\footnote{Here we bypass a slight subtlety. In addition to $t$ one usually introduces $x=n g_{\text{s}} \in [0, t]$, and only then takes the large $N$ limit \cite{m08, asv11}. Here we immediately jump to $t$ as our variable, not to clutter notation too much.}
\begin{align}
\label{eq:cubicstringequationthooftlimit}
\frac{1}{2} R (t) \left(\sqrt{1 - \lambda^2\, R (t) - \lambda^2\, R (t-g_{\text{s}})} + \sqrt{1 - \lambda^2\, R (t) - \lambda^2 R (t+g_{\text{s}})} \right) = t.
\end{align}
\noindent
This is a nonlinear equation which can be treated with resurgent asymptotic methods.

Both the solution to the above string-equation and its associated free energy \eqref{eq:freeenergyfromstringequation} result in two-parameter, resonant, resurgent transseries of the form \eqref{eq:transseriesexpansionGeneric}-\eqref{eq:transseriesexpansionGeneric-sectors}. They read:
\begin{align}
R (t, g_{\text{s}}; \sigma_1,\sigma_2) &= \sum\limits_{n=0}^{\infty} \sum\limits_{m=0}^{\infty} \sigma_1^n \sigma_2^m\, \text{e}^{-\left(n-m\right)\frac{A(t)}{g_{\text{s}}}}\, R^{(n,m)} (t, g_{\text{s}}), \\
F (t, g_{\text{s}}; \sigma_1,\sigma_2) &= \sum\limits_{n=0}^{\infty} \sum\limits_{m=0}^{\infty} \sigma_1^n \sigma_2^m\, \text{e}^{-\left(n-m\right)\frac{A(t)}{g_{\text{s}}}}\, F^{(n,m)} (t, g_{\text{s}}).
\label{eq:FTSforCUBIC}
\end{align}
\noindent
We have listed some explicit sectors in appendix~\ref{app:CubicMMData}. It is also customary to express generic transseries sectors as functions of the lowest-order sector \cite{biz80, iz92}
\begin{equation}
\label{eq:definitionrcubic}
2t = 2 R^{(0,0)}_0(t)\, \sqrt{1-2\lambda^2 R^{(0,0)}_0(t)},
\end{equation}
\noindent
which we shall henceforth denote as $r \equiv r(t) := R^{(0,0)}_0(t)$.

\begin{figure}
\centering
\includegraphics[scale=1.0]{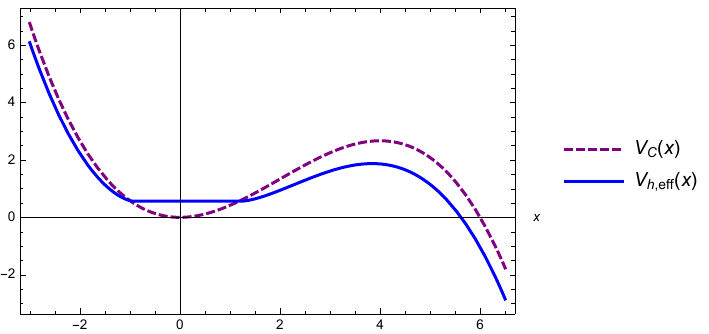}
\caption{Plot of the holomorphic effective potential ({\color{blue}blue}) for the cubic matrix model, on the real line with $r=0.2$ and $\lambda=\frac{1}{2}$. For comparison, we also plot the cubic potential $V_{\text{cubic}}(x)$.}
\label{fig:CeffectivePotential}
\end{figure}

Let us now see how to reproduce the transseries structure of this model together with its attached Borel residues using the matrix-integral formulae established in section~\ref{sec:ResurgenceInMatrixModels}. First recall some explicit spectral-curve quantities of the cubic model, in its one-cut phase \eqref{eq:hermitianspectralcurve}. The moment function was already given in \eqref{eq:MCubicExample}, and an illustration of the corresponding spectral curve in figure~\ref{fig:cubicspectralcurvejustphysicalaction} (or, more schematically, in figure~\ref{fig:settingcutandsaddle}). The endpoints of the cut may be written as
\begin{equation}
a = \frac{1}{\lambda} \left( 1 - 2 \lambda \sqrt{r} - \sqrt{1 - 2 \lambda^2 r} \right), \qquad b = \frac{1}{\lambda} \left( 1 + 2 \lambda \sqrt{r} - \sqrt{1 - 2 \lambda^2 r} \right),
\end{equation}
\noindent
with the (single) nonperturbative saddle located at  $x^{\star} = \frac{2}{\lambda} - \frac{1}{2} \left(a+b\right)$. The general setting of eigenvalue integration-contours is the one of figures~\ref{fig:settingcutandsaddleStandardtunneling} and~\ref{fig:settingcutandsaddleGeneraltunneling}, which was further specified for the cubic model in figure~\ref{fig:cubicexampletransition}. The holomorphic effective potential reads
\begin{align}
V_{\text{h;eff}} (t) &= \frac{1}{24} \sqrt{\left(t-a\right) \left(t-b\right)}\, \Big\{ \lambda \left( 4 t^2 + 2 t \left( a + b \right) - 3 a^2 - 2 a b - 3 b^2 \right) + 6 \left( a + b - 2 t \right) \bigg\} + \nonumber\\
&+ \frac{1}{8} \left(b-a\right)^2 \Big\{ \lambda \left(a+b\right) - 2 \Big\}\, \log  \left( \sqrt{a-t}+\sqrt{b-t} \right),
\end{align}
\noindent
is illustrated in figure~\ref{fig:CeffectivePotential}, and immediately leads to the cubic instanton action $A(t)$ via \eqref{eq:InstantonActionmsw07}. This sums up the required input for our matrix integrals, which we may next evaluate and compare against direct transseries calculations. As usual, we start at the boundary of the transseries and then move inwards.

\paragraph{Boundary of Resonant Transseries:}

Let us first check both the forward ``bridge'' relation \eqref{eq:forwardresurgentEVtunneling} with $\ell=1$, as well as its backward sibling \eqref{eq:backwardresurgentEVtunneling}. Using our matrix-integral formulae \eqref{eq:resultedgeforward} and \eqref{eq:resultedgebackward} we find to first order in $g_{\text{s}}$
\begin{equation}
\rme^{\frac{A(t)}{g_{\text{s}}}}\, \frac{\mathcal{Z}^{(1|0)} (t, g_{\text{s}})}{\mathcal{Z}^{(0|0)} (t, g_{\text{s}})} = \rmi \cdot \frac{\mathcal{Z}^{(0|1)} (t, g_{\text{s}})}{\mathcal{Z}^{(0|0)} (t, g_{\text{s}})}\, \rme^{-\frac{A(t)}{g_{\text{s}}}} = - \sqrt{g_{\text{s}}}\, \frac{\rmi \lambda^2 \sqrt{r}}{4\sqrt{2\pi} \left( 1-3\lambda^2  r \right)^{5/4}}.
\end{equation}
\noindent
On the other hand, on the transseries side we have
\begin{align}
\text{e}^{-\frac{A(t)}{g_{\text{s}}}}\, \mathsf{S}_{(0,0)\rightarrow (1,0)}\, \frac{Z^{(1,0)}(t, g_{\text{s}})}{Z^{(0,0)}(t, g_{\text{s}})} &= \mathsf{S}_{(0,0)\rightarrow (1,0)}\, \rme^{-\frac{A(t)}{g_{\text{s}}}}\, F^{(1,0)}(t, g_{\text{s}}), \\
\text{e}^{\frac{A(t)}{g_{\text{s}}}}\, \mathsf{S}_{(0,0)\rightarrow (0,1)}\, \frac{Z^{(0,1)}(t, g_{\text{s}})}{Z^{(0,0)}(t, g_{\text{s}})} &= \mathsf{S}_{(0,0)\rightarrow (0,1)}\, \rme^{\frac{A(t)}{g_{\text{s}}}}\, F^{(0,1)}(t, g_{\text{s}}).
\end{align}
\noindent
Comparing against cubic matrix-model transseries data in appendix~\ref{app:CubicMMData}, we find \textit{exact} matches at the functional level. On top of that we can also match numerical pre-factors,
\begin{equation}
\label{eq:resulttrivialStokesCubic}
\mathsf{S}_{(0,0)\rightarrow (1,0)} = \rmi \cdot \mathsf{S}_{(0,0)\rightarrow (0,1)} = - \frac{\rmi}{\sqrt{2\pi}}.
\end{equation}
\noindent
These are indeed Stokes data of the cubic model; see appendix~\ref{app:CubicMMData} for another \textit{precise} match.

\paragraph{Bulk of Resonant Transseries:}

Our first incursion into the ``bulk'' of the transseries lattice is, as expected, with the $(1|1)$ sector---as described by our matrix-integral formula \eqref{eq:generalfinalresultoneonesector}. Plugging-in spectral data for the cubic model, we arrive at
\begin{align}
\frac{\mathcal{Z}^{(1|1)} (t, g_{\text{s}})}{\mathcal{Z}^{(0|0)} (t, g_{\text{s}})} &\simeq \frac{\rmi}{2\pi g_{\text{s}}} \left(r \sqrt{1-2\lambda^2 r}\, \log \left( \frac{2 \sqrt{6  \lambda^4 r^2 - 5 \lambda^2 r + 1} - 5 \lambda^2 r + 2}{\lambda^2 r} \right) + \frac{2 \sqrt{1 - 3 \lambda^2 r}}{3 \lambda^2} \right) - \nonumber \\
&- g_{\text{s}}\, \frac{\rmi \left(57\lambda^4 r^2 - 60\lambda^2 r + 8\right)}{192\pi r \left(1-3\lambda^2 r\right)^{5/2}} + \cdots.
\end{align}
\noindent
Comparison of matrix-integral results with transseries sectors now occurs via our ``bridge'' relation \eqref{eq:generalexpressionforeigenvaluetunnelingtwoparameters}; which for this sector effectively encodes the free-energy transseries contributions in \eqref{eq:oneonesectortransseries}. Using results for the free-energy transseries sector $F^{(1,1)} (t, g_{\text{s}})$ in appendix~\ref{app:CubicMMData} we find \textit{exact} agreement at functional level. On top of that, the prediction for the Borel-residue pre-factors also \textit{precisely} matches against cubic transseries data.

\paragraph{Resonance and Logarithms:}

We are left with the $(2|1)$ configuration. As explained in subsection~\ref{subsec:resonanceandlogs}, the ``bridge'' relation we now want to establish is \eqref{eq:resurgenceeigenvaluetunneling21simplified} (where, for the cubic matrix model, $\alpha = \frac{1}{2}$). Using our matrix-integral formula \eqref{eq:generlaformula21} for the cubic model it follows
\begin{align}
2\, \frac{\mathcal{Z}^{(2|1)} (t, g_{\text{s}})}{\mathcal{Z}^{(0|0)} (t, g_{\text{s}})} - 2\, \frac{\mathcal{Z}^{(1|0)} (t, g_{\text{s}})}{\mathcal{Z}^{(0|0)} (t, g_{\text{s}})}\, \frac{\mathcal{Z}^{(1|1)} (t, g_{\text{s}})}{\mathcal{Z}^{(0|0)} (t, g_{\text{s}})} &\simeq \\
&
\hspace{-150pt}
\simeq \sqrt{g_{\text{s}}}\, \frac{1}{(2\pi)^{3/2}}\, \frac{\lambda^2 \sqrt{r}}{4 \left(1 - 3 \lambda^2 r\right)^{5/4}} \left( 2 \gamma_{\text{E}} + \log \left( \frac{256}{g_{\text{s}}^2}\, \frac{\left(1 - 3 \lambda^2 r\right)^5}{\lambda^8 r^2} \right) \right) \rme^{-\frac{A(t)}{g_{\text{s}}}} + O (g_{\text{s}}^{3/2}). \nonumber
\end{align}
\noindent
Matching against transseries data in appendix~\ref{app:CubicMMData} via \eqref{eq:resurgenceeigenvaluetunneling21simplified} we once again find an \textit{exact} functional match. On top of this, we analytically predict the non-trivial Stokes vector
\begin{equation}
\boldsymbol{S}_{(0,-1)} = \frac{\rmi}{\sqrt{2\pi}} \left( \gamma_{\text{E}} + \log 96\sqrt{3} \right) \left[\begin{array}{c}
2\\
1
\end{array}\right],
\end{equation}
\noindent
associated to the Borel residue $\mathsf{S}_{(1,1)\to(1,0)}$. This may be double-scaled to Painlev\'e~I Stokes data (see the upcoming subsection~\ref{subsec:DSLexample}) where it then \textit{precisely} matches against the result in \cite{bssv22}.

\paragraph{Higher-Step Stokes Data:}

Finally, we use the Stokes-data consistency-conditions discussed in appendix~\ref{app:HigherOrderStokes}, in particular formula \eqref{eq:higherorderStokes}, in order to calculate the two-step Stokes vector $\boldsymbol{S}_{(1,-1)}$. All required input is the ``canonical'' Stokes coefficient in \eqref{eq:resulttrivialStokesCubic}, from where one then obtains
\begin{equation}
\boldsymbol{S}_{(1,-1)} = \frac{\rmi}{4} \left[\begin{array}{c}
2\\
0
\end{array}\right].
\end{equation}
\noindent
This \textit{precisely} coincides with the result obtained from comparison against the Painlev\'e~I double-scaling limit in \eqref{eq:higherstepstokescubic}, further matching against \cite{bssv22}.

\subsection{The Quartic Matrix Model}\label{subsec:quarticexample}

Our next example is the quartic hermitian matrix model, which is defined by \eqref{eq:partitionfunctionhermitianmatrix} with potential
\be
\label{eq:quarticPotential}
V_{\text{quartic}}(x) = \frac{1}{2}x^2 - \frac{\lambda}{24}x^4
\ee
\noindent
(where we will again henceforth drop ``quartic'' subscripts). The orthogonal-polynomial recursion-coefficients $\left\{ r_{n} \right\}$ for the quartic potential \eqref{eq:quarticPotential} satisfy the string equation \cite{biz80, m04}
\begin{align}
\label{eq:stringequationquarticmatrixmodel}
r_n \left( 1 - \frac{\lambda}{6}\, \big( r_{n-1} + r_n + r_{n+1} \big) \right) = n g_{\text{s}}.
\end{align}
\noindent
In the large $N$ 't~Hooft limit, the ``continuous'' recursion coefficients \eqref{eq:rn-to-Rtgs} $r_n \mapsto R(t)$ satisfy the corresponding finite-difference equation
\begin{align}
\label{eq:quarticstringequationthooftlimit}
R (t) \left( 1 - \frac{\lambda}{6}\, \Big( R (t-g_{\text{s}}) + R (t) + R (t+g_{\text{s}}) \Big) \right) = t.
\end{align}
\noindent
This is a nonlinear equation which can be treated with resurgent asymptotic methods.

Both the solution to the above string-equation and its associated free energy \eqref{eq:freeenergyfromstringequation} result in two-parameter, resonant, resurgent transseries of the form \eqref{eq:transseriesexpansionGeneric}-\eqref{eq:transseriesexpansionGeneric-sectors}. They read \cite{m08, asv11}
\begin{align}
R (t, g_{\text{s}}; \sigma_1,\sigma_2) &= \sum\limits_{n=0}^{\infty} \sum\limits_{m=0}^{\infty} \sigma_1^n \sigma_2^m\, \text{e}^{-\left(n-m\right)\frac{A(t)}{g_{\text{s}}}}\, R^{(n,m)} (t, g_{\text{s}}), \\
F (t, g_{\text{s}}; \sigma_1,\sigma_2) &= \sum\limits_{n=0}^{\infty} \sum\limits_{m=0}^{\infty} \sigma_1^n \sigma_2^m\, \text{e}^{-\left(n-m\right)\frac{A(t)}{g_{\text{s}}}}\, F^{(n,m)} (t, g_{\text{s}}).
\label{eq:FTSforQUARTIC}
\end{align}
\noindent
We have listed some explicit sectors in appendix~\ref{app:QuarticMMData}. Analogously to the cubic case, we express generic transseries sectors as functions of the lowest-order sector $R^{(0,0)}_0(t)$, which now is
\begin{equation}
\label{eq:definitionrquartic}
r = \frac{1}{\lambda} \left(1-\sqrt{1-2\lambda t}\right).
\end{equation}

\begin{figure}
\centering
\includegraphics[scale=1.0]{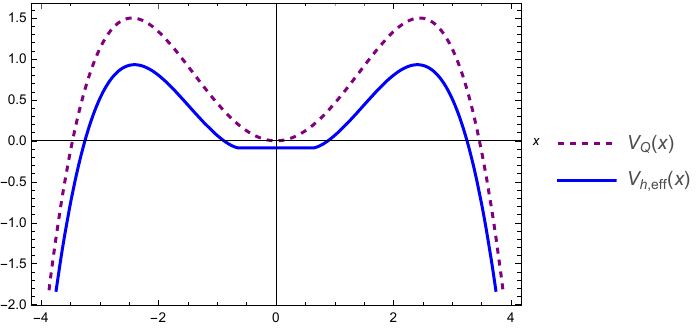}
\caption{The holomorphic effective potential ({\color{blue}blue}) for the quartic matrix model, upon the real line with $t=0.1$ and $\lambda=1$. For comparison we also plot the original potential $V_{\text{quartic}}(x)$.}
\label{fig:QeffectivePotential}
\end{figure}

Let us next reproduce the transseries structure of this model together with its attached Borel residues using the matrix-integral formulae established in section~\ref{sec:ResurgenceInMatrixModels}. First recall some explicit spectral-curve quantities of the quartic model in its one-cut phase \eqref{eq:hermitianspectralcurve}. The moment function is now
\begin{equation}
M(z) = 1 - \frac{\lambda}{3} \alpha^2 - \frac{\lambda}{6} z^2,
\end{equation}
\noindent
and the endpoints of the cut are symmetric, at $a=-2\alpha$ and $b=2\alpha$, with
\begin{equation}
\label{eq:definitionalpha}
\alpha^2 = \frac{1}{\lambda} \left(1-\sqrt{1-2\lambda t}\right).
\end{equation}
\noindent
Note that $r = \alpha^2$. The nonperturbative saddle-point structure is slightly more complicated than for the cubic case. There are now two symmetric nonperturbative saddles located at
\begin{equation}
x_1^{\star} = \sqrt{\frac{2}{\lambda} \left(3 - \lambda\alpha^2\right)}, \qquad x_2^{\star} = - \sqrt{\frac{2}{\lambda} \left(3 - \lambda\alpha^2\right)}.
\end{equation}
\noindent
The general setting of eigenvalue integration-contours is the one of figures~\ref{fig:settingcutandsaddleStandardtunneling} and~\ref{fig:settingcutandsaddleGeneraltunneling}, which was further specified for the quartic model in figure~\ref{fig:quarticexampletransition}. Note how one now has to deal with two, symmetric nonperturbative saddle points. This implies some of the formulae in section~\ref{sec:ResurgenceInMatrixModels} will require slight upgrades, which we explicitly discuss below. The holomorphic effective potential reads
\begin{align}
\label{eq:holomorphiceffectivequarticpotential}
V_{\text{h;eff}}(t) = \sqrt{t^2-4\alpha^2} \left\{ \frac{1}{2} \left( 1 - \frac{\lambda}{6} \alpha^2 \right) t - \frac{\lambda}{24} t^3 \right\} - 2 \alpha^2 \left\{ 1-\frac{1}{2} \lambda \alpha^2 \right\} \log \left( t + \sqrt{t^2-4\alpha^2} \right),
\end{align}
\noindent
is illustrated in figure~\ref{fig:QeffectivePotential}, and immediately leads to the quartic instanton action $A(t)$ via \eqref{eq:InstantonActionmsw07} (which is independent of the chosen $x^{\star}_{1,2}$). This sums up the required input for our matrix integrals, which we next evaluate and compare against direct transseries calculations. We start at the boundary of the transseries and then move inwards.

\paragraph{Boundary of Resonant Transseries:}

Checking forward and backward ``bridges'', \eqref{eq:forwardresurgentEVtunneling} and \eqref{eq:backwardresurgentEVtunneling}, first seems to require evaluation of the matrix-integrals in \eqref{eq:resultedgeforward} and \eqref{eq:resultedgebackward}. But for the quartic matrix model, and in contrast with our earlier cubic example, an additional subtlety appears. We now have two symmetric nonperturbative saddle points, $x_1^{\star} \equiv x^{\star}$ and $x_2^{\star} = - x^{\star}$, and we need to consider combinations of both. This will result in (small) changes to the actual matrix integrals one needs to evaluate within the aforementioned ``bridge'' relations.

Let us first address the forward contribution, \textit{i.e.}, let us look back at the $\ell$-th eigenvalue configuration in \eqref{eq:forwardcontributionstunneled}-\eqref{eq:integralkforwardcontributions}. It turns out that the leading nonperturbative contributions (\textit{i.e.}, those to lowest order in $g_{\text{s}}$) now arise from configurations where eigenvalues are most evenly distributed in-between the two saddles, $\pm x^{\star}$. To see this, one first realizes that there is a \textit{change} in eigenvalue combinatorics which depends on the parity of $\ell$. There are two possibilities:
\begin{itemize}
\item Consider even $\ell$. The split of these eigenvalues into the two saddles occurs as expected,
\begin{align}
\label{eq:forwardsaddleconfigurationeven}
\left( x_1, \ldots, x_\ell \right)\, \leftrightarrow\, \Big( \underbrace{x^{\star}, \ldots, x^{\star}}_{\ell/2},\, \underbrace{-x^{\star}, \ldots, -x^{\star}}_{\ell/2} \Big),
\end{align}
\noindent
over all its permutations. This configuration hence yields a symmetry factor of ${{\ell}\choose{\ell/2}}$.
\item Consider odd $\ell$. The split into the two saddles now occurs in two possible ways,
\begin{align}
\label{eq:oddkfirstconfiguration}
\left( x_1, \ldots, x_\ell \right)\, &\leftrightarrow\, \Big( \underbrace{x^{\star}, \ldots, x^{\star}}_{(\ell-1)/2},\, \underbrace{-x^{\star}, \ldots, -x^{\star}}_{(\ell+1)/2} \Big), \\
\label{eq:oddksecondconfiguration}
\left( x_1, \ldots, x_\ell \right)\, &\leftrightarrow\, \Big( \underbrace{x^{\star}, \ldots, x^{\star}}_{(\ell+1)/2},\, \underbrace{-x^{\star}, \ldots, -x^{\star}}_{(\ell-1)/2} \Big),
\end{align}
\noindent
and respective permutations. Both configurations have the same symmetry factor ${{\ell+1}\choose{(\ell+1)/2}}$.
\end{itemize} 
\noindent
Once the tunneled eigenvalues are distributed in these ways, the lowest $g_{\text{s}}$-order will become associated to the ``mixed saddles''. In fact, differences of the type $(x_i-x_j)$ in the matrix-integral Vandermonde-determinant acquire $\sqrt{g_{\text{s}}}$ pre-factors under saddle-point evaluation. When eigenvalues sit in the same nonperturbative saddle, these will result in powers of $g_{\text{s}}$. But if we can distribute them across distinct saddles, that will \textit{lower} this corresponding $g_{\text{s}}$ power. As such, what the above distributions imply is that \eqref{eq:resultedgeforward} is not the \textit{lowest} $g_{\text{s}}$-order contribution to the $\ell$-eigenvalue configuration; one must instead address the above ``mixed saddles'' configurations. The calculation is nevertheless straightforward as a saddle-point evaluation (see \cite{msw07} as well), and we find:
\begin{align}
\label{eq:resultallforwardsectors}
\frac{\mathcal{Z}^{(\ell|0)}(t, g_{\text{s}})}{\mathcal{Z}^{(0|0)}(t, g_{\text{s}})} \simeq g_{\text{s}}^{\frac{1}{2}\left\lfloor\frac{\ell^2+1}{2}\right\rfloor}\, \rme^{-\frac{\ell}{g_{\text{s}}} A(t)}\, \CN_{\ell}\, \CT_{\ell}^{\text{forw}} (\lambda, \alpha) + \cdots,
\end{align}
\noindent
where $\lfloor\bullet\rfloor$ denotes the integer part, where we have the numerical pre-factor
\begin{align}
\label{eq:forwardbackwardnumericalprefactor}
\CN_{\ell} = -\frac{1}{\ell!}\, \frac{1}{\pi^{\ell/2}} \times 
\begin{cases}
{{\ell}\choose{\ell/2}}\, G_2\left(\frac{\ell+4}{2}\right)^2 2^{-\frac{\ell(\ell+2)}{4}} & \ell \text{ even},\\
{{\ell+1}\choose{(\ell+1)/2}}\, G_2\left(\frac{\ell+3}{2}\right) G_2\left(\frac{\ell+5}{2}\right) 2^{-\frac{(\ell+1)^2+6}{4}} & \ell \text{ odd},
\end{cases}
\end{align}
\noindent
and where the forward transseries structure is
\begin{align}
\label{eq:forwardtransseriesstructure}
\CT_{\ell}^{\text{forw}} (\lambda, \alpha) = \alpha^{\ell^2} \times 
\begin{cases}
\left( -\frac{3\lambda^3}{2 \left( 3 - \lambda \alpha^2 \right)^{3/2} \left( 3 - 3 \lambda \alpha^2 \right)^{5/2}} \right)^{\left(\frac{\ell}{2}\right)^2} & \ell \text{ even},\\
\left( -\frac{24 \sqrt{3 - \lambda \alpha^2}}{\lambda \sqrt{3 - 3 \lambda \alpha^2}} \right)^{\frac{\ell^2+1}{4}} \left( \frac{\lambda}{2 \sqrt{3 - 3 \lambda \alpha^2}\, \sqrt{3 - \lambda \alpha^2}}\right)^{\ell^2} \frac{\sqrt{\lambda}}{\sqrt{3 - 3 \lambda \alpha^2}} & \ell \text{ odd}.
\end{cases}
\end{align}

It is now with the above results which we explicitly check the ``bridge'' between eigenvalue-tunneling and resurgent-transseries (just as in subsection~\ref{subsec:eigenvalueTunnelingContours}). Being the lowest $g_{\text{s}}$ contributions, the simplest first check is comparing starting genus $\beta_{nm}^{[k]}$, as in \eqref{eq:phinm-beta-starting-g}, between the above predictions and the transseries computations in appendix~\ref{app:QuarticMMData}. The predictions are immediate to read-off from \eqref{eq:resultallforwardsectors}
\begin{align}
\beta_{\ell,0}^{[0],\mathcal{Z}} = \frac{1}{2} \left\lfloor\frac{\ell^2+1}{2}\right\rfloor \qquad \Rightarrow \qquad
\begin{tabular}{c|ccccccc}
$\ell$ & $0$ & $1$ & $2$ & $3$ & $4$ & $5$ & $6$ \\ \hline
$\beta_{\ell,0}^{[0],\mathcal{Z}}$ & $0$ & $1/2$ & $1$ & $5/2$ & $4$ & $13/2$ & $9$ \\
\end{tabular}.
\end{align}
\noindent
Comparing against quartic matrix-model transseries data in appendix~\ref{app:QuarticMMData} we find an \textit{exact} match, as well as for the functional form of explicit nonperturbative coefficients (\textit{e.g.}, we have listed explicit such transseries coefficients for $\ell=1,\ldots,6$ in appendix~\ref{app:QuarticMMData}). Using the forward ``bridge''  \eqref{eq:forwardresurgentEVtunneling} we may further explicitly compare (or, to some extent, analytically compute) Borel residues. For instance, in the $\ell=3$ example the above formulae yield at leading order
\begin{align}
\rme^{-3\frac{A(t)}{g_{\text{s}}}}\, \mathsf{S}_{(0,0)\rightarrow (3,0)}\, \frac{Z^{(3,0)}(t, g_{\text{s}})}{Z^{(0,0)}(t, g_{\text{s}})} = -\rmi\, g_{\text{s}}^{5/2}\, \frac{3^{5/2}\,\lambda^{7}\, r^{9/2}}{64\, \pi^{3/2} \left( 3-\lambda r \right)^{13/4} \left( 3-3\lambda^3 \right)^{25/4}}\, \rme^{-3\frac{A(t)}{g_{\text{s}}}}.
\end{align}
\noindent
Proceeding analogously for the first $6$ Borel residues, one quickly finds
\begin{align}
\label{eq:forwardBorelResiduesquartic}
\mathsf{S}_{(0,0)\rightarrow (\ell,0)} = -\left(\rmi\,\sqrt{\frac{3}{\pi\lambda}}\right)^\ell, \qquad \ell=1,\ldots,6.
\end{align} 
\noindent
Via \eqref{eq:forwardbackwardBorelResidues} this immediately translates to the ``canonical'' Stokes vector extensively studied in the literature \cite{msw07}, which is the well-known result
\begin{equation}
\label{eq:forwardtrivialstokesquartic}
\boldsymbol{S}_{(1,0)} = \rmi\, \sqrt{\frac{3}{\pi\lambda}} \left[\begin{array}{c}
1\\
0
\end{array}\right].
\end{equation}

The purely anti-eigenvalue matrix-integral in the backward ``bridge'' \eqref{eq:backwardresurgentEVtunneling} is computed in complete analogy to what was described above for the forward direction. There is an extra minus sign in front of the holomorphic effective potential which eventually yields
\begin{align}
\frac{\mathcal{Z}^{(0|\bar{\ell})}(t, g_{\text{s}})}{\mathcal{Z}^{(0|0)}(t, g_{\text{s}})} \simeq g_{\text{s}}^{\frac{1}{2}\left\lfloor\frac{\bar{\ell}^2+1}{2}\right\rfloor}\, \rme^{\frac{\bar{\ell}}{g_{\text{s}}} A(t)}\, \CN_{\bar{\ell}}\, \CT_{\bar{\ell}}^{\text{back}} (\lambda, \alpha) + \cdots,
\end{align}
\noindent
where the numerical pre-factor $\CN_{\bar{\ell}}$ is the same as before, in \eqref{eq:forwardbackwardnumericalprefactor}, and where the backward transseries structure is given by
\begin{align}
\label{eq:backwardtransseriesstructure}
\CT_{\bar{\ell}}^{\text{back}} (\lambda, \alpha) = \alpha^{\bar{\ell}^2} \times 
\begin{cases}
\left( \frac{3\lambda^3}{2 \left( 3 - \lambda  \alpha^2 \right)^{3/2} \left( 3 - 3 \lambda \alpha^2 \right)^{5/2}} \right)^{\left(\frac{\bar{\ell}}{2}\right)^2} & \bar{\ell} \text{ even},\\
 \left( \frac{24 \sqrt{3 - \lambda \alpha^2}}{\lambda \sqrt{3 - 3 \lambda \alpha^2}} \right)^{\frac{\bar{\ell}^2+1}{4}} \left( \frac{\lambda}{2 \sqrt{3 - 3 \lambda \alpha^2}\, \sqrt{3 - \lambda \alpha^2}}\right)^{\bar{\ell}^2} \frac{\sqrt{\lambda}}{\sqrt{3 - 3 \lambda \alpha^2}} & \bar{\ell} \text{ odd}.
\end{cases}
\end{align}
\noindent
The backward ``bridge'' relation hence boils down to
\begin{equation}
\text{e}^{+\bar{\ell} \frac{A(t)}{g_{\text{s}}}}\, \mathsf{S}_{(0,0)\rightarrow (0,\bar{\ell})}\, \frac{Z^{(0,\bar{\ell})}(t, g_{\text{s}})}{Z^{(0,0)}(t, g_{\text{s}})} \simeq g_{\text{s}}^{\frac{1}{2}\left\lfloor\frac{\bar{\ell}^2+1}{2}\right\rfloor}\, \rme^{\frac{\bar{\ell}}{g_{\text{s}}} A(t)}\, \CN_{\bar{\ell}}\, \CT_{\bar{\ell}}^{\text{back}} (\lambda, \alpha).
\end{equation}
\noindent
Comparing against quartic matrix-model transseries data in appendix~\ref{app:QuarticMMData} we find an \textit{exact} match, where we now explicitly listed transseries coefficients for $\bar{\ell}=1,\ldots,4$. The corresponding Borel residues are found to be
\begin{align}
\label{eq:backwardBorelResiduesquartic}
\mathsf{S}_{(0,0)\rightarrow (0,\bar{\ell})} = -\left(\sqrt{\frac{3}{\pi\lambda}}\right)^{\bar{\ell}}, \qquad \bar{\ell}=1,\ldots,4.
\end{align}
\noindent
Via \eqref{eq:forwardbackwardBorelResidues} this further translates to the Stokes vector
\begin{equation}
\label{eq:backwardtrivialStokesquartic}
\boldsymbol{S}_{(0,1)} = \sqrt{\frac{3}{\pi\lambda}} \left[\begin{array}{c}
0\\
1
\end{array}\right].
\end{equation}
\noindent
This result \textit{precisely} matches the literature \cite{bssv22}.

\paragraph{Bulk of Resonant Transseries:}

Enter the ``bulk'' of the transseries lattice with the $(1|1)$ sector (at least up to our $\sim g_{\text{s}}$ order). The subtlety with symmetric nonperturbative saddles $x_2^{\star} = - x_1^{\star}$ keeps haunting us, as we can now distribute our one eigenvalue and one anti-eigenvalue over these two saddles. There are two possibilities:
\begin{itemize}
\item When the eigenvalue sits at $x_1^{\star}$ and the anti-eigenvalue at $x_2^{\star}$, evaluation of \eqref{eq:finalresultseparatedeigenvalues} with $(\ell|\bar{\ell})=(1|1)$ is straightforward. There is now no singularity in the integrand (as the integration contours do not intersect), in which case we just need to perform a standard saddle-point integration in the respective steepest-descent contours,
\begin{align}
\label{eq:higherorder11}
\int_{\CC_1^{\star}} \frac{\text{d}x}{2\pi} \int_{\bar{\CC}_2^{\star}} 
\frac{\text{d}\bar{x}}{2\pi}\, \frac{1}{(x-\bar{x})^2}\, \rme^{-\frac{1}{g_{\text{s}}}\left( V(x)-V(\bar{x}) \right)} \ev{\frac{\det \left(x-M\right)^2}{\det \left(\bar{x}-M\right)^2}}_{N}.
\end{align}
\noindent
There is an additional global factor of $2$ (as we could have initially swapped eigenvalue and anti-eigenvalue), in which case we find the contribution
\begin{equation}
g_{\text{s}}\, \frac{3\rmi\lambda\sqrt{3-\lambda r}}{16\pi \left(3-3\lambda r\right)^{5/2}}.
\end{equation}
\item When both eigenvalue and anti-eigenvalue sit on the same saddle, we just use our earlier result from formula \eqref{eq:generalfinalresultoneonesector}. There is an additional symmetry factor of $2$ (as there are two nonperturbative saddles), in which case we find
\begin{align}
&
\frac{1}{g_{\text{s}}}\, \frac{\rmi}{2\pi\lambda} \left(\sqrt{3-3\lambda r}\, \sqrt{3-\lambda r} - \lambda r \left(2-\lambda r\right) \log \left( \frac{3-2r\lambda + \sqrt{3-3\lambda r}\,  \sqrt{3-\lambda r}}{\lambda r} \right) \right) + \nonumber \\
&
+ g_{\text{s}}\, \frac{3\rmi \left(36-117\lambda r+54\lambda^2 r^2-7\lambda^3 r^3\right)}{16\pi r\left(3-3\lambda r\right)^{5/2}\left(3-\lambda r\right)^{3/2}}.
\end{align}
\end{itemize}
\noindent
Taking both above contributions together finally yields
\begin{align}
\frac{\mathcal{Z}^{(1|1)} (t, g_{\text{s}})}{\mathcal{Z}^{(0|0)} (t, g_{\text{s}})} &\simeq \frac{1}{g_{\text{s}}}\, \frac{\rmi}{2\pi\lambda} \left( \sqrt{3-3\lambda r}\, \sqrt{3-\lambda r} - \lambda r \left(2-\lambda r\right) \log \left( \frac{3-2r\lambda+\sqrt{3-3\lambda r}\,  \sqrt{3-\lambda r}}{\lambda r} \right) \right) + \nonumber \\
&+ g_{\text{s}}\, \frac{3\rmi \left(\lambda r \left(4\lambda^2 r^2-30\lambda r+63 \right) - 18\right)}{8\pi r \left(3-\lambda r\right)^{3/2} \left(3-3\lambda r\right)^{5/2}} + \cdots.
\end{align}
\noindent
Comparison of this matrix-integral result with the transseries occurs via the ``bridge'' relation \eqref{eq:generalexpressionforeigenvaluetunnelingtwoparameters}; which for this sector of the quartic matrix model effectively encodes the free-energy transseries contributions (via $Z = \exp F$) as
\begin{align}
\mathsf{S}_{(0,0)\rightarrow (0,1)}\, \mathsf{S}_{(0,1)\rightarrow (1,1)}\, \frac{Z^{(1,1)}}{Z^{(0,0)}} &\simeq \frac{\mathcal{Z}^{(1|1)}}{\mathcal{Z}^{(0|0)}} \simeq \\
&
\hspace{-50pt}
\simeq \mathsf{S}_{(0,0)\rightarrow (0,1)}\, \mathsf{S}_{(0,1)\rightarrow (1,1)} \left\{ \frac{1}{g_{\text{s}}}\, F^{(1,1)}_{-1} + g_{\text{s}} \left( F^{(1,1)}_{1} + F^{(1,0)}_{1/2}\, F^{(0,1)}_{1/2} \right) + \cdots \right\}. \nonumber
\end{align}
\noindent
Comparing with the quartic transseries data \eqref{eq:qdata10}-\eqref{eq:qdata11} yields \textit{exact} agreement.

\paragraph{Resonance and Logarithms:}

We are left with the $(2|1)$ configuration and, in particular, checking our result \eqref{eq:finalresult21} within the quartic matrix model. As usual for this example, we first need to take into account the possible (anti) eigenvalue configurations associated to the two symmetric saddles. Distribution of our two eigenvalues and one anti-eigenvalue leads to three possibilities (upon permutation symmetry):
\begin{itemize}
\item The two eigenvalues sit in opposing saddles and the anti-eigenvalue occupies one of those saddles (with associated symmetry factor of $2$). Not being quite the same calculation which led to \eqref{eq:finalresult21}, it does proceed exactly along the same lines as explained in subsection~\ref{subsec:resonanceandlogs}. We end-up with a contribution of
\begin{equation}
\frac{\rmi}{(2\pi)^2} \sqrt{\frac{2\pi g_{\text{s}}}{V^{\prime\prime}_{\text{eff}} (x^{\star})}}\, \frac{\left(b-a\right)}{4}\, \rme^{-\frac{1}{g_{\text{s}}} \left( V_{\text{eff}}(x^{\star}) - V_{\text{eff}}(b)\right)}\, \log \left(\frac{x^{\star}}{b}\right)^2.
\end{equation}
\item The two eigenvalues and one anti-eigenvalue all occupy the same saddle. This exactly leads to our expression \eqref{eq:finalresult21}, only with an additional symmetry factor of $2$.
\item The two eigenvalues sit in the same saddle, with anti-eigenvalue on the opposing one. This is a pure saddle-point calculation, which one may readily estimate will have lowest contribution at order $g_{\text{s}}^{3/2}$. We discard such contribution as \eqref{eq:finalresult21} is an order-$g_{\text{s}}^{1/2}$ calculation.
\end{itemize}
\noindent
Taking all above possibilities together, we find the result (for the quartic model $\alpha = \frac{\lambda}{12}$)
\begin{align}
\mathsf{S}_{(0,0)\rightarrow (0,1)}\, \mathsf{S}_{(0,1)\rightarrow (1,1)} \left( \mathsf{S}_{(1,1)\to(1,0)} + \mathsf{S}_{(1,1)\to(2,1)}\, \frac{\lambda}{12}\, \log \frac{f(t)}{g_{\text{s}}^2} \right) F^{(1,0)} (t, g_{\text{s}}) &\simeq \\
&
\hspace{-345pt}
\simeq \sqrt{g_{\text{s}}}\, \sqrt{\frac{3}{\lambda}}\, \frac{1}{\pi^{3/2}}\, \frac{(\lambda r)^{3/2}}{r \left(3-3\lambda r\right)^{5/4} \left(3-\lambda r\right)^{1/4}} \left\{ 2\gamma_{\text{E}} + \log \left( \frac{16}{9 g_{\text{s}}^2}\, \frac{\left(3-\lambda r\right)^3 \left(3-3\lambda r\right)^5}{\lambda^6 r^4} \right) \right\} + O (g_{\text{s}}^{3/2}). \nonumber
\end{align}
\noindent
Comparison with transseries data in appendix~\ref{app:QuarticMMData} yields \textit{exact} agreement. Furthermore, this yields an analytical result for a rather non-trivial Stokes coefficient,
\begin{equation}
\boldsymbol{S}_{(0,-1)} = \rmi\, \sqrt{\frac{\lambda}{12\pi}} \left( \gamma_{\text{E}} + \log 96\sqrt{3} \right) \left[\begin{array}{c}
2\\
1
\end{array}\right],
\end{equation}
\noindent
which is associated to the Borel residue $\mathsf{S}_{(1,1)\to(1,0)}$. As in the earlier cubic example, also herein this Stokes coefficient may be double-scaled to Painlev\'e~I Stokes data (see the next subsection~\ref{subsec:DSLexample}) where it then \textit{precisely} matches against the result in \cite{bssv22}.

\paragraph{Higher-Step Stokes Data:}

Finally, following what we already did in the cubic case, let us use the Stokes-data consistency-conditions discussed in appendix~\ref{app:HigherOrderStokes}, in particular formula \eqref{eq:higherorderStokes}, in order to calculate the two-step Stokes vector $\boldsymbol{S}_{(1,-1)}$. This just requires the quartic ``canonical'' Stokes vector in \eqref{eq:forwardtrivialstokesquartic}, and the result reads
\begin{equation}
\boldsymbol{S}_{(1,-1)} = \frac{\rmi}{4} \left[\begin{array}{c}
2\\
0
\end{array}\right].
\end{equation}
\noindent
This \textit{precisely} coincides with the result obtained from comparing against the Painlev\'e~I double-scaling limit in \eqref{eq:higherstepStokesquartic}, further matching against \cite{bssv22}.

\subsection{Matrix Model Double-Scaling Limits}\label{subsec:DSLexample}

Having established cubic and quartic matrix models, we finally address their (common) Painlev\'e~I double-scaling limit. This is realized in steps. First, rescale matrix-model free-energy transseries-parameters $\sigma_1, \sigma_2$ by a factor $C$ fixed such that transseries sectors will double scale to those of Painlev\'e~I;
\begin{equation}
\label{eq:rescalecubicBorelResiduesForDSL}
F^{(n,m)} \mapsto  C^{n+m}\, F^{(n,m)}, \qquad \mathsf{S}_{(n,m)\rightarrow (p,q)} \mapsto C^{n+m-p-q}\, \mathsf{S}_{(n,m)\rightarrow (p,q)}.
\end{equation}
\noindent
This transformation is consistent and does not change the full transseries, as can be checked via \eqref{eq:forwardStokesAutomorphism}-\eqref{eq:backwardStokesAutomorphism}. The factor $C$ for cubic and quartic matrix models is given by\footnote{Notice how the quartic factor is $\lambda$-dependent whereas this is not the case for the cubic. This is due to different integration choices in \cite{msw08} and \cite{asv11}, and we have kept their conventions for consistency.}
\begin{equation}
C_{\text{cubic}} = -\frac{\sqrt{2}}{3^{1/4}}, \qquad C_{\text{quartic}} = -\frac{2\cdot 3^{1/4}}{\sqrt{\lambda}}.
\end{equation}
\noindent
For the cubic matrix model, the double-scaling limit is now implemented via
\begin{align}
t &= \frac{1}{3\sqrt{3}\lambda^2} + \kappa\, \frac{g_{\text{s}}^{4/5}}{2^{1/5}\, 3^{1/10} \lambda^{2/5}}, \\
3\lambda^2 R_{\text{cubic}} (t, g_{\text{s}}) &= 1 - \left(12\lambda^4 g_{\text{s}}^2\right)^{1/5} u_{\text{PI}}(\kappa), \\
F_{\text{cubic}}(t,g_{\text{s}}) &= F_{\text{PI}}(\kappa),
\end{align}
\noindent
and taking the limit towards criticality. In this limit the cubic string equation \eqref{eq:stringequationcubicmatrixmodel} becomes the Painlev\'e~I equation
\begin{equation}
\label{eq:PainleveI}
u_{\text{PI}}(\kappa)^2-\frac{1}{6} u_{\text{PI}}^{\prime\prime}(\kappa) = \kappa,
\end{equation}
\noindent
with the free energy now given by $F_{\text{PI}}^{\prime\prime}(\kappa) = -u_{\text{PI}}(\kappa)$. An analogous procedure for the quartic matrix model can be found in, \textit{e.g.},  \cite{asv11}. The double-scaling of the full transseries is further discussed in appendix~\ref{app:MinimalStringsData}, with explicit comparisons to cubic and quartic transseries \cite{msw08, asv11}. Moreover, one may relate (cubic or quartic) matrix model Borel residues with the ones of Painlev\'e~I via \cite{asv11}
\begin{equation}
\label{eq:BorelResiduesCubicToPI}
\mathsf{S}_{(n,m)\rightarrow (p,q)}^{\text{MM}} = C^{p+q-n-m}\,\mathsf{S}_{(n,m)\rightarrow (p,q)}^{\text{PI}}.
\end{equation}
\noindent
Let us explicitly do this with the Borel residues of the cubic and of the quartic matrix model, which we computed earlier in this section. Explicitly they become
\begin{align}
\mathsf{S}_{(0,0)\rightarrow (1,0)} &= \rmi\,\frac{\sqrt[4]{3}}{2\sqrt{\pi}},  \qquad \qquad
\mathsf{S}_{(0,0)\rightarrow (0,1)} = \frac{\sqrt[4]{3}}{2\sqrt{\pi}}, \\
\mathsf{S}_{(1,1)\rightarrow (1,0)} &= \frac{\rmi}{\sqrt[4]{3}\sqrt{\pi}} \left( 2 \gamma_{\text{E}} + 2 \log 96\sqrt{3} \right). 
\end{align}
\noindent
Translated to Stokes vectors, we arrive at
\begin{align}
\boldsymbol{S}_{(1,0)} &= -\rmi\,\frac{\sqrt[4]{3}}{2\sqrt{\pi}} \left[\begin{array}{c}
1\\
0
\end{array}\right], & 
\boldsymbol{S}_{(0,1)} &= - \frac{\sqrt[4]{3}}{2\sqrt{\pi}} \left[\begin{array}{c}
0\\
1
\end{array}\right], \\
\boldsymbol{S}_{(0,-1)} &= - \frac{\rmi}{\sqrt[4]{3}\sqrt{\pi}} \left( \gamma_{\text{E}} + \log 96\sqrt{3} \right) \left[\begin{array}{c}
2\\
1
\end{array}\right], & 
\boldsymbol{S}_{(1,-1)} &= \frac{\rmi}{4} \left[\begin{array}{c}
2\\
0
\end{array}\right].
\end{align}
\noindent
These (non-trivial) Stokes data \textit{precisely} match the recent results in \cite{bssv22}. 


\acknowledgments
We would like to thank
Paolo Gregori,
Pavel Putrov,
Noam Tamarin,
for useful discussions, comments and/or correspondence. The results in this paper were reported by RS at the ``First Internal ReNewQuantum Workshop'' held in August at the Centre for Quantum Mathematics, University of Southern Denmark, and by MS at the ``Physical Resurgence: On Quantum, Gauge, and Stringy'' workshop held in September at the Isaac Newton Institute for Mathematical Sciences, University of Cambridge, to which the authors would like to thank for hospitality. RS and MS would further like to thank the Isaac Newton Institute for Mathematical Sciences, Cambridge, for support and hospitality during the programme ``Applicable Resurgent Asymptotics: Towards a Universal Theory'' where work on this paper was undertaken. RS and MS would also like to thank CERN TH-Division and the University of Geneva for extended hospitality, where parts of this work were conducted. MS is supported by the LisMath Doctoral program and FCT-Portugal scholarship SFRH/PD/BD/135525/ 2018. This research was supported in part by CAMGSD/IST-ID and via the FCT-Portugal grants UIDB/04459/2020, UIDP/04459/2020, PTDC/MAT-OUT/28784/2017. This work was partially supported by EPSRC grant EP/R014604/1. This paper is partly a result of the ERC-SyG project, Recursive and Exact New Quantum Theory (ReNewQuantum) funded by the European Research Council (ERC) under the European Union's Horizon 2020 research and innovation programme, grant agreement 810573.

\newpage

\appendix

\section{On Computing Integrals and Their Toy Examples}\label{app:DerivativeTrick}

This appendix covers more details concerning the derivative-method introduced in subsection~\ref{subsec:bulkoneonesector}. Within toy examples, we better explore when this method is applicable, and we further support its use with numerical checks. We also briefly discuss their associated integration contours.

\subsection*{The Gaussian-Integral Toy Model}\label{app:GaussianToyModel}

Let us address our derivative-method in the context of a Gaussian toy-model integral,
\begin{equation}
\label{eq:gaussiantoyintegral}
\CI = \pvint_{\mathbb{R}} \text{d}x\, \pvint_{\rmi\mathbb{R}} \text{d}\bar{x}\, \frac{1}{\left(x-\bar{x}\right)^2}\, \rme^{-\frac{1}{g_{\text{s}}} \left(\frac{1}{2}x^2-\frac{1}{2}\bar{x}^2\right)},
\end{equation}
\noindent
which is essentially the Gaussian version of \eqref{eq:toymodelexample11} (and where we recall the principal-value prescription in \eqref{eq:principalvalueprescription}). Our main question from subsection~\ref{subsec:bulkoneonesector} was when can one exchange principal-value integrals with the primitives of the derivative-method. We illustrate this issue by first directly employing the derivative method (and failing) and then by further reducing the order of the pole via partial integrations (and succeeding).

\paragraph{Direct Evaluation:}

Evaluating the integral \eqref{eq:gaussiantoyintegral} via our derivative-method starts with
\begin{equation}
\frac{1}{g_{\text{s}}^2} \left(\frac{1}{2}x^2-\frac{1}{2}\bar{x}^2\right)^2 \int^{+\infty}_{1} \text{d}u\, \int^{+\infty}_{u} \text{d}v\, \rme^{- \frac{v}{g_{\text{s}}} \left(\frac{1}{2}x^2-\frac{1}{2}\bar{x}^2\right)} = \rme^{-\frac{1}{g_{\text{s}}}\left(\frac{1}{2}x^2-\frac{1}{2}\bar{x}^2\right)},
\end{equation}
\noindent
in which case we may rewrite $\CI$ as
\begin{equation}
\CI = \pvint_{\mathbb{R}} \text{d}x\, \pvint_{\rmi\mathbb{R}} \text{d}\bar{x}\, \int^{+\infty}_{1} \text{d}u\, \int^{+\infty}_{u} \text{d}v\, \frac{1}{g_{\text{s}}^2} \left(\frac{x+\bar{x}}{2}\right)^2 \rme^{- \frac{v}{g_{\text{s}}} \left(\frac{1}{2}x^2-\frac{1}{2}\bar{x}^2\right)}.
\end{equation}
\noindent
Next, let us switch the principal-value integrals with the $u$, $v$ integrals. Explicitly incorporating \eqref{eq:principalvalueprescription} in the above and blindly swapping integrals would yield
\begin{equation}
\CI = \lim\limits_{\epsilon\to 0}\, \int^{+\infty}_{1} \text{d}u\, \int^{+\infty}_{u} \text{d}v \int\limits_{(-\infty,-\epsilon)\cup (\epsilon,+\infty)} \text{d}x \int\limits_{(-\rmi\infty,-\rmi\epsilon)\cup (\rmi\epsilon,+\rmi\infty)} \text{d}\bar{x}\, \frac{1}{g_{\text{s}}^2} \left(\frac{x+\bar{x}}{2}\right)^2 \rme^{- \frac{v}{g_{\text{s}}} \left(\frac{1}{2}x^2-\frac{1}{2}\bar{x}^2\right)}.
\end{equation}
\noindent
But this result will be divergent. In fact, changing the order of integration is only allowed if the Fubini theorem is fulfilled, and whether this is possible may be checked via the Tonelli theorem on absolute values. In our case this immediately translates to
\begin{align}
\label{eq:gaussiantoycheck}
\int^{+\infty}_{1} \text{d}u\, \int^{+\infty}_{u} \text{d}v \int\limits_{(-\infty,-\epsilon)\cup (\epsilon,+\infty)} \text{d}x \int\limits_{(-\rmi\infty,-\rmi\epsilon)\cup (\rmi\epsilon,+\rmi\infty)}\text{d}\bar{x}\, \left|\left(\frac{x+\bar{x}}{2}\right)^2\, \rme^{- \frac{v}{g_{\text{s}}} \left(\frac{1}{2}x^2-\frac{1}{2}\bar{x}^2\right)}\right| &= \\
&
\hspace{-200pt}
= - g_{\text{s}} \int^{+\infty}_{1} \text{d}u\, \left|\epsilon^2\,\text{E}_{1/2} \left(\frac{\epsilon^2 u}{2 g_{\text{s}}}\right)^2 \right|\, \nless\infty, \quad \forall\epsilon>0, \nonumber
\end{align}
\noindent
with $\text{E}_{n}(x)$ the exponential integral function. For very small $\epsilon$ this integral diverges logarithmically. What this implies is that if the derivative-method is to work, it cannot be used blindly and some extra step is required in the calculation.

\paragraph{Evaluation After Partial Integration:}

Let us instead consider the integration
\begin{equation}
\label{eq:gaussiantoymodelpartialintegration}
\CI = -\frac{1}{g_{\text{s}}} \pvint_{\mathbb{R}} \text{d}x\, \pvint_{\rmi\mathbb{R}} \text{d}\bar{x}\, \frac{x}{x-\bar{x}}\, \rme^{-\frac{1}{g_{\text{s}}}\left(\frac{1}{2}x^2-\frac{1}{2}\bar{x}^2\right)},
\end{equation}
\noindent
which follows from \eqref{eq:gaussiantoyintegral} \textit{after performing\footnote{Note that one has to be careful with partial integration of principal-value integrals. In this case one can check that the boundary terms close to $x=0=\bar{x}$ do not give additional contributions.} a partial integration}. Moving-on to our derivative method, it now starts with
\begin{equation}
- \frac{1}{g_{\text{s}}} \left(\frac{1}{2}x^2-\frac{1}{2}\bar{x}^2\right) \int^{+\infty}_{1} \text{d}v\, \rme^{- \frac{v}{g_{\text{s}}} \left(\frac{1}{2}x^2-\frac{1}{2}\bar{x}^2\right)} = \rme^{-\frac{1}{g_{\text{s}}}\left(\frac{1}{2}x^2-\frac{1}{2}\bar{x}^2\right)}.
\end{equation}
\noindent
Plugging this into the above \eqref{eq:gaussiantoymodelpartialintegration} we find, after switching the order of integrations,
\begin{equation}
\CI = \frac{1}{g_{\text{s}}^2}\, \lim\limits_{\epsilon\to 0}\, \int^{+\infty}_{1} \text{d}v \int\limits_{(-\infty,-\epsilon)\cup (\epsilon,+\infty)} \text{d}x \int\limits_{(-\rmi\infty,-\rmi\epsilon)\cup (\rmi\epsilon,+\rmi\infty)} \text{d}\bar{x}\, x\, \frac{x+\bar{x}}{2}\, \rme^{- \frac{v}{g_{\text{s}}} \left(\frac{1}{2}x^2-\frac{1}{2}\bar{x}^2\right)}.
\end{equation}
\noindent
This result is no longer divergent. In fact, changing the order of integration is now allowed as, applying our previous argument,
\begin{align}
\int^{+\infty}_{1} \text{d}v \int\limits_{(-\infty,-\epsilon)\cup (\epsilon,+\infty)} \text{d}x \int\limits_{(-\rmi\infty,-\rmi\epsilon)\cup (\rmi\epsilon,+\rmi\infty)} \text{d}\bar{x}\, \left|x\, \frac{x+\bar{x}}{2}\, \rme^{- \frac{v}{g_{\text{s}}} \left(\frac{1}{2}x^2-\frac{1}{2}\bar{x}^2\right)}\right| &= \\
&
\hspace{-100pt}
= \frac{1}{2}\, g_{\text{s}} \left|\epsilon^2\, \text{E}_{1/2} \left(\frac{\epsilon^2}{2 g_{\text{s}}}\right)^2 \right|\, <\infty, \quad \forall\epsilon>0. \nonumber
\end{align}
\noindent
From here the $\epsilon\to 0$ limit follows and we would be done. For the harder examples in subsection~\ref{subsec:bulkoneonesector} we still need to continue with the derivative-method. This then entails further changing the principal-value integrals with the $v$-integral, alongside the limit, which is validated by the Lebesgue theorem of dominated convergence holding due to uniform convergence of our integral kernel, \textit{i.e.}, we have
\begin{equation}
\int\limits_{(-\infty,-\epsilon)\cup (\epsilon,+\infty)} \text{d}x \int\limits_{(-\rmi\infty,-\rmi\epsilon)\cup (\rmi\epsilon,+\rmi\infty)} \text{d}\bar{x}\, x\, \frac{x+\bar{x}}{2}\, \rme^{- \frac{v}{g_{\text{s}}} \left(\frac{1}{2}x^2-\frac{1}{2}\bar{x}^2\right)} = \frac{\epsilon^4}{2}\, \text{E}_{-1/2} \left(\frac{\epsilon^2 v}{2 g_{\text{s}}}\right) \text{E}_{1/2} \left(\frac{\epsilon^2 v}{2 g_{\text{s}}}\right),
\end{equation}
\noindent
which obeys
\begin{equation}
\left|\, \left( \frac{\epsilon^4}{2}\, \text{E}_{-1/2} \left(\frac{\epsilon^2 v}{2 g_{\text{s}}}\right) \text{E}_{1/2} \left(\frac{\epsilon^2 v}{2 g_{\text{s}}}\right) \right) - \frac{\pi g_{\text{s}}^2}{v^2}\, \right|\, \xrightarrow{\epsilon\to 0} 0,
\end{equation}
\noindent
where $\int_{1}^{+\infty} \rmd v/v^2$ is finite. Hence changing order of integrations is fully validated, and our ``derivative-method calculation'' goes through,
\begin{align}
\CI &= \frac{1}{g_{\text{s}}^2} \int^{+\infty}_{1} \text{d}v\, \pvint_{\mathbb{R}} \text{d}x\, \pvint_{\rmi\mathbb{R}} \text{d}\bar{x}\, x\, \frac{x+\bar{x}}{2}\, \rme^{- \frac{v}{g_{\text{s}}} \left(\frac{1}{2}x^2-\frac{1}{2}\bar{x}^2\right)} = \nonumber \\
&= \frac{1}{g_{\text{s}}^2} \int^{+\infty}_{1} \text{d}v \int_{\mathbb{R}} \text{d}x \int_{\rmi\mathbb{R}} \text{d}\bar{x}\, x\, \frac{x+\bar{x}}{2}\, \rme^{- \frac{v}{g_{\text{s}}} \left(\frac{1}{2}x^2-\frac{1}{2}\bar{x}^2\right)} = \int^{+\infty}_{1} \text{d}v\, \frac{\rmi\pi}{v^2} = \rmi\pi,
\end{align}
\noindent
where we have dropped principal-value integrals in the second line. The resulting integral can now indeed be solved directly by saddle-point evaluation. Let us further comment that a brief numerical exploration of the integral \eqref{eq:gaussiantoyintegral} undoubtedly confirms this result, $\CI = \rmi\pi$.

\subsection*{The $(1|1)$-Integral Toy Model}\label{app:11QuarticToyModel}

\begin{figure}
\centering
\begin{tikzpicture}[scale=0.8]
\draw[->, line width=2pt] (-4, 0) -- (4, 0);
\draw[->, line width=2pt] (0,-4) -- (0,4);
\fill[cornellred] (-2, 0) circle (4pt);
\fill[cornellred] (2, 0) circle (4pt);
\fill[cornellred] (0, 0) circle (4pt);
\node at (-3.8, 3.8) {$\mathbb{C}$};
\draw[line width=2pt] (-4, 3.5) -- (-3.46, 3.5);
\draw[line width=2pt] (-3.5, 3.5) -- (-3.5, 4);
\node[blue] at (3, 2) {$\gamma$};
\node[orange] at (3, -0.4) {$\bar{\gamma}$};
\draw[->, blue, line width=2pt] (4, -4) -- (0,0) -- (4,4);
\draw[->, orange, line width=2pt] (0, -4) --(0,0)-- (4, 0);
\end{tikzpicture}
\hspace{0.6cm}
\begin{tikzpicture}[scale=0.8]
\draw[->, line width=2pt] (-4, 0) -- (4, 0);
\draw[->, line width=2pt] (0,-4) -- (0,4);
\fill[cornellred] (-2, 0) circle (4pt);
\fill[cornellred] (2, 0) circle (4pt);
\fill[cornellred] (0, 0) circle (4pt);
\node at (-3.8, 3.8) {$\mathbb{C}$};
\draw[line width=2pt] (-4, 3.5) -- (-3.46, 3.5);
\draw[line width=2pt] (-3.5, 3.5) -- (-3.5, 4);
\node[blue] at (1, -1) {$\gamma_{\text{res}}$};
\node[blue] at (3, 4) {$\gamma^{\star}$};
\node[orange] at (3, -0.4) {$\bar{\gamma}$};
\draw[->, blue, line width=2pt] (2, -0.1) -- (1.7, -1/2) -- (0.3,-1/2) -- (0,0) -- (0.3,1/2) -- (1.7,1/2) -- (2, 0.1);
\draw[->, blue, line width=2pt] (4, -4) -- (2,-1/2) -- (2,1/2) -- (4,4);
\draw[->, orange, line width=2pt] (0, -4) --(0,0)-- (4, 0);
\end{tikzpicture}
\caption{Integration contours $\gamma$ and $\bar{\gamma}$ for (anti) eigenvalue tunneling, as in figure~\ref{fig:settingcutandsaddleGeneraltunneling}; and where the saddle-points of the quartic potential have been further highlighted as the red dots. The left-plot shows the original contours, whereas the right-plot shows the usual deformation which turns the original contour $\gamma$ into a residue contribution $\gamma_{\text{res}}$ alongside a steepest-descent path $\gamma^{\star}$. It is important to note that such a deformation does not cross the pole at $x=\bar{x}$. The contours $\gamma^{\star}$ and $\bar{\gamma}$ are the ones used in the integral \eqref{eq:quartictoymodelintegral}.}
\label{fig:ContourDeformationQuarticToyModel}
\end{figure}
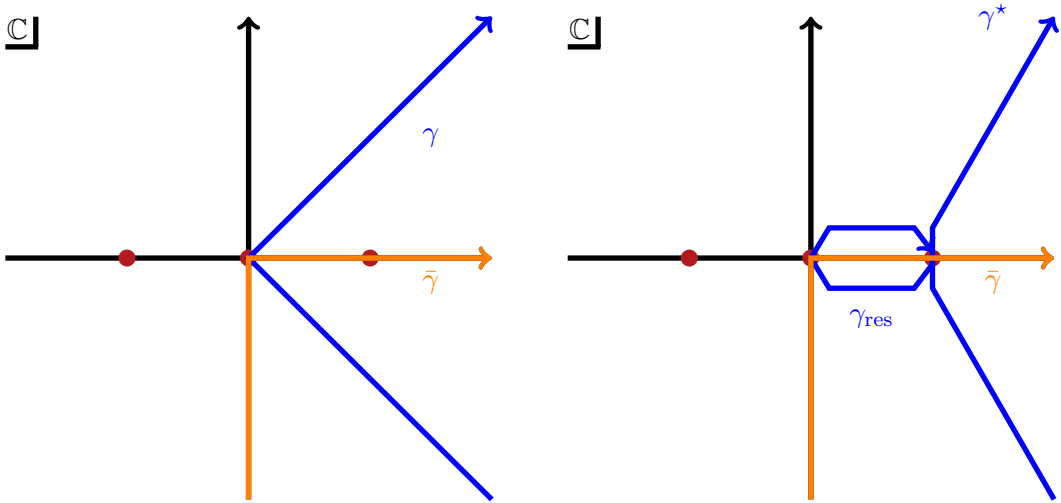

Let us go back to the toy example of subsection~\ref{subsec:bulkoneonesector} illustrating the $(1|1)$ matrix integral, which was given in \eqref{eq:toymodelexample11}. For simplicity we shall herein drop the $1/(2\pi)^2$ normalization and focus on the example of a quartic potential (with positive $\lambda$),
\begin{equation}
\label{eq:quarticappendixV}
V (x) = \frac{1}{2}x^2-\frac{\lambda}{24}x^4.
\end{equation}
\noindent
We also consider slightly different integration contours, \textit{i.e.}, we consider the integral
\begin{equation}
\label{eq:quartictoymodelintegral}
\CI = \pvint_{\gamma^{\star}} \text{d}x\, \pvint_{\bar{\gamma}} \text{d}\bar{x}\, \frac{1}{(x-\bar{x})^2}\, \rme^{-\frac{1}{g_{\text{s}}} \left(V(x)-V(\bar{x})\right)},
\end{equation}
\noindent
where the principal-value is the usual \eqref{eq:principalvalueprescription}, where the contour $\gamma^{\star}$ is steepest-descent through the nonperturbative saddle $x^{\star}=\sqrt{6/\lambda}$, and where the contour $\bar{\gamma}$ goes through $0$, as illustrated in figure~\ref{fig:ContourDeformationQuarticToyModel}. Note that herein we have dropped\footnote{Such residue contribution can always be added at any later stage in the computation.} the residue contribution---we will only be interested in the steepest-descent contour $\gamma^{\star}$ for the current discussion. There are two ways to evaluate this integral. On the one hand, we can solve it numerically. On the other hand, we can use our derivative method (upon resummation) and then compare to check its validity.

Let us begin with the derivative method. Having dropped the residue contribution, we readily evaluate the integral $\CI$ on the steepest-descent contour $\gamma^{\star}$. To first order, this exact same calculation has already been done in subsection~\ref{subsec:bulkoneonesector}. Herein it is straightforward to go to higher orders, where we obtain an asymptotic series,
\begin{equation}
\label{eq:QuarticToyModelAsymptoticSeries}
\pvint_{\gamma^{\star}} \text{d}x\, \pvint_{\bar{\gamma}} \text{d}\bar{x}\, \frac{1}{(x-\bar{x})^2}\, \rme^{-\frac{1}{g_{\text{s}}}\left(V(x)-V(\bar{x})\right)} \simeq \sum_{g=1}^{\infty} A_{g}(\lambda)\, g_{\text{s}}^g = \frac{\pi\lambda\rmi}{24}\, g_{\text{s}} + \frac{5\pi\lambda^3\rmi}{288}\, g_{\text{s}}^3 + \cdots.
\end{equation}
\noindent
We have computed terms in this series up to order $40$ in $g_{\text{s}}$. 

Having obtained $\CI$ from the derivative method, let us just outline the numerical\footnote{The integrations were performed with the \texttt{NIntegrate} function in \textit{Mathematica}, with built-in error estimate.} evaluation. To do this properly, one needs to be careful in performing the principal-value prescription \eqref{eq:principalvalueprescription}; in particular the parametrization of the contour integral must generically be piecewise-smooth but, importantly, smooth \textit{inside} the disk which has been removed around the pole. Further, the same-size disk (\textit{i.e.}, the same $\epsilon$) must be chosen for both contours in \eqref{eq:principalvalueprescription}.

In order to perform a meaningful comparison between the result obtained from direct numerical integration and the one from the derivative-method, we need to Borel--Laplace resum the asymptotic series \eqref{eq:QuarticToyModelAsymptoticSeries} (which has asymptotic growth $A_g \sim \Gamma(g-1)/\left(3/2\right)^g$). Once this is done we compare to the numerical evaluation of \eqref{eq:quartictoymodelintegral}. We performed two tests. Table~\ref{tab:comparingtoymodellessprec} has lateral Borel--Laplace resummation close to but above the real line; whereas table~\ref{tab:comparingtoymodelmoreprec} has higher numerical precision to illustrate how the results converge properly. The results show very good agreement and validate how the derivative-method indeed produces the correct asymptotic series associated to the integral $\CI$.

\begin{table}
\centering
\begin{tabular}{c|cc}
$g_{\text{s}}$ & Resummed Asymptotic Series & Numerical Principal Value \\
\hline\hline
$\frac{1}{10}\rme^{\frac{\rmi}{100}}$ & $-0.0030627445...+0.0563580235...\rmi$ & $-0.003062{\color{red}6033...} + 0.056353{\color{red}3856...}\rmi$\\
$\frac{1}{12}\rme^{\frac{\rmi}{100}}$ & $-0.0015144998...+0.0461713206...\rmi$ & $-0.001514{\color{red}4358...}+ 0.046167{\color{red}5600...}\rmi$\\
$\frac{1}{14}\rme^{\frac{\rmi}{100}}$ & $-0.0008464636...+0.0390336330...\rmi$ & $-0.000846{\color{red}4267...} + 0.039030{\color{red}1472...}\rmi$\\
\hline
$\frac{1}{10}\rme^{\frac{\rmi}{30}}$ & $-0.0044849274...+0.0560706365...\rmi$ & $-0.00448{\color{red}47002...} + 0.05606{\color{red}60117...}\rmi$\\
\end{tabular}
\caption{Comparison of the resummed asymptotic series for the integral \eqref{eq:quartictoymodelintegral}, with the results of direct numerical integration close (and above) the real line. In red we highlight the digits of precision for the numerical integration obtained from the \texttt{NIntegrate} built-in error. Here we chose $\epsilon=10^{-5}$, $\lambda=4$, and \texttt{WorkingPrecision} of 12. The agreement is very good.}
\label{tab:comparingtoymodellessprec}
\end{table}

\begin{table}
\centering
\begin{tabular}{c|cc}
$g_{\text{s}}$ & Resummed Asymptotic Series & Numerical Principal Value \\
\hline\hline
$\frac{1}{10}\rme^{\frac{\rmi}{100}}$ & $-0.0030627445...+0.0563580235...\rmi$ & $-0.0030627{\color{red}413...} + 0.0563580{\color{red}193...}\rmi$\\
$\frac{1}{12}\rme^{\frac{\rmi}{100}}$ & $-0.0015144998...+0.0461713206...\rmi$ & $-0.0015144{\color{red}995...}+ 0.0461671{\color{red}283...}\rmi$\\
\end{tabular}
\caption{Comparison of the resummed asymptotic series for the integral \eqref{eq:quartictoymodelintegral}, with the results of direct numerical integration. We now chose $\epsilon=10^{-7}$, $\lambda=4$. Again in red we highlight the digits of precision for the numerical integration obtained from the \texttt{NIntegrate} built-in error, albeit herein the \texttt{WorkingPrecision} of the numerical integration has been improved to 15. Indeed as expected, we hence find an even better agreement of the results.}
\label{tab:comparingtoymodelmoreprec}
\end{table}

\subsection*{Integration Contours and Multi-Sheeted Integrands}\label{app:detailsintegrationcontours}

Finally, let us make some comments concerning the behavior of all these integration contours---when are they just saddle-contours and when are they not. We continue with our favorite example, \eqref{eq:toymodelexample11} or \eqref{eq:quartictoymodelintegral},
\begin{equation}
\label{eq:TheN1M1Integral}
\CI = \pvint_{\CC_y} \text{d}y\, \pvint_{\CC_x} \text{d}x\, \frac{1}{(x-y)^2}\, \text{e}^{-\frac{1}{g_s} \left(V(x)-V(y)\right)},
\end{equation}
\noindent
where $\CC_x$ is a steepest-descent contour through a saddle $x^{\star}$ of $V(x)$, whereas $\CC_y$ is a steepest-ascent contour associated to the same saddle. Let us consider \eqref{eq:TheN1M1Integral} as an iterated integral, and start by understanding the result of the first integration,
\begin{equation}
\widetilde{\CI} (y) = \int_{\CC_x} \text{d}x\, \frac{1}{(x-y)^2}\, \text{e}^{-\frac{1}{g_s}\left( V(x)-V(y) \right)}.
\end{equation}
\noindent
Note how when $y$ crosses $\CC_x$ the function $\widetilde{\CI}(y)$ is \textit{discontinuous}, with discontinuity given by
\begin{equation}
\Res_{x=y} \left( \frac{1}{(x-y)^2}\, \text{e}^{-\frac{1}{g_s} \left( V(x)-V(y) \right)} \right) = -\frac{2\pi\rmi}{g_s}\, V^{\prime} (y).
\end{equation}
\noindent
This discontinuity only vanishes when $y$ sits at a saddle-point of the potential. In that case, $\CC_y$ will avoid the discontinuous jump of the integrand $\widetilde{\CI}(y)$ as it precisely crosses $\CC_x$ at $x^{\star}$.

Avoiding such discontinuities of the integrand $\widetilde{\CI} (y)$ leads to a classification of admissible integration contours. There are two basic rules for such contours:
\begin{itemize}
\item Admissible integration contours at infinity must behave like steepest-descent contours. This is required in order to find convergent integrals.
\item Admissible integration contours must only cross at saddle-points of the potential. This is required in order to avoid the discontinuity described above.
\end{itemize}
\noindent
These requirements become very clear in the example of the quartic potential \eqref{eq:quarticappendixV}. In this case, both original contours $\gamma$ and $\bar{\gamma}$ in figure~\ref{fig:ContourDeformationQuarticToyModel} are clearly admissible: they both end along steepest-descent directions and only cross at the saddle-point at the origin. Admissibility remains true even after deformation, illustrating how more general contours than just steepest-descent contours can in fact occur. The same holds for the configurations in the main body of the paper.

\section{Resurgence of Partition Function and Free Energy}\label{app:resurgenceZvsF}

This appendix discusses a subtlety in the transseries constructions of matrix-model free energy and its partition function. If we recall how to obtain the free energy out of the string-equation solution in \eqref{eq:freeenergyfromstringequation}, it is clear how the free energy is only defined up to an additive factor. This is clearly also true for its transseries (where some toy-models with this feature were discussed in \cite{abs18}). What this implies is that the free-energy transseries is generically writeable as (compare with the generic \eqref{eq:transseriesexpansionGeneric}, and with the free energies of cubic \eqref{eq:FTSforCUBIC} and quartic \eqref{eq:FTSforQUARTIC} matrix models)
\begin{equation}
F_{\text{total}} (t, g_{\text{s}}; \sigma_0, \sigma_1,\sigma_2) = \sigma_0\, \widetilde{F} + \sum\limits_{n=0}^{\infty} \sum\limits_{m=0}^{\infty} \sigma_1^n \sigma_2^m\, \text{e}^{-\left(n-m\right)\frac{A(t)}{g_{\text{s}}}}\, F^{(n,m)} (t, g_{\text{s}}),
\end{equation}
\noindent
where $\widetilde{F}$ is a trivial (non-asymptotic) constant sector. Schematically this can also be written as $F_{\text{total}} = F_{\text{trivial}} + F_{\text{non-trivial}}$ where $F_{\text{trivial}}$ is just the additive constant $\sigma_0\, \widetilde{F}$ and $F_{\text{non-trivial}}$ denotes the free energies we addressed in the main body of the paper, \textit{e.g.}, \eqref{eq:FTSforCUBIC} or \eqref{eq:FTSforQUARTIC}. This addition further implies that resurgence relations can end on the trivial sector but they cannot start on it \cite{abs18}. It further leads to the (small) change at the level of Stokes automorphisms (compare with \eqref{eq:forwardStokesAutomorphism} and \eqref{eq:backwardStokesAutomorphism}, respectively)
\begin{align}
\label{eq:forwardStokesAutomorphism-F}
\underline{\mathfrak{S}}_0 F^{(n,m)} &= F^{(n,m)} - \sum_{\ell=1}^{\infty} \rme^{-\ell \frac{A}{g_{\text{s}}}} \sum_{p=0}^{\min(n+\ell,m)} \left( \mathsf{S}_{(n,m)\to(n+\ell-p,m-p)}\, F^{(n+\ell-p,m-p)} + \right. \nonumber \\
&
\left. + \delta_{n+\ell-p}\, \delta_{m-p}\, \mathsf{S}_{(n,m)\to\widetilde{0}}\, \widetilde{F} \right), \\
\label{eq:backwardStokesAutomorphism-F}
\underline{\mathfrak{S}}_\pi F^{(n,m)} &= F^{(n,m)} - \sum_{\ell=1}^{\infty} \rme^{+\ell \frac{A}{g_{\text{s}}}} \sum_{p=0}^{\min(n,m+\ell)} \left( \mathsf{S}_{(n,m)\to(n-p,m+\ell-p)}\, F^{(n-p,m+\ell-p)} + \right. \nonumber \\
&
\left. + \delta_{n-p}\, \delta_{m+\ell-p}\, \mathsf{S}_{(n,m)\to\widetilde{0}}\, \widetilde{F} \right).
\end{align}

That such a contribution is really there is not too hard to check. For example, consider the backward Stokes transition on the $(1,0)$ sector,
\begin{align}
\underline{\mathfrak{S}}_\pi F^{(1,0)} &= F^{(1,0)} - \rme^{\frac{A}{g_{\text{s}}}} \left( \mathsf{S}_{(1,0)\to\widetilde{0}}\, \widetilde{F} + \mathsf{S}_{(1,0)\to(1,1)}\, F^{(1,1)} \right) - \nonumber \\
&- \rme^{2 \frac{A}{g_{\text{s}}}} \left( \mathsf{S}_{(1,0)\to(1,2)}\, F^{(1,2)} + \mathsf{S}_{(1,0)\to(0,1)}\, F^{(0,1)} \right) - \cdots,
\end{align}
\noindent
and focus on the leading two terms in the discontinuity (\textit{i.e.}, the first line). Let us explicitly consider the example of Painlev\'e~I (simply because it is much easier to produce large-order data) and numerically investigate the contribution of these two terms at large order. The Borel residue $\mathsf S_{(1,0)\to (1,1)} = \frac{\sqrt[4]{3}}{2\sqrt{\pi}}$ is known; \textit{e.g.}, \cite{msw07}. Then, using the numerical methods described in \cite{bssv22}, we can immediately obtain the value of $\mathsf{S}_{(1,0)\to\widetilde{0}}$ (to extract this number, we have calculated the $(1,0)$ and $(1,1)$ sectors up to order $100$). With a precision of $11$ digits, we find
\begin{equation}
\mathsf{S}_{(1,0)\to\widetilde{0}} \approx -0.21434568952... = -\frac{S_{(1,0)\to (1,1)}}{\sqrt{3}} = - \frac{1}{2\sqrt[4]{3}\sqrt{\pi}},
\end{equation}
\noindent
where we have guessed the analytical closed-form expression from the numerical number. This we consider strong evidence for the appearance of the $\widetilde{F}$ sector in the resurgence relations.

Having established the (small) novelty in the basic resurgence properties of the free energy, let us move to the partition function. Akin to the free energy, it will now factorize as $Z_{\text{total}} = Z_{\text{trivial}} \cdot Z_{\text{non-trivial}}$, where $Z_{\text{trivial}} = \exp \sigma_0\, \widetilde{F}$ and $Z_{\text{non-trivial}}$ denotes the partition function addressed in the main text. Let us focus on the total contribution and consider its perturbative sector $Z^{(0,0)}$ undergoing a backward Stokes transition. Here, \eqref{eq:forwardStokesAutomorphism-F}-\eqref{eq:backwardStokesAutomorphism-F} do not allow for any novelties due to $\widetilde{F}$ and we find \eqref{eq:purelybackwardsdiscontinuity} as usual,
\begin{equation}
\text{Disc}_{\pi}\, Z^{(0,0)} = \sum_{\ell=1}^{\infty} \rme^{\ell \frac{A}{g_{\text{s}}}}\, \mathsf{S}_{(0,0)\to(0,\ell)}\, Z^{(0,\ell)}.
\end{equation}
\noindent
The novelties appear when undergoing a forward discontinuity on top of the above one. Explicitly writing sectors up to the $(1,1)$ sector, one arrives at\footnote{Note how the last term in \eqref{eq:diskadditionalterm} actually contains \textit{three} Borel residues rather than \textit{two}. This is an artifact of effectively dealing with a \textit{three}-parameter transseries albeit written with only \textit{two} parameters (see \cite{abs18}).}
\begin{align}
\label{eq:diskadditionalterm}
\text{Disc}_{0}\, \text{Disc}_{\pi}\, Z^{(0,0)} &= \mathsf{S}_{(0,0)\to(0,1)}\, \mathsf{S}_{(0,1)\to(1,1)}\, Z^{(1,1)} + \mathsf{S}_{(0,0)\to(0,1)}\, \mathsf{S}_{(0,1)\to\widetilde{0}}\, Z^{(0,0)}\, \widetilde{F} + \nonumber \\
&+ \mathsf{S}_{(0,0)\to(0,1)}\, \mathsf{S}_{(0,0)\to(1,0)}\, \mathsf{S}_{(0,1)\to\widetilde{0}}\, \rme^{-\frac{A}{g_{\text{s}}}}\, Z^{(1,0)} \widetilde{F} + \cdots.
\end{align}
\noindent
This result should be compared with \eqref{eq:TwoAutomorphismsAllSectors}. There are now clear extra contributions which were (harmlessly) ignored in subsection~\ref{subsec:eigenvalueTunnelingContours} and which we need to address. Let us understand how these novelties also (consistently) appear at the level of (anti) eigenvalue contour deformations.

The complete Stokes automorphisms \eqref{eq:forwardStokesAutomorphism-F}-\eqref{eq:backwardStokesAutomorphism-F} are just the usual ones \eqref{eq:forwardStokesAutomorphism}-\eqref{eq:backwardStokesAutomorphism}---which have been described by (anti) eigenvalue tunneling in the main text---alongside an additional contribution taking us to the $\widetilde{F}$ trivial sector. On top of this, there are situations where such a novelty does not even appear: \textit{e.g.}, starting from the partition-function perturbative-sector, either forward or backward Stokes transitions will see no additional terms. As already mentioned in subsection~\ref{subsec:eigenvalueTunnelingContours}, the $\widetilde{F}$ contributions may be regarded as associated to the tunneling of nonperturbative (anti) eigenvalues \textit{back to the perturbative cut}, and this can now be illustrated by explicitly checking the above discontinuity \eqref{eq:diskadditionalterm} with (anti) eigenvalue contours. Up to the $(1,1)$ contribution, all we need to consider are contours with $1$ eigenvalue and $1$ anti-eigenvalue. Those have been already explicitly spelled out in \eqref{eq:disc0Z11CstarCbarstar}, where the subsequent contributions originating from nonperturbative anti-eigenvalues returning to the perturbative saddle were dropped. Herein we resuscitate such contributions in our familiar quartic example. In this case, the full content of \eqref{eq:disc0Z11CstarCbarstar} reads (notice we have dropped all exponential factors in the underbraces):
\begin{align}
&\ZPertpe\,\ZSadBarpe\,-\,\ZPertme\,\ZSadBarme\,= \\
&= \left(\,\ZPertme\,+\,\ZSuperSaddles\,\right) \left(\,\ZSadBarme\,-2\,\ZPertBarme\,\right) -\,\ZPertme\,\ZSadBarme\,= \nonumber \\
&=\underbrace{\ZSuperSaddles\,\ZSadBarme}_{\sim \mathsf{S}_{(0,0)\to(0,1)}\, \mathsf{S}_{(0,1)\to(1,1)}\, Z^{(1,1)}} - \underbrace{2\,\ZPertme\,\ZPertBarme}_{\sim \mathsf{S}_{(0,0)\to(0,1)}\, \mathsf{S}_{(0,1)\to\widetilde{0}}\, Z^{(0,0)}\, \widetilde{F}} - \underbrace{2\,\ZSuperSaddles\,\ZPertBarme}_{\sim \mathsf{S}_{(0,0)\to(0,1)}\, \mathsf{S}_{(0,0)\to(1,0)}\, \mathsf{S}_{(0,1)\to\widetilde{0}}\, Z^{(1,0)}\, \widetilde{F}}. \nonumber
\end{align}
\noindent
The above contour deformations exactly match the resurgence discontinuity, illustrating the ``perturbative return'' of anti-eigenvalues, which we consider strong evidence for our picture. The incorporation of all these trivial (additive for the free energy, multiplicative for the partition function) sectors is therefore straightforward, in which case we have not discussed them any further in subsection~\ref{subsec:eigenvalueTunnelingContours}. In fact, it is worth stressing that these trivial contributions do not produce any instanton terms. Further, as our matches between resurgent transseries and matrix integrals always end-up being done at free energy level, any additive terms do not carry any new information (hence are better off omitted). Of course explicitly writing the \textit{total} partition function from the resurgent transseries formulation still requires taking this sector into account.

\section{On Higher-Order Stokes Data from Matrix Integrals}\label{app:HigherOrderStokes}

There is one final subtlety to discuss, which concerns Stokes data. As we already alluded to in the main text, the Borel plane has non-trivial branched structure due to its many singularities, in which case the Borel residues (Stokes data) one computes are dependent upon the specific sequence of Stokes transitions one considers in each case. Consistency conditions on all these Stokes data do exist, which we shall now discuss.

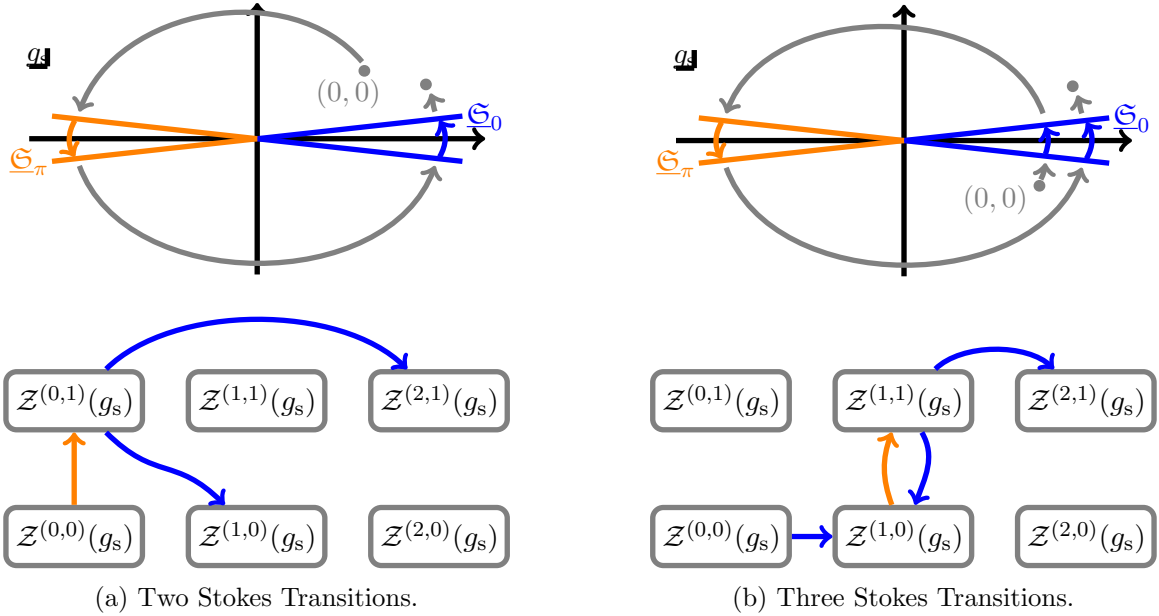
\begin{figure}
\centering
\begin{subfigure}[b]{0.45\textwidth}
\centering
\begin{tikzpicture}[
	 scale =0.6, line width=2
	]
	\draw[->] (-5,0) -- (5, 0);
	\draw[->] (0,-3) -- (0, 3);
	\draw (-5, 1.6) -- (-4.6, 1.6);
	\draw (-4.6, 1.6) -- (-4.6, 2);
	\node at (-4.8, 1.8) {$g_{\text{s}}$};
	\draw[blue] (0,0) -- (4.5, -0.5);
	\draw[blue] (0,0) -- (4.5, 0.5);
	\draw[->, blue] (4, -0.4) to[bend right =30] (4, 0.4);
	\node[blue] at (5, 0.5) {$\underline{\mathfrak{S}}_0$};
	\draw[color=orange] (0,0) -- (-4.5, -0.5);
	\draw[color=orange] (0,0) -- (-4.5, 0.5);
	\draw[->, color=orange] (-4, 0.4) to[bend right =30] (-4, -0.4);
	\node[color=orange] at (-5, -0.5) {$\underline{\mathfrak{S}}_{\pi}$};
	\draw[gray, ->] (2.3, 1.7) to[out=130, in=70] (-3.9, 0.6);
	\draw[gray, ->] (3.9, 0.6) to[out=110, in=290] (3.8, 1);
	\draw[gray, ->] (-3.9, -0.6) to[out=290, in=250] (3.9, -0.6);
	\filldraw[gray] (3.7, 1.2) circle (2pt);
	\filldraw[gray] (2.35, 1.5) circle (2pt);
	\node[gray] at (2, 1) {$(0,0)$};
	\end{tikzpicture}\\
\begin{tikzpicture}[
	grayframe/.style={
		rectangle,
		draw=gray,
		text width=4em,
		align=center,
		rounded corners,
		minimum height=2em
	}, scale =0.6, line width=2
	]
	\foreach \n in {0,...,2}{
	  \foreach \m in {0,..., 1}{
	\node[grayframe] (\n_\m) at (4*\n,3*\m) {$\mathcal{Z}^{(\n, \m)}(g_{\text{s}})$};
	}}
		\draw[color=blue, ->] (0_1) to [out=45, in=135, looseness=0.8] (2_1);
		\draw[color=orange, ->] (0_0) -- (0_1);
	\draw[color=blue, ->] (0_1) to [out=315, in=135, looseness=1.3] (1_0);
	\end{tikzpicture}\\
\caption{Two Stokes Transitions.}
\label{fig:transitioningfrom00to21a}
\end{subfigure}
\hspace{1cm}
\begin{subfigure}[b]{0.45\textwidth}
\centering
\begin{tikzpicture}[
	 scale =0.6, line width=2
	]
	\draw[->] (-5,0) -- (5, 0);
	\draw[->] (0,-3) -- (0, 3);
	\draw (-5, 1.6) -- (-4.6, 1.6);
	\draw (-4.6, 1.6) -- (-4.6, 2);
	\node at (-4.8, 1.8) {$g_{\text{s}}$};
	\draw[blue] (0,0) -- (4.5, -0.5);
	\draw[blue] (0,0) -- (4.5, 0.5);
	\draw[->, blue] (4, -0.4) to[bend right =30] (4, 0.4);
	\node[blue] at (5, 0.5) {$\underline{\mathfrak{S}}_0$};
	\draw[color=orange] (0,0) -- (-4.5, -0.5);
	\draw[color=orange] (0,0) -- (-4.5, 0.5);
	\draw[->, color=orange] (-4, 0.4) to[bend right =30] (-4, -0.4);
	\node[color=orange] at (-5, -0.5) {$\underline{\mathfrak{S}}_{\pi}$};
	\filldraw[gray] (2.97, -1) circle (2pt);
	\draw[gray, ->] (3, -0.86) to[bend right =10] (3.1, -0.5);
	\draw[->, blue] (3.1, -0.3) to[bend right =20] (3.1, 0.3);
	\draw[gray, ->] (3.05, 0.6) to[out=110, in=70] (-3.9, 0.6);
	\draw[gray, ->] (-3.9, -0.6) to[out=290, in=250] (3.902, -0.6);
	\draw[gray, ->] (3.9, 0.6) to[out=110, in=290] (3.8, 1);
	\filldraw[gray] (3.7, 1.2) circle (2pt);
	\node[gray] at (2, -1.3) {$(0,0)$};
	\end{tikzpicture}\\\vspace{0.505cm}
\begin{tikzpicture}[
	grayframe/.style={
		rectangle,
		draw=gray,
		text width=4em,
		align=center,
		rounded corners,
		minimum height=2em
	}, scale =0.6, line width=2
	]
	\foreach \n in {0,...,2}{
	  \foreach \m in {0,..., 1}{
	\node[grayframe] (\n_\m) at (4*\n,3*\m) {$\mathcal{Z}^{(\n, \m)}(g_{\text{s}})$};
	}}
		\draw[color=blue, ->] (1_1) to [out=45, in=135, looseness=0.8] (2_1);
	\draw[color=blue, ->] (0_0) -- (1_0);
	\draw[color=blue, ->] (1_1) to [out=300, in=70] (1_0);
	\draw[color=orange, ->] (1_0) to [out=110, in=250] (1_1);
	\end{tikzpicture}\\
\caption{Three Stokes Transitions.}
\label{fig:transitioningfrom00to21b}
\end{subfigure}
\caption{Illustration of two paths which end at the $(2|1)$ (anti) eigenvalue configuration. Both $(2,1)$ and $(1,0)$ sectors appear in the end result, but because the paths we took on the resurgence lattice are distinct, so will be the combinations of Borel residues appearing in the final formula. These paths also end up as being distinct upon the $g_{\text{s}}$-plane, as illustrated in the top-figures.}
\label{fig:transitioningfrom00to21}
\end{figure}

One example of this ``path dependency'' stems from \eqref{eq:generalsectorfromeigenvaluetunneling}, which implies we have reached the $(\ell|\bar{\ell})$ sector by jumping $\ell$ horizontal steps and $\bar{\ell}$ vertical steps on the resurgence lattice. But we could have reached that same node differently---for instance iterating one-step jumps---in which case we would necessarily obtain a different combination of Borel residues appearing in \eqref{eq:generalexpressionforeigenvaluetunnelingtwoparameters}. Let us illustrate this conundrum when reaching for the $(2|1)$ (anti) eigenvalue configuration, along the two different resurgence-lattice-paths illustrated in figure~\ref{fig:transitioningfrom00to21}. Whereas we end up with the same sectors, we reach them with different combinations of Borel residues. For these two possibilities of figure~\ref{fig:transitioningfrom00to21}, we find the following results:
\begin{itemize}
\item Consider the motion depicted in figure~\ref{fig:transitioningfrom00to21a}, where we cross the $\theta=\pi$ and $\theta=0$ Stokes lines once. Notice how upon completion of the motion we are not required to complete a full circle; rather just make sure we start and end somewhat above\footnote{Let us stress here that when calculating Borel residues this choice of phase of $g_{\text{s}}$ is quite relevant. It will influence the numerical value of some Borel residues \cite{bssv22}. Here we made the standard choice in the literature \cite{bssv22}.} the positive real line. Further notice how this is the same motion as in figure~\ref{fig:resonantresurgentlatticeTwoAutomorphisms}, which was discussed in the main text. This results in the precise contents of \eqref{eq:generalexpressionforeigenvaluetunnelingtwoparameters-ell1},
\begin{align}
\label{eq:resurgenceeigenvaluetunneling21}
\rme^{-\frac{A}{g_{\text{s}}}}\, \mathsf{S}_{(0,0)\to(0,1)} \left( \mathsf{S}_{(0,1)\to(2,1)}\, \frac{Z^{(2,1)}}{Z^{(0,0)}} + \mathsf{S}_{(0,1)\to(1,0)}\, \frac{Z^{(1,0)}}{Z^{(0,0)}} \right) \simeq \frac{\mathcal{Z}^{(2|1)}}{\mathcal{Z}^{(0|0)}}.
\end{align}
\noindent
Note how the involved forward Borel residues represent two-step motions. 
\item Next, consider the motion depicted in figure~\ref{fig:transitioningfrom00to21b}, where we cross the $\theta=0$ forward discontinuity twice. This implies we are effectively performing three one-step jumps on the resurgence lattice. Further notice how we are now required to complete just above a full circle. We find the result
\begin{align}
\label{eq:resurgenteigenvaluetunnleing21onestepscopy}
\rme^{-\frac{A}{g_{\text{s}}}}\, \mathsf{S}_{(0, 0)\rightarrow (1,0)}\, \mathsf{S}_{(1,0)\rightarrow (1,1)} \left( \mathsf{S}_{(1,1)\to(2,1)}\, \frac{Z^{(2,1)}}{Z^{(0,0)}} + \mathsf{S}_{(1,1)\to(1,0)}\, \frac{Z^{(1,0)}}{Z^{(0,0)}} \right) \simeq 2\, \frac{\mathcal{Z}^{(2|1)}}{\mathcal{Z}^{(0|0)}}.
\end{align}
\noindent
Note how all involved Borel residues now correspond to single-step motions.
\end{itemize}

Consistency of the whole resurgence framework of course demands consistency of these two formulae. But extra structure emerges upon comparison of both results: as we shall see, in this case, the computation of the Stokes vector $\boldsymbol{S}_{(1,-1)}$ follows ``for free'', \textit{i.e.}, without any need to address the matrix-integrals in the above right-hand sides. Let us make the comparison in steps. First recall how the $(2,1)$ sector has logarithmic components dictated by \eqref{eq:formof21sector-2}. In this expression, its $(1,0)$ sector has a (starting genus) $\beta$-factor of $1/2$, which implies it is an asymptotic series in half-integer powers of $g_{\text{s}}$. The same holds true for the $(2,1)[0]$ sector. Next, let us clarify the difference between the $g_{\text{s}}$ rotation in \eqref{eq:resurgenceeigenvaluetunneling21} and in \eqref{eq:resurgenteigenvaluetunnleing21onestepscopy} (as illustrated in the top-plots of figure~\ref{fig:transitioningfrom00to21}).  In figure~\ref{fig:transitioningfrom00to21a} we almost completed a full circle, which we will denote as returning to $g_{\text{s}}$, whereas in figure~\ref{fig:transitioningfrom00to21b} we went slightly over a full circle, which we will denote as arriving at $g_{\text{s}}^+$. In light of what we just said concerning the structure of $(2,1)$ and $(1,0)$ sectors, this then translates to the relations
\begin{align}
Z^{(1,0)} \left(t, g_{\text{s}}^{+}\right) &= - Z^{(1,0)} \left(t, g_{\text{s}}\right),\\
Z^{(2,1)} \left(t, g_{\text{s}}^{+}\right) &= - Z^{(2,1)} \left(t, g_{\text{s}}\right) - 2\pi\rmi\alpha\, Z^{(1,0)} \left(t, g_{\text{s}}\right).
\end{align}
\noindent
If we further re-express the Borel residues in \eqref{eq:resurgenceeigenvaluetunneling21} and in \eqref{eq:resurgenteigenvaluetunnleing21onestepscopy} in terms of their Stokes data (using formulae \eqref{eq:BorelResidue01-11,11-01sector}-\eqref{eq:BorelResidue01-21,11-21sector}-\eqref{eq:BorelResidue01to10sector}) the consistency of these two equations (whose right-hand side is unchanged in the rotation) yields, upon comparison, the Stokes data relation
\begin{equation}
-\rmi\pi\alpha \left(\boldsymbol{S}_{(1,0)}\cdot\left[\begin{array}{c}
1\\
0
\end{array}\right]\right)\left(\boldsymbol{S}_{(1,0)}\cdot\left[\begin{array}{c}
2\\
1
\end{array}\right]\right) = \boldsymbol{S}_{(1,-1)}\cdot\left[\begin{array}{c}
1\\
0
\end{array}\right].
\end{equation}
\noindent
Using the known vectorial-structure of two-parameter, resonant Stokes data \cite{bssv22}, it follows
\begin{equation}
\label{eq:higherorderStokes}
\boldsymbol{S}_{(1,-1)} = -\frac{\rmi\pi\alpha}{2} \left(\boldsymbol{S}_{(1,0)}\cdot\left[\begin{array}{c}
1\\
0
\end{array}\right]\right)\left(\boldsymbol{S}_{(1,0)}\cdot\left[\begin{array}{c}
2\\
1
\end{array}\right]\right) \cdot \left[\begin{array}{c}
2\\
0
\end{array}\right].
\end{equation}
\noindent
The Stokes vector on the right-hand side is the well-known ``canonical'' coefficient.

\section{Data for Resonant Resurgent Transseries}\label{app:TransseriesData}

This appendix collects resurgent-transseries data for cubic \cite{msw08, kmr10} and quartic \cite{m08, asv11, csv15} matrix models, as well as for their common double-scaling limit towards the Painlev\'e~I equation \cite{msw07, gikm10, kmr10, asv11, bssv22}. These data were extensively used throughout section~\ref{sec:ExamplesAndChecks}.

\subsection*{Cubic Matrix Model Data}\label{app:CubicMMData}

The cubic matrix model \eqref{eq:partitionfunctionhermitianmatrix} has potential \eqref{eq:cubicPotential}, string equation \eqref{eq:cubicstringequationthooftlimit}, and its free energy follows from \eqref{eq:freeenergyfromstringequation} with transseries \eqref{eq:FTSforCUBIC}. Explicitly, as in \eqref{eq:transseriesexpansionGeneric}-\eqref{eq:transseriesexpansionGeneric-sectors}-\eqref{eq:phinm-beta-starting-g}, this is
\begin{align}
F (t, g_{\text{s}}; \sigma_1,\sigma_2) &= \sum\limits_{n=0}^{\infty} \sum\limits_{m=0}^{\infty} \sigma_1^n \sigma_2^m\, \text{e}^{-\left(n-m\right)\frac{A(t)}{g_{\text{s}}}}\, F^{(n,m)} (t, g_{\text{s}}), \\
F^{(n,m)}(t, g_{\text{s}}) &= \sum\limits_{k=0}^{\min(n, m)} F^{(n,m)[k]}(t, g_s)\, \log^k \frac{f(t)}{g_{\text{s}}^2}, \\
F^{(n,m)[k]}(t, g_{\text{s}}) &\simeq \sum\limits_{g=\beta_{nm}^{[k]}}^{\infty} F^{(n,m)[k]}_g (t)\, g_{\text{s}}^{g}.
\end{align}
\noindent
In these expressions, instanton action $A(t)$ and function $f(t)$ are determined via
\begin{equation}
2-5\lambda^2 r = \lambda^2 r\, \cosh \left(A^{\prime}(t)\right), \qquad f(t) = \frac{\left(1-3\lambda^2 r\right)^5}{108 \lambda^8 r^2},
\end{equation}
\noindent
where we defined $r$ in equation \eqref{eq:definitionrcubic}. The free-energy coefficients satisfy the symmetry \eqref{eq:Phinmk=alphaPhinm0},
\begin{align}
F^{(n,m)[k]}(t, g_\text{s}) = \frac{1}{k!}\, \Big( \alpha \left(n-m\right) \Big)^k\, F^{(n-k,m-k)[0]}(t, g_{\text{s}}) \quad \text{ with } \quad \alpha=\frac{1}{2},
\end{align}
\noindent
and a few values for the starting genus are given in table~\ref{tab:startinggenusCubicF}.

\begin{table}
\centering
\begin{tabular}{c|cccc}
\diagbox{$m$}{$n$} & 0 & $1$ & $2$ & $3$\\ \hline
0 & $-2$ & $1/2$ & $1$ & $3/2$\\
$1$ & $1/2$ & $-1$ & $3/2$ & $2$\\
$2$ & $1$ & $3/2$ & 0 & $3/2$\\
$3$ & $3/2$ & $2$ & $3/2$ & $1$\\
\end{tabular}
\caption{Starting genus $\beta_{n,m}^{[0]}$ for the free energy of the cubic matrix model.}
\label{tab:startinggenusCubicF}
\end{table}

Let us list a few such transseries coefficients. One first has to compute the solution to the string-equation transseries \eqref{eq:cubicstringequationthooftlimit}. For the perturbative sector one finds the first few coefficients
\begin{align}
R^{(0,0)}_1(t) &= 0, & R^{(0,0)}_2(t) &= \frac{\lambda^4 r \left(5-9\lambda^2 r\right)}{8 \left(1-3\lambda^2 r\right)^4}, \\
R^{(0,0)}_3(t) &= 0, & R^{(0,0)}_4(t) &= - \frac{3\lambda^8 r \left(\lambda^2 r \left(9\lambda^2 r \left(18\lambda^2 r+113\right)-1316\right)+385\right)}{128 \left(3\lambda^2 r-1\right)^9}.
\end{align}
\noindent
For the $(1,0)$ instanton sector, we find\footnote{Herein we have set the integration constant for the lowest $g_{\text{s}}$-order contribution to $1$, while we set the integration constants of higher-order contributions to $0$. These results may also be compared to the ones in \cite{msw08}, with the change of conventions:
\begin{align}
\sigma_1 &\mapsto \sqrt{g_{\text{s}}}\, \sigma_1 \left( 108 \lambda^2 g_{\text{s}}^2 \right)^{- \frac{1}{2} \sigma_1\sigma_2}, \\
\sigma_2 &\mapsto \sqrt{g_{\text{s}}}\, \sigma_2 \left( 108 \lambda^2 g_{\text{s}}^2 \right)^{\frac{1}{2} \sigma_1\sigma_2}. 
\end{align}
} for instance
\begin{align}
R^{(1,0)}_{1/2}(t) &= \frac{\sqrt{r}}{\sqrt[4]{1-3\lambda^2 r}}, & R^{(1,0)}_{3/2}(t) &= \frac{3\lambda^2 r \left(4-3\lambda^2 r\right)-8}{96\sqrt{r} \left(1-3\lambda^2 r\right)^{11/4}}, \\
R^{(0,1)}_{1/2}(t) &= \frac{\sqrt{r}}{\sqrt[4]{1-3\lambda^2 r}}, & R^{(0,1)}_{3/2}(t) &= -\frac{3\lambda^2 r \left(4-3\lambda^2 r\right)-8}{96 \sqrt{r} \left(1-3\lambda^2 r\right)^{11/4}}.   
\end{align}
\noindent
For the $(1,1)$ sector, one obtains the following coefficients
\begin{align}
R^{(1,1)}_1(t) &= \frac{2-3\lambda^2 r}{\left(1-3\lambda^2 r\right)^{3/2}}, & R^{(1,1)}_2(t) &= 0, \\
R^{(1,1)}_3(t) &= \frac{\lambda^4 \left(3\lambda^2 r \left(9\lambda^2 r-8\right) \left(243\lambda^2 r-94\right)+160\right)}{128 \left(1-3\lambda^2 r\right)^{13/2}}, & R^{(1,1)}_4(t) &= 0.
\end{align}
\noindent
Notice how every second coefficient above vanishes, which is consistent with the expected expansion. It is in the $(2,1)$ sector that logarithms appear for the first time, where we have
\begin{equation}
R^{(2,1)}_{-1}(t) = \frac{\sqrt{r} \left(-2 \log r + 5 \log \left(1-3\lambda^2 r\right)\right)}{2 \sqrt[4]{1-3\lambda^2 r}}.
\end{equation}
\noindent
For the free energy the exact same can be done. A few transseries coefficients are as follows
\begin{align}
F^{(1,1)}_{-1}(t) &= \frac{2 \sqrt{1-3\lambda^2 r}}{3\lambda^2} + & & \nonumber \\
& & & 
\hspace{-225pt}
+ r \sqrt{1-2\lambda^2 r} \left( \log r - \log \left(-5\lambda^2 r+2\sqrt{1-3\lambda^2 r} \sqrt{1-2\lambda^2 r} + 2\right) \right), \\ 
F^{(1,1)}_{0}(t) &= 0, & F^{(1,1)}_{1}(t) &= -\frac{63 \lambda ^4 r^2-60 \lambda ^2 r+8}{96 r \left(1-3 \lambda ^2 r\right)^{5/2}}, \\
F^{(1,0)}_{1/2} (t) &= \frac{\lambda^2 \sqrt{r}}{4\left(1-3\lambda^2 r\right)^{5/4}}, & F^{(0,1)}_{1/2} (t) &= \frac{\lambda^2 \sqrt{r}}{4\left(1-3\lambda^2 r\right)^{5/4}}, \\
F^{(2,1)[0]}_{1/2} (t) &= \frac{\lambda^2 \sqrt{r}}{8\left(1-3\lambda^2 r\right)^{5/4}}\, \log \frac{\left(1-3\lambda^2 r\right)^5}{108 r^2}. & &
\end{align}

Finally, let us list some Borel residues and Stokes data for the cubic matrix model computed via the double-scaling \eqref{eq:BorelResiduesCubicToPI}, and where for Painlev\'e~I we are collecting data from \cite{bssv22}. We have
\begin{align}
\mathsf{S}_{(0,0)\rightarrow (1,0)} &= -\frac{\rmi}{\sqrt{2\pi}}, & \mathsf{S}_{(0,0)\rightarrow (0,1)} &= -\frac{1}{\sqrt{2\pi}},\\
\mathsf{S}_{(1,1)\rightarrow (1,0)} &= -\rmi\, \sqrt{\frac{2}{\pi}} \left( \gamma_{\text{E}} + \log 96\sqrt{3} \right). & &
\end{align}
\noindent
We also have the following Stokes vectors
\begin{align}
\boldsymbol{S}_{(1,0)} &= \frac{\rmi}{\sqrt{2\pi}} \left[\begin{array}{c}
1\\
0
\end{array}\right], & \boldsymbol{S}_{(0,1)} &= \frac{1}{\sqrt{2\pi}}\left[\begin{array}{c}
0\\
1
\end{array}\right],\\
\boldsymbol{S}_{(0,-1)} &= \rmi\, \sqrt{\frac{1}{2\pi}} \left(\gamma_{\text{E}} + \log 96\sqrt{3} \right) \left[\begin{array}{c}
2\\
1
\end{array}\right], &
\boldsymbol{S}_{(1,-1)} &= \frac{\rmi}{4}\left[\begin{array}{c}
2\\
0
\end{array}\right].
\label{eq:higherstepstokescubic}
\end{align}

\subsection*{Quartic Matrix Model Data}\label{app:QuarticMMData}

The quartic matrix model \eqref{eq:partitionfunctionhermitianmatrix} has potential \eqref{eq:quarticPotential}, string equation \eqref{eq:quarticstringequationthooftlimit}, and its free energy follows from \eqref{eq:freeenergyfromstringequation} with transseries \eqref{eq:FTSforQUARTIC}. Explicitly, as in \eqref{eq:transseriesexpansionGeneric}-\eqref{eq:transseriesexpansionGeneric-sectors}-\eqref{eq:phinm-beta-starting-g}, it is exactly the same structure as we just illustrated for the cubic matrix model. The differences of course start with the instanton action $A(t)$ and function $f(t)$, which are now determined via
\begin{align}
\label{eq:quarticmatrixmodelinstantonactions}
A(t) &= \frac{1}{2\lambda} \left(3-3\lambda r\right)^{\frac{1}{2}} \left(3-\lambda r\right)^{\frac{1}{2}} - \frac{1}{2}\, r \left(2-\lambda r\right) \text{arccosh} \left(\frac{3-2\lambda r}{\lambda r}\right), \\
f(t) &= \frac{(3-\lambda r)^3 (3-3\lambda r)^5}{15552 \lambda^6 r^4},
\end{align}
\noindent
and where we defined $r$ in equation \eqref{eq:definitionrquartic}. The free-energy coefficients satisfy the symmetry \eqref{eq:Phinmk=alphaPhinm0} with $\alpha=\frac{\lambda}{12}$, and a few values for the starting genus are given in table~\ref{tab:startinggenusQuarticF}.

\begin{table}
\centering
\begin{tabular}{c|cccc}
\diagbox{$m$}{$n$} & 0 & $1$ & $2$ & $3$\\ \hline
0 & $-2$ & $1/2$ & $1$ & $3/2$\\
$1$ & $1/2$ & $-1$ & $3/2$ & $2$\\
$2$ & $1$ & $3/2$ & 0 & $3/2$\\
$3$ & $3/2$ & $2$ & $3/2$ & $1$\\
\end{tabular}
\caption{Starting genus $\beta_{n,m}^{[0]}$ for the free energy of the quartic matrix model.}
\label{tab:startinggenusQuarticF}
\end{table}

Listing a few transseries coefficients for the free energy, start with the $(1,0)$ instanton sector to find
\begin{align}
\label{eq:qdata10}
F^{(1,0)}_{1/2}(t) &= \frac{(\lambda r)^{3/2}}{2 r\left(3-3\lambda r\right)^{5/4}\left(3-\lambda r\right)^{1/4}}, \\
F^{(1,0)}_{3/2}(t) &= \frac{(\lambda r)^{3/2}\left(-27-\frac{675}{2}\lambda r+243\lambda^2 r^2-45\lambda^3 r^3\right)}{8 r^2 \left(3-3\lambda r\right)^{15/4}\left(3-\lambda r\right)^{7/4}}.
\end{align}
\noindent
For the $(1,1)$ and $(2,1)$ sectors (where $g_{\text{s}}$-logarithms first appear in the latter), we have
\begin{align}
\label{eq:qdata11}
F^{(1,1)}_{-1}(t) &= \frac{1}{6} \left( \sqrt{3-3\lambda r}\, \sqrt{3-\lambda r} - \lambda r \left(2-\lambda r\right) \log \left(\frac{3-2 r\lambda+\sqrt{3-3\lambda r}\, \sqrt{3-\lambda r}}{\lambda r} \right) \right), \\
F^{(1,1)}_{1}(t) &= \frac{\lambda\left(-9+\frac{63}{2}\lambda r-18\lambda^2 r^2+3\lambda^3 r^3\right)}{4 r \left(3-3\lambda r\right)^{5/2}\left(3-\lambda r\right)^{3/2}}, \\
F^{(2,1)[0]}_{1/2}(t) &= \frac{\lambda (\lambda r)^{3/2}}{24 r\left(3-3\lambda r\right)^{5/4} \left(3-\lambda r\right)^{1/4}}\, \log \frac{f(t)}{g_{\text{s}}^2}.
\end{align}
\noindent
For the partition function the exact same can be done. Recall from subsection~\ref{subsec:resurgenttranssriesandresonance} how not to deal with additional factors of $\exp \frac{1}{g_{\text{s}}^2} F_{-2}^{(0,0)}$ we will normalize by the perturbative sector. Listing the lowest $g_{\text{s}}$-order terms, we find
\begin{align}
\frac{Z^{(1, 0)}}{Z^{(0, 0)}}\Bigg|_{1/2} &= \frac{\lambda^{3/2}\sqrt{r}}{2\left(3-3\lambda r\right)^{5/4}\left(3-\lambda r\right)^{1/4}}, & \frac{Z^{(2, 0)}}{Z^{(0, 0)}}\Bigg|_{1} &= \frac{\lambda^4 r^2}{8\left(3-3\lambda r\right)^{5/2}\left(3-\lambda r\right)^{3/2}}, \\
\frac{Z^{(3, 0)}}{Z^{(0, 0)}}\Bigg|_{5/2} &= -\frac{3 \lambda^{17/2}r^{9/2}}{64\left(3-\lambda r\right)^{13/4}\left(3-3\lambda 3\right)^{25/4}}, & \frac{Z^{(4, 0)}}{Z^{(0, 0)}}\Bigg|_{4} &= \frac{9 \lambda^{14} r^8}{1024\left(3-\lambda r\right)^{6}\left(3-3\lambda r\right)^{10}}, \\
\frac{Z^{(5, 0)}}{Z^{(0, 0)}}\Bigg|_{13/2} &= \frac{81 \lambda^{43/2}r^{25/2}}{16384\left(3-\lambda r\right)^{37/4}\left(3-3\lambda r\right)^{65/4}}, & \frac{Z^{(6, 0)}}{Z^{(0, 0)}}\Bigg|_{9} &= \frac{729 \lambda^{30} r^{18}}{524288\left(3-\lambda r\right)^{27/2}\left(3-3\lambda r\right)^{45/2}}.
\end{align}
\noindent
Entries along the backward direction easily follow from the above ones, \textit{e.g.},
\begin{align}
\frac{Z^{(0, 1)}}{Z^{(0, 0)}}\Bigg|_{1/2} = \frac{Z^{(1, 0)}}{Z^{(0, 0)}}\Bigg|_{1/2}, \quad \frac{Z^{(0, 2)}}{Z^{(0, 0)}}\Bigg|_{1} = \frac{Z^{(2, 0)}}{Z^{(0, 0)}}\Bigg|_{1}, \quad 
\frac{Z^{(0, 3)}}{Z^{(0, 0)}}\Bigg|_{5/2} = -\frac{Z^{(3, 0)}}{Z^{(0, 0)}}\Bigg|_{5/2}, \quad \frac{Z^{(0, 4)}}{Z^{(0, 0)}}\Bigg|_{4} = \frac{Z^{(4, 0)}}{Z^{(0, 0)}}\Bigg|_{4}.
\end{align}

Finally, let us list some Borel residues and Stokes data for the quartic matrix model, in consistency with the Painlev\'e~I double-scaling limit from \cite{bssv22}. We have
\begin{align}
\mathsf{S}_{(0,0)\rightarrow (1,0)} &= -\rmi\,\sqrt{\frac{3}{\pi\lambda}}, & \mathsf{S}_{(0,0)\rightarrow (0,1)} &= -\sqrt{\frac{3}{\pi\lambda}}, \\
\mathsf{S}_{(1,1)\rightarrow (1,0)} &= -\rmi\sqrt{\frac{\lambda}{3\pi}} \left( \gamma_{\text{E}} + \log 96\sqrt{3} \right). & &
\end{align}
\noindent
We also have the following Stokes vectors
\begin{align}
\boldsymbol{S}_{(1,0)} &= \rmi\,\sqrt{\frac{3}{\pi\lambda}}\left[\begin{array}{c}
1\\
0
\end{array}\right], & 
\boldsymbol{S}_{(0,1)} &= \sqrt{\frac{3}{\pi\lambda}}\left[\begin{array}{c}
0\\
1
\end{array}\right],\\
\boldsymbol{S}_{(0,-1)} &= \rmi\, \sqrt{\frac{\lambda}{12\pi}} \left( \gamma_{\text{E}} + \log 96\sqrt{3} \right)\left[\begin{array}{c}
2\\
1
\end{array}\right], & 
\boldsymbol{S}_{(1,-1)} &= \frac{\rmi}{4}\left[\begin{array}{c}
2\\
0
\end{array}\right].
\label{eq:higherstepStokesquartic}
\end{align}

\subsection*{Double-Scaling Limit Data}\label{app:MinimalStringsData}

Finally, let us address the double-scaling limit of cubic and quartic matrix models towards the Painlev\'e~I equation. This equation is \eqref{eq:PainleveI}, and its two-parameter resonant transseries solution---for the so-called specific heat $u(\kappa)$---has been studied at length in the literature; \textit{e.g.}, \cite{gikm10, asv11, bssv22}. The free energy follows from the specific heat by double integration,
\begin{equation}
F^{\prime\prime}(\kappa) = -u(\kappa).
\end{equation}
\noindent
With the convenient labeling $x=\kappa^{-5/4}$, the transseries for the free energy reads
\begin{equation}
F(x;\sigma_1,\sigma_2) = - \frac{4}{15 x^2} + \left(\frac{1}{60}+\frac{2}{3}\sigma_1^2\sigma_2^2\right) \log x + \sum\limits_{n=0}^{\infty} \sum\limits_{m=0}^{\infty} \sigma_1^{n} \sigma_2^{m}\, \rme^{-\left(n-m\right)\frac{A}{x}}\, F^{(n,m)}(x),
\end{equation}
\noindent
where the instanton action is $A=\frac{8\sqrt{3}}{5}$ and where we have the following sectorial logarithmic dependence
\begin{equation}
F^{(n,m)}(x) = \sum\limits_{k=0}^{\min(n,m)} F^{(n,m)[k]}(x)\, \log^k x.
\end{equation}
\noindent
These logarithmic sectors are of course related to each other, via
\begin{equation}
F^{(n,m)[k]}(x) = \frac{1}{k!} \left(-\frac{2\left(n-m\right)}{\sqrt{3}}\right)^k F^{(n-k,m-k)[0]}(x),
\end{equation}
\noindent
with an extra ``backward-forward'' symmetry which in the lowest sector reads
\begin{equation}
F^{(n,m)[0]}_{g} (x) = (-1)^{g+\lfloor n/2 \rfloor}\, F^{(m,n)[0]}_{g} (x), \qquad n>m.
\end{equation}
\noindent
Each sector is an asymptotic divergent series of the form
\begin{equation}
F^{(n,m)[0]}(x) \simeq \sum\limits_{g=0}^{\infty} F_{g}^{(n,m)[k]}\, x^{g+\beta_{nm}}, \qquad \beta_{nm} = \frac{n+m}{2} - \left\lfloor\frac{\min(n,m)-m\,\delta_{n,m}}{2}\right\rfloor.
\end{equation}

Listing a few free-energy transseries coefficients, we have for the perturbative sector
\begin{equation}
F^{(0,0)}(x) = \frac{7}{5760}\, x^2 + \frac{245}{331776}\, x^4 + \cdots.
\end{equation}
\noindent
For nonperturbative sectors, we list a few
\begin{align}
F^{(1,0)[0]}(x) &= -\frac{1}{12}\, x^{1/2} + \frac{37}{768\sqrt{3}}\, x^{3/2} + \cdots, \\
F^{(2,0)[0]}(x) &= -\frac{1}{288}\, x + \frac{109}{27648\sqrt{3}}\, x^2 + \cdots, \\
F^{(2,1)[0]}(x) &= -\frac{71}{864}\, x^{3/2} + \frac{2999}{18432\sqrt{3}}\, x^{5/2} + \cdots, \\
F^{(1,1)[0]}(x) &= \frac{16}{5}\, \frac{1}{x} + \frac{5}{96}\, x + \frac{15827}{1474560}\, x^3 + \cdots, \\
F^{(2,2)[0]}(x) &= \frac{1555}{20736}\, x^2 + \frac{5288521}{95551488}\, x^4 + \cdots .
\end{align}

\newpage

\bibliographystyle{plain}

\end{document}